\documentclass[12pt]{article}
\usepackage{amsmath}
\usepackage{amssymb}
\usepackage{euscript}

\usepackage{epsfig}

\usepackage{cite}

\usepackage{alltt}

\usepackage{epic}

\def\Order#1{${\cal O}(#1$)}

\def\bbeta{\bar{\beta}}

\newcommand{\Rcal}{{\cal R}}
\newcommand{\Rpar}{\Rcal_{\wp}}

\begin{document}                     

\allowdisplaybreaks

\begin{titlepage}
\begin{flushright}
{\bf  CERN-TH/2001-040\\
      UTHEP-01-0102}
\end{flushright}

\begin{center}
{\Large
The Monte Carlo Program {\tt KoralW} version {\tt 1.51} and \\
The Concurrent Monte Carlo {\tt KoralW$\&$YFSWW3}\\
with All Background Graphs and \\ 
First-Order Corrections to $W$-Pair Production$^\dag$
}
\end{center}

\begin{center}
  {\bf   S. Jadach$^{a,b}$,}
  {\bf   W. P\l{}aczek$^{e,a}$,}
  {\bf   M. Skrzypek$^{b,a}$,}
  {\bf   B.F.L. Ward$^{c,d,a}$}
  {\em and}
  {\bf   Z. W\c{a}s$^{b,a}$ }
{\em $^a$CERN, Theory Division, CH-1211 Geneva 23, Switzerland,}\\
{\em $^b$Institute of Nuclear Physics,
  ul. Kawiory 26a, 30-055 Cracow, Poland,}\\
{\em $^c$Department of Physics and Astronomy,\\
  The University of Tennessee, Knoxville, TN 37996-1200, USA}\\
{\em $^d$SLAC, Stanford University, Stanford, CA 94309, USA}\\
{\em $^e$  Institute of Computer Science, Jagellonian University,\\
        ul. Nawojki 11, Cracow, Poland}
\end{center}

\begin{abstract}
The version {\tt 1.51} of the Monte Carlo (MC) program {\tt KoralW}
for all $e^+e^-\to f_1\bar f_2 f_3\bar f_4$ processes is presented.
The most important change from the previous version {\tt 1.42} 
is the facility for
writing MC events on the mass storage device
and reprocessing them later on.
In the reprocessing parameters
of the Standard Model may be modified
in order to fit them to experimental data.
Another important new feature is the possibility
of including complete ${\cal O}(\alpha)$ corrections
to double-resonant $W$-pair component-processes in addition to all
background (non-$WW$) graphs.
The inclusion 
is done with the help of the {\tt YFSWW3} 
MC event generator
for fully exclusive differential distributions (event per event).
Technically, it is done in such a way that {\tt YFSWW3} runs
concurrently with {\tt KoralW} as a separate slave process,
reading momenta of the MC event generated by {\tt KoralW} and returning 
the correction weight to {\tt KoralW}.
The latter introduces the ${\cal O}(\alpha)$ correction using this weight,
and finishes processing the event (rejection due to total MC weight, hadronization,
etc.).
The communication between KoralW and {\tt YFSWW3}
is done with the help of the FIFO facility of the UNIX/Linux operating system.
This does not require any 
modifications of the FORTRAN source codes. 
From the user's point of view, the resulting Concurrent MC event
generator {\tt KoralW$\&$YFSWW3}
looks as a regular single MC event generator
with all the standard features.
\end{abstract}

\vspace{-1mm}
\begin{center}
{\it To be submitted to Computer Physics Communications}
\end{center}

\footnoterule
\noindent
{\footnotesize
\begin{itemize}
\item[${\dag}$]
Work supported in part by the Polish Government grant KBN 5P03B09320,
the US DoE contracts DE-FG05-91ER40627 and DE-AC03-76SF00515,
the European Commission 5-th framework contract HPRN-CT-2000-00149,
NATO grant ST.CLG.977761
and the Polish-French Collaboration within IN2P3 through LAPP Annecy.
\end{itemize}
}
\begin{flushleft}
{\bf CERN-TH/2001-040\\
     April 2001 
}
\end{flushleft}
\end{titlepage}
\tableofcontents
\newpage

\noindent{\bf NEW VERSION SUMMARY}
\vspace{10pt}

\noindent{\sl Title of the program:} {\tt KoralW}, version {\tt 1.51}.

\vskip 0.2cm
\noindent{\sl Reference to original program:}
Comput. Phys. Commun. {\bf 94} (1996) 215; {\bf 119} (1999) 272


\vskip 0.2cm
\noindent{\sl Computer:} 
any computer with the FORTRAN 77 compiler under UNIX or Linux 
operating system

\vskip 0.2cm
\noindent{\sl Operating system:}
UNIX, Linux version 6.x and 7.x

\vskip 0.2cm
\noindent{\sl Programming language used:}
FORTRAN 77

\vskip 0.2cm
\noindent{\sl High-speed storage required:}  $<$ 25 MB





\vskip 0.2cm
\noindent{\sl Nature of the physical problem:}\\
The precise study of $W$-pair production and decay at LEP2 
requires both non-double-resonant and {${\cal O}(\alpha)$} corrections. 
So far each of these corrections is available as a separate Monte Carlo
program and there is no Monte Carlo that could simulate in a complete
way both
effects at the same time.
Such a MC event generator would be of importance for example 
for apparatus simulations  or Monte Carlo based fits.
The previous version of {\tt KoralW} \cite{koralw:1998} 
included {\em all} non-double-resonant
corrections to {\em all} double-resonant four-fermion processes in $e^+e^-$
annihilation. The present version {\tt 1.51} allows, for the first time ever,
the inclusion on an {\em event-per-event basis} of the respective {${\cal
O}(\alpha)$} corrections generated at the same time by the independently
running Monte Carlo program {\tt YFSWW3} \cite{yfsww3:2001},
as well as a reweighting of any earlier generated events with modified 
parameter sets.

\vskip 0.2cm
\noindent{\sl Method of solution:}\\
The Monte Carlo method used to simulate the all four-fermion final-state
processes in the $e^+e^-$ collisions in the presence of multiphoton
initial-state radiation has not changed since version 1.42 \cite{koralw:1998}.
Adding the {${\cal
O}(\alpha)$} corrections generated by {\tt YFSWW3} is done at the level
of the UNIX/Linux operating system with the help of the FIFO mechanism
(``named pipes''). 

\vskip 0.2cm
\noindent{\sl Restrictions on the complexity of the problem:}\\
For {\tt KoralW} as in version 1.42 \cite{koralw:1998}; for {\tt
KoralW$\&$YFSWW3} as in \cite{koralw:1998} and \cite{yfsww3:2001}.

\vskip 0.2cm
\noindent{\sl Typical running time:}\\
%
Approximate times on a PC Pentium III @ 800 MHz for cuts as described in
this article:\\
5 minutes per 1000 constant-weight CCall events ({\tt KoralW} stand-alone)\\
50 minutes  per 1000 constant-weight CCall events with the {${\cal O}(\alpha)$}
    correction (the CMC {\tt KoralW$\&$YFSWW3}, max.\ weight for rejection
increased by a factor 2).

\newpage

\section{Introduction}

After many years of fruitful
operation, the LEP experiments have been closed down and the LEP2 data
analysis is approaching its final stage. 
In the area of $W$-pair physics the experimental precision is very high:
in order to match it,  
theoretical calculations must include not only tree-level four-fermion
Born contributions, 
with the numerically leading higher-order effects (mostly QED), 
but also the complete {${\cal O}(\alpha)$}
electroweak (EW) corrections to $W$-pair production \cite{LEP2YR:2000}.
This applies not only to inclusive quantities such as the total cross section
but also to various differential distributions, such as the angular or
invariant mass ones.

To date there is, however, no single Monte Carlo (MC) event generator
that would include simultaneously the complete
four-fermion background for massive fermions
and the {${\cal O}(\alpha)$} EW corrections to $W$-pair mediated
processes in {\em all} possible $W$-pair decay channels. 
For example {\tt KoralW} 
\cite{koralw:1995a,koralw:1995b,koralw:1998,skrzypek:2000}
can generate {\em all} four-fermion final states with the fully massive 
phase space and the complete Born-level four-fermion massive matrix
element generated by 
the GRACE2 package \cite{GRACE2}.
Apart from {\tt KoralW} there exists a number of other MC
programs for all four-fermion processes at the Born level
\cite{alpha,Accomando:1997es,wwf,wwgenpv,Montagna:2000ev,%
grc4f,Pukhov:1999gg,Kanaki:2000ms,%
lpww02,wopper,Berends:2000fj,erato:1997}.
The complete ${\cal O}(\alpha)$
corrections to the signal $e^+e^-\to W^+W^-\to 4f$ process
are implemented only in two MC programs:
{\tt YFSWW3} 
\cite{yfsww2:1996,yfsww3:1998,yfsww3:1998b,yfsww3:2000a,yfsww3:2001}
and {\tt RacoonWW}
\cite{Denner:2000bm,Denner:2000bj}.
{\tt YFSWW3} includes the library of electroweak corrections from 
Refs.~\cite{fleischer:1989,kolodziej:1991,fleischer:1993,fleischer:1995}. 
The {\tt RacoonWW} program, in addition to ${\cal O}(\alpha)$ corrections
to the $WW$ process, can also calculate the four-fermion corrections in
the massless fermion approximation and single, hard, non-collinear, 
real photon radiation  
in all four-fermion processes.
The massless fermion approximation prevents {\tt RacoonWW} from being
fully exclusive%
\footnote{The zero-fermion-mass approximation enforces the use of the
  inclusive treatment of the fermion and collinear photons (structure 
  functions). 
  This is experimentally not realistic for the final states with muons
  or soft electrons.}.
Also, {\tt RacoonWW} seems to still have some problem with providing 
constant-weight events in the full operational mode.

Concerning the efficient use of the {\tt KoralW} and {\tt YFSWW3} MC 
event generators,
the critical open question is:
How is it possible to combine their results,
so that for every interesting physical
observable we get a prediction that includes
the complete ${\cal O}(\alpha)$ Standard Model (SM) corrections 
for the $W$-pair production and decay process,
keeping sufficient control on the smaller contributions from the 
``background diagrams''?
Before we answer this question, let us mention important practical limitations
and requirements.
For the purpose of the LEP2 data analysis it is of paramount importance that 
the results of {\tt YFSWW3} and {\tt KoralW} are combined for the 
{\em fully exclusive}
distributions, in other words on an {\em event-per-event} basis.
It is not sufficient to combine the predictions of two separate MC runs of
{\tt YFSWW3} and {\tt KoralW} programs for {\em inclusive}
observables such as an integrated cross section, asymmetries, 
single-dimensional angular or $W$-mass distributions.
Such a procedure is not sufficient for full detector simulations and
for data analysis, in which the important SM parameter, the mass of the $W$,
is fitted to experimental data using a series of the (fully exclusive)
MC events!

Apparently, we are asking whether the {\tt KoralW} and {\tt YFSWW3} 
MC programs could be merged into a single new MC event generator,
that is a single MC program with a single source code 
and a single executable object in the machine processor.
In principle, this could be done, but not within the time  
left for the LEP2 data analysis.
Nevertheless, the situation is not completely hopeless and
there seem to be some sensible ways out.
One possible solution is to combine {\tt KoralW} and {\tt YFSWW3} into 
a single tool
using events stored on the mass data storage, which we shall call 
a ``disk file'' or simply a ``disk''.
Storing events is done routinely for the purpose of the data analysis anyway.
In this scenario, constant-weight
events generated with {\tt KoralW} are stored on the disk and  later on
read by {\tt YFSWW3} and finally corrected for the missing 
${\cal O}(\alpha)$ terms
with the help of a special correction weight.
The resulting events would be variable-weight (weighted) events.
This kind of organization is not completely trivial and requires certain
``tuning'' of both programs; see below for the details.
Note also that; for the purpose of fitting the $W$ mass, the events 
generated by {\tt KoralW}
and stored on the disk can be corrected in a similar way by {\tt KoralW},
with the weight corresponding to a change of the mass of $W$,
or due to any change of the other SM parameters (any change of the input data 
of {\tt KoralW}).
Coming back to the above procedure of combining {\tt KoralW} and {\tt YFSWW3},
we see two important disadvantages:
(i) running two separate MC programs that communicate through a disk file is 
inconvenient and 
(ii) the correction weight of {\tt YFSWW3} may have a long tail, so that
it would be  difficult or impossible to produce constant-weight 
(unweighted) events
through a rejection technique.

In this article we present another solution to the above
problems and the corresponding MC tool, based on the 
{\tt KoralW} and {\tt YFSWW3} MC programs, which is able to provide
constant-weight events,
implements the ${\cal O}(\alpha)$ corrections
for $W$-pair production process and includes all of the background diagrams.
The present new version {\tt 1.51} of the {\tt KoralW} program
provides a programming framework for this new solution.
Contrary to the previous solution where {\tt KoralW} and {\tt YFSWW3}
were communicating through a disk file,
here, variable-weight events from {\tt KoralW}
are sent immediately, in real time, as an input to {\tt YFSWW3},
using the ``named pipe'' of the FIFO mechanism in the UNIX/Linux 
operating system.
{\tt YFSWW3} calculates the ${\cal O}(\alpha)$ correction weight and sends 
it back to  {\tt KoralW} with the help of another ``named pipe'' of the FIFO.
Afterward, {\tt KoralW} performs the final rejection according to the total 
MC weight and invokes hadronization, etc.

The important advantage of the above method is that it provides the
constant-weight events with the ${\cal O}(\alpha)$ corrected $W$-pair
process quite efficiently,
including all the background diagrams, the higher-order ISR corrections, 
the hadronization, etc.
For the FIFO-based solution,
as compared with the disk-mediated solution mentioned earlier,
{\em no} additional modifications
of the FORTRAN source codes of both programs are necessary. 
Since {\tt KoralW} and {\tt YFSWW3} run as two separate, 
concurrent processes, which communicate with one another, we call this solution
a ``Concurrent Monte Carlo (CMC)  {\tt KoralW$\&$YFSWW3}''.
From the user's point of view, it acts like a single MC program.
To our knowledge, this could be the first important practical application,
albeit rather simple,
of the concept of ``concurrency'' in the area of the high energy physics
Monte Carlo event generators.
We shall also discuss very briefly possible future extensions/improvements 
of the above technique.

The modifications of the {\tt YFSWW3} program necessary for this technique
are discussed in detail in \cite{yfsww3:2001} and in this paper we shall 
describe them only to a minimum necessary extent.

The second group of modifications included in {\tt KoralW} version {\tt 1.51} 
is motivated by the use of {\tt KoralW} to study the background to
two-fermion processes due to the emission of a second fermion pair. 
In short, the modifications provide a number
of approximate matrix elements,
denoted according to ref.~\cite{2f-LEP2YR:2000} as ISNS, FSNS, etc.,
and a new ``extrapolation procedure'' better suited for 
the $t$-channel-dominated
photonic radiation. 

The layout of the paper is as follows. 
In Section 2, we discuss various ways of merging {\tt KoralW} with
{\tt YFSWW3}. 
In particular we show how to do it by means of the
FIFO (``named pipes'') mechanism and discuss how to reweight previously
generated events from tapes. 
In the Section 3, we provide some numerical tests of the CMC {\tt
KoralW$\&$YFSWW3}.
In Section 4, we describe 
in detail all modifications of {\tt KoralW} version {\tt 1.51} related 
to the $WW$ physics and the reweighting procedures, the
two-fermion physics and miscellaneous topics, respectively.
In Section 5, we explain how to 
install the version {\tt 1.51} of {\tt KoralW} and
in Section 6 we describe briefly the organization of the source code.
In Section 7, we describe in detail various demo programs included in 
the package, with special emphasis on the practical use of 
the FIFO (``named pipes'') mechanism
and the construction of the Concurrent Monte Carlo.
We summarize the paper in Section 8. In an appendix, we describe
new and modified program parameters.

\section{Concurrent merge of {\tt KoralW} and {\tt YFSWW3}}

In this section we shall describe in detail the method of 
combining results of the MC event generators {\tt KoralW} and {\tt YFSWW3}
at the level of the {\em fully exclusive} differential distribution,
such that the resulting distribution features the ${\cal O}(\alpha)$ 
corrections
for the $WW$ signal process and  the background graphs of 
the four-fermion process (with the ISR corrections).
As already indicated, this can be done either by using a series of MC events 
stored on the disk
or through the concurrent use of {\tt KoralW} and {\tt YFSWW3},
which effectively act together as a single MC event generator.
In both scenarios
the underlying methodology of constructing MC correction weights is the same.
It will be described in detail in the following.

Let us remind the reader that
{\tt KoralW} is a dedicated MC event generator with a powerful
four-fermion phase-space generator capable of generating every possible 
four-fermion final state in the complete phase space for massive fermions 
(including electrons)
\cite{koralw:1995a,koralw:1995b,koralw:1998,skrzypek:2000}
with the importance sampling due to all possible singularities
in the Feynman diagrams.
{\tt KoralW} uses the exact massive Born-level matrix element generated 
by the GRACE2 package \cite{GRACE2}.

On the other hand, the {\tt YFSWW3} MC event generator
\cite{yfsww3:2001,yfsww2:1996,yfsww3:1998,yfsww3:1998b,yfsww3:2000a}
is the MC program dedicated to the $W$-pair production and decay process.
It includes the complete ${\cal O}(\alpha)$ EW library of real and virtual 
corrections to the $W$-pair production process
of Refs.~\cite{fleischer:1989,kolodziej:1991,fleischer:1993,fleischer:1995},
along with the multiple photon radiation from the $W$-pair (WSR). 
In the following we will often use the notation Non-Leading (NL) 
${\cal O}(\alpha)$ corrections to denote the
 remaining part of the ${\cal O}(\alpha)$ EW
corrections after subtraction of the ``trivial'' 
leading universal corrections: Initial State Radiation (ISR) and Coulomb
correction (Cc).

Certain features of  {\tt KoralW} and {\tt YFSWW3} are critical on
the possibility of combining their results
for fully exclusive differential distributions.
The most important is that both programs do implement well-defined,
fully exclusive, distributions for four final-state fermions and $n$ photons,
normalized with respect to the standard Lorentz-invariant phase space (LIPS)
\begin{equation}
  d\Phi_{4+n}(P;q_1,q_2,q_3,q_4,k_1,\dots,k_n),
\end{equation}
as defined in the PDG~\cite{PDG:2000}.
The four-momenta of the final-state fermions are $q^\mu_i, i=1,...,4$ and
of the photons are $k^\mu_j$, $j=1,\dots,n$.
In spite of the fact that both programs use the leading-logarithmic (LL) 
models for the higher-order
ISR and the final-state radiation (FSR),
they do not employ an inclusive approach in which the collinear photon
is irreversibly associated with the parent fermion (structure-function 
approach).
The massive kinematics of all fermions allows for full coverage of the phase space.
It is also helpful that both programs implement the same CC03 matrix element
in the 't~Hooft--Feynman gauge, which
coincides with the gauge-invariant Leading Pole Approximation (LPA) of the
complete four-fermion Born-level matrix element and has, therefore,
a well defined physical meaning, see \cite{yfsww3:2000a,yfsww3:2001} for 
more discussion.
The ${\cal O}(\alpha)$ NL effect comes as a correction to the above 
CC03 distribution; see below for more details.

Even before the advent of the CMC {\tt KoralW$\&$YFSWW3},
both programs (in their unpublished versions) 
gradually acquired the capability
of writing the four-momenta of generated events, together with some auxiliary 
information,
to an external device and reading them back in order to calculate 
the correction weight.
The correction weight was that corresponding to the modification of
the fully exclusive differential distribution:
\begin{equation}
  \rho(q_1,q_2,q_3,q_4,k_1,\dots,k_n)
  = d\sigma / d\Phi_{4+n}(P;q_1,q_2,q_3,q_4,k_1,....k_n),
\end{equation}
due to a change of the input parameters of the MC event generators.
The most important change was due to a variation of the input $W$-mass
and was instrumental in the fitting of the $W$-mass to the LEP experimental 
data using a series of the MC events from {\tt KoralW} or {\tt YFSWW3}.
It is quite likely that
this kind of facility will be a standard feature for any future MC event 
generator aimed at precision measurements in the future experiments.
The analysis of the LEP data points out in that direction for future 
developments.

As indicated in the Introduction, the next non-trivial step was to
provide one of the programs, {\tt KoralW} or {\tt YFSWW3}, with the capability 
of correcting the fully exclusive distributions using the correction weight
calculated by another program.
In practice, it has turned out that correcting the events produced with 
{\tt KoralW} 
using the correction weight of {\tt YFSWW3} leads to a lower rejection rate
for the final MC weight than the other way around, 
i.e.\ correcting the events of {\tt YFSWW3} with the weight of {\tt KoralW}.
The reason is that {\tt KoralW} has a better importance sampling for 
the background processes (some of them dominated by the $t$-channel exchange).
{\tt KoralW} does not have a sufficiently good
importance sampling for WSR. However, this seems to matter less
than the lack of importance sampling for the background processes,
particularly the ones with electron(s) in the final state.
The main difficulty in developing the above cross-correcting
capabilities of one MC program by another
was to match correctly the relative normalizations of the fully
exclusive differential distributions in both programs in order to define
the correction weight properly; see below.

Finally, after both programs have evolved to acquire subprograms
for reading and writing events on the external device and calculating the
properly normalized
correction weights for use by the same or other programs,
it was a purely technical exercise to organize both programs in such a way
that cross-correcting one program by another
could be done ``in flight'' by running {\em simultaneously } 
{\tt KoralW} and {\tt YFSWW3}, as two independent concurrent processes.
The communication between concurrent processes was organized
under UNIX/Linux quite easily with the help of the standard FIFO facility 
of the ``named pipes'',
using the already existing subprograms for reading/writing on/from 
the external device.
In this way, a new solution has emerged, which we call the Concurrent Monte
Carlo (CMC) {\tt KoralW$\&$YFSWW3}.
From the users point of view it acts as a single Monte Carlo
program with all its regular features.
The main advantage of the CMC {\tt KoralW$\&$YFSWW3} with respect to the 
previous
solution (with disk files) is that it can produce efficiently the 
constant-weight MC events.
These events can be fed into a detector MC simulation program%
\footnote{
  It is very important to provide constant-weight events as an input for 
  the full detector simulation, because the detector simulation is
  very slow and the detector-simulated events are rather voluminous.
  It would be rather wasteful, in terms of CPU time and data storage, 
  to feed into the detector simulation the variable-weight events
  from the physics MC event generator.}
and stored on the disk.
There is no problem with applying to these events a correction weight due
to a change of the $W$-mass or other SM parameters. 

In the first part of this section, we shall discuss the theoretical
foundations and requirements for this kind of a merge of two
different programs. 
Then, we shall go into the details of the technical realization of the 
CMC {\tt KoralW$\&$YFSWW3} solution.
The actual modifications of the {\tt KoralW} will be presented in detail
in one of the later sections.

\subsection{Correction weights -- theoretical discussion}
In the following we shall discuss the physics meaning of the 
correction weights.
After introducing our basic notation and terminology we shall briefly characterize
fully exclusive differential distributions in {\tt KoralW} and {\tt YFSWW3} and
define a common reference differential distribution with which
we shall define correction weights in both programs.

\subsubsection{Notation and terminology}
In the following considerations we shall use the differentials  $d\sigma$,
which we always understand as the fully exclusive
differential distributions normalized to the $(n+4)$-particle 
Lorentz-invariant phase space
\begin{equation}
  d\sigma = \rho(q_1,...,q_4,k_1,...,k_n)\; d \Phi_{n+4}(p_1+p_2;q_1,...,q_4,k_1,...,k_n).
\end{equation}

In the following we shall need a clearly defined terminology for the various
contributions and (perturbative) corrections in $d\sigma$.
We shall often use objects like $d\sigma_{Y_a}^{{\cal O}(\alpha)+ISR_{23}}$.
What is the meaning of the symbols used as superscripts and subscripts?
Here is the complete list, with explanations:
\begin{itemize}
\item $K$ and $Y$: 
  Denotes the origin from the MC program: {\tt KoralW} and {\tt YFSWW3}, 
  respectively.
\item CC03:
  In {\tt YFSWW3}, we implement the Leading Pole Approximation (LPA) 
  to define the double-resonant, 
  $WW$, component of the $e^-e^+\to 4f$ process;
  in particular CC03 denotes here the tree-level (Born) part of LPA,
  which coincides with the CC03 matrix element in the 't~Hooft--Feynman
  gauge.
  We shall sometimes use $Y_a$ or CC03$_a$
    to underline that we use {\tt YFSWW3} with the version LPA$_a$,
    see Refs.~\cite{yfsww3:2000a,yfsww3:2001}.
\item ${\cal O}(\alpha)$:
  Denotes the LPA with the complete on-shell ${\cal O}(\alpha)$ corrections 
  for the $e^-e^+\to W^-W^+$ process;
  we understand that this includes the CC03 tree-level, one-loop ${\cal O}(\alpha)$
  virtual corrections and the exact QED matrix element for photon emission from $WW$ (WSR).
  The one-loop EW 
    corrections to decays
    are included in the  present version of {\tt YFSWW3} as an overall
    factor and real photon emission is implemented 
    using the approximate treatment 
    of {\tt PHOTOS}~\cite{photos:1994}.
\item $ISR_{23}$:
  This means that we include the \Order{\alpha^2} and \Order{\alpha^3}
  {\em missing} QED ISR correction 
  (which has not been already included through exponentiation);
  it is always understood that it is done in the LL approximation;
  a similar subscript like $0123$ is self-explanatory;
  in particular subscript $0$ means ${\cal O}(\alpha^0)$ exponentiation
  (the Born-level matrix element convoluted with the photon emission in the 
   soft-photon approximation).
\item $Cc$:
  This means that, close to the $WW$-threshold, we include properly the 
  non-relativistic Coulomb effect,
  which far from the threshold is matched correctly with the ${\cal O}(\alpha)$ WSR.
  If not stated otherwise, we use a variant of $Cc$ with the ``screening''
    of Ref.~\cite{Chapovsky:1999kv},
    which incorporates the numerically leading part of the non-factorizable
    QED interference between two $W$ decays.
\item $4f$:
  This denotes the tree-level matrix element for $e^-e^+\to 4f$ 
  (the constant $W$ width, massive fermions),
  which can be split into the double-resonant CC03 part
  and the four-fermion-correction due to background diagrams.
\end{itemize}

\subsubsection{General strategy}
As already indicated, we can either correct $d\sigma_{Y_a}$
of {\tt YFSWW3} using the correction of {\tt KoralW} according to
\begin{equation}
  \begin{split}
    d\sigma_{Y+\delta K} &= d\sigma_{Y_a}^{{\cal O}(\alpha)+ISR_{23}+Cc} +\Delta d\sigma_{K},\\
    \Delta d\sigma_{K}   &= d\sigma_{K}^{4f+ISR_{123}+Cc} -d\sigma_{K}^{{\rm CC03}+ISR_{123}+Cc},
  \end{split}
  \label{y/k}
\end{equation}
or correct $d\sigma_{K}$ of {\tt KoralW} using the correction of {\tt YFSWW3} according to
\begin{equation}
  \begin{split}
  d\sigma_{K+\delta Y} &= d\sigma_{K}^{4f+ISR_{123}+Cc} +\Delta d\sigma_{Y_a},\\
  \Delta d\sigma_{Y_a} &= d\sigma_{Y_a}^{{\cal O}(\alpha)+ISR_{23}+Cc} 
                         -d\sigma_{Y_a}^{{\rm CC03}+ISR_{123}+Cc}.
  \end{split}
  \label{k/y}
\end{equation}
Before we enter into more of the details of how the above
is implemented in terms of the MC correction weight, let us briefly 
characterize the fully
differential exclusive distributions
generated by {\tt YFSWW3} and {\tt KoralW}.

\subsubsection{Differential distributions of {\tt YFSWW3} and {\tt KoralW}}
Let us now explicitly define the component distributions in
Eqs.~(\ref{y/k}) and (\ref{k/y}).
The fully exclusive differential
distribution of {\tt KoralW} is given in%
\footnote{ There is a misprint in the second line of Eq.~(4) of 
  Ref.~\cite{koralw:1998}: the $\Theta_\epsilon^{cm}$ should be replaced by
  $\theta(k^0 - \epsilon\sqrt{s}/2)$.}
Eq.~(4) of Ref.~\cite{koralw:1998}:
\begin{equation}
\begin{split}
  d\sigma_{K}^{4f+ISR_{123}+Cc}& =
  d \Phi_{n+4}(p_1+p_2;q_1,...,q_4,k_1,...,k_n)\;
  \frac{1}{n!}\;
  \prod_{i=1}^n  \tilde{S}_I(p_1,p_2,k_i) \theta(k_i^0 -k_{\epsilon})
\\& 
  e^{Y(p_1,p_2,k_{\epsilon})}\;
  \Bigg[
         \bbeta^{(3)}_{0,ISR}(\{p,q\}^{\Rcal})
        +\!\sum_{i=1}^n
        \frac{\bbeta^{(3)}_{1,ISR}( \{p,q\}^{\Rcal}, k_i)}
             {\tilde{S}_I(k_i)}
\\&\qquad\qquad
        +\!\sum_{ i>j }^n
        \frac{\bbeta^{(3)}_{2,ISR}( \{p,q\}^{\Rcal}, k_i, k_j)}
             {\tilde{S}_I(k_i)\tilde{S}(k_j)}
        +\sum_{ i>j>l }^n
        \frac{\bbeta^{(3)}_{3,ISR}( \{p,q\}^{\Rcal}, k_i, k_j, k_l)}
             {\tilde{S}_I(k_i) \tilde{S}_I(k_j) \tilde{S}(k_l) }
  \Bigg],
\end{split}
\label{master}
\end{equation}
where
\begin{equation}
  \tilde{S}_I(p_1,p_2,k) = 
      -\frac{\alpha}{4\pi^2} \left( \frac{p_1}{kp_1} - \frac{p_2}{kp_2} \right)^2.
\label{eq;S-tilde-I}
\end{equation}
For the definition of the YFS form factor $Y(p_1,p_2,k_{\epsilon})$, IR-finite $\bar\beta$'s
and other elements in the above distribution we refer the reader to
Ref.~\cite{koralw:1998}.
Let us only mention that the tree-level four-fermion matrix element is
hidden in $\bar\beta_{i,ISR}, i=0,1,2,3$.

The analogous distribution for {\tt YFSWW3}, see Ref.~\cite{yfsww3:2001},
is more complicated not only because it features photon emission from $W$'s,
but also because it includes summation over photon partitions, 
that is over photon associations to either ISR or WSR.
This trick is useful for efficient
introduction of the LL corrections beyond \Order{\alpha}.
As we shall see later, it is
relevant for the discussion of the reweighting procedures.
For $n$ photons, a single partition is represented by 
the vector $\wp=(\wp_1,...,\wp_n)$, $\wp_i=I,W$ ($I$ for ISR and $W$ for WSR).
The sum over partitions is weighted by the partition weight
$ p_\wp =  {\cal N} \prod_{i=1}^n \tilde{S}_{\wp_i}(k) $
where ${\cal N}$ is adjusted such that $\sum_\wp p_\wp =1$ and $\tilde{S}_W(k)$ for WSR
is defined analogously to $\tilde{S}_I(k)$ of Eq.~(\ref{eq;S-tilde-I}).
The exclusive differential distribution of {\tt YFSWW3} reads as follows:
\begin{equation}
  \begin{split}
  &  d\sigma_{Y_a}^{{\cal O}(\alpha)+ISR_{23}+Cc} =
    d \Phi_{n+4}(p_1+p_2;q_1,...,q_4,k_1,...,k_n)\;
    \frac{1}{n!}
    \prod_{i=1}^n \tilde{S}(\{p,Q\},k_i)\; \theta(k_i^0 - k_{\epsilon})
\\& \quad
  e^{ Y'(p_1,p_2,Q_1,Q_2,k_{\epsilon})}\;
  \left(1 + \delta_C\right)
  \sum_\wp \; p_\wp
  \left(1 + \delta_{An}^{\rm TGC}\right)
  \Bigg\{
  \bbeta^{(1)}_{0}\left(\{p,Q,q\}^\Rpar\right)
\\& \quad
  +\sum_{i=1}^{n} 
   \frac{\bbeta^{(1)}_{1}\left(\{p,Q,q,k_i\}^\Rpar\right)}
        {\tilde{S}\left(\{p,Q,k_i\}^\Rpar\right)}
  +\Delta\bbeta^{(3)}_{0,ISR}\left(\{p,Q,q\}^\Rpar\right)
  +\sum_{\wp_i=I}
   \frac{\Delta\bbeta^{(3)}_{1,ISR}\left(\{p,Q,q\}^\Rpar,k_i\right)}
       {\tilde{S}(\{p,Q\}^\Rpar,k_i)}
\\&\quad
  +\frac{1}{2}\sum_{\wp_{i,j}=I}
  \frac{\bbeta^{(3)}_{2,ISR} \left(\{p,Q,q\}^\Rpar,k_i,k_j\right)}
       {\tilde{S}_I(\{p\}^\Rpar,k_i) \tilde{S}_I(\{p\}^\Rpar,k_j)}
  +\frac{1}{6}\sum_{\wp_{i,j,l}=I}
  \frac{\bbeta^{(3)}_{3,ISR}\left(\{p,Q,q\}^\Rpar,k_i,k_j,k_l\right)}
       {\tilde{S}_I(\{p\}^\Rpar,k_i)\tilde{S}_I(\{p\}^\Rpar,k_j) \tilde{S}_I(\{p\}^\Rpar,k_l)}
  \Bigg\},
  \end{split}
  \label{Master}
\end{equation}
where radiation from the $W$-pair is included in
\begin{equation}
  \tilde{S}(p_1,p_2,Q_1,Q_2,k) =
  \tilde{S}\left(\{p,Q\},k_i\right) = -\frac{\alpha}{4\pi^2}
  \left( \frac{p_1}{kp_1} - \frac{p_2}{kp_2} 
    - \frac{Q_1}{kQ_1} + \frac{Q_2}{kQ_2} \right)^2.
\end{equation}
Furthermore, $Q_1=q_1+q_2$ and $Q_2=q_3+q_4$ denote the four-momenta of the
$W^-$ and $W^+$, respectively.
The  YFS form factor $Y'(p_1,p_2,Q_1,Q_2,k_{\epsilon})$ also includes the WSR.
The complete ${\cal O}(\alpha)$ corrections for the $W$ pair production process
reside in the $\bar\beta^{1}_{0,1}$-functions.
For the $ISR_{123}$ corrections, the sum over real photons extends only
over ISR photons.
See Ref.~\cite{yfsww3:2001} for an explanation of the rest of the notation.

\subsubsection{The need of the reference differential distribution}
The MC programs {\tt KoralW} and {\tt YFSWW3} generate the events according 
to the distributions
of Eqs.~(\ref{master})--(\ref{Master}).
In fact, both of these distributions come in several variants, 
with certain physical effects
and higher-order radiative corrections switched on/off.
It is therefore possible, in principle, 
to calculate separately any component in 
the distributions of Eqs.~(\ref{y/k}) and (\ref{k/y}).

In the real MC programs it is, however, difficult or impossible to find a single subprogram,
that provides numerically the distributions $\rho_{n+4}=d\sigma/d\Phi_{n+4}$ 
exactly as defined by Eqs.~(\ref{master})--(\ref{Master}), without any additional factor.
In the typical MC program one always deals with the MC weights, 
which include not only
expressions like Eqs.~(\ref{master})--(\ref{Master}), but also some factors 
representing technicalities of the MC algorithm.
The MC weight is $w=d\sigma_T/d\sigma_{Primary}$, where $d\sigma_T$
is the ``target'' distribution defined by the physics model and
$d\sigma_{Primary}$ is the multidifferential distribution actually generated
in a given MC program using elementary (primitive) MC techniques.
$d\sigma_{Primary}$ is different in {\tt KoralW} and {\tt YFSWW3}
and the MC weight $w$ can be different in two MC programs, 
even if $d\sigma_T$ is the same.

Because of the above specific normalization properties of the MC weights,
it is not trivial to construct the correcting weight, which is calculated 
in one MC program and used in another one.
The basic practical methodology relies on introducing
a certain ``reference'' differential density $d\sigma^R$,
which is identically the same for {\tt KoralW} and {\tt YFSWW3},
and there is a MC weight
in both programs representing $d\sigma^R$.

In principle the $d\sigma^R$ is a dummy quantity that serves only the
purpose of fixing the absolute normalization between the two programs. 
For $d\sigma^R$ understood like this, we could, of course, pick
``any'' distribution --
even one that does not coincide with any meaningful physical model.
However, we shall also need another auxiliary distribution,
$d\sigma^{Common}_{Max}$, which is the maximal
common part of the best distributions of the two programs to be merged.
In other words,
the distribution $d\sigma^{Common}_{Max}$ should include all
components/corrections that are present in both {\tt KoralW} and {\tt YFSWW3}
and should not include any correction that is present in only one of them.
The simplest possible approach is to choose
the reference distribution $d\sigma^R$ to be equal to $d\sigma^{Common}_{Max}$.
Accordingly, our choice of $d\sigma^R$ is the following:
\begin{equation}
  d\sigma^R \equiv d\sigma^{Common}_{Max} = d\sigma^{\rm CC03+ISR_{123}+Cc}.
  \label{eq:dsig-refer}
\end{equation}
As we remember, CC03 we understand in the gauge-invariant way in 
terms of the LPA.

How is the common reference $d\sigma^R$ realized in {\tt KoralW} and {\tt YFSWW3}?
In {\tt YFSWW3} the starting point is the
differential distribution of Eq.~(\ref{Master}) --
in order to realize the universal $d\sigma^R$, the radiation
from the $W$-pair must be switched off.
Only one partition $\wp=(I,I,...,I)$ remains.
Also the perturbative series of the $\bar\beta$-functions must be truncated to
the CC03 matrix element with the ISR up to third order
in the LL approximation and the (screened) Coulomb correction.
For {\tt KoralW} $d\sigma^R$ is equal to the $d\sigma_{K}$ of Eq.~(\ref{master}) with
the four-fermion Born matrix element simplified to the CC03 level, i.e.
$d\sigma^R=d\sigma_{K}\big|_{\rm CC03}$.

We also have to remove another possible source of the difference in $d\sigma^R$
as implemented in both programs.
The function $\bar\beta_0$ is really identical in both programs
(for any fermionic four-momenta) only in the case without additional 
photon radiation.
Otherwise, in the presence of additional photons, attention has to be paid to
the so-called ``extrapolation/reduction procedure''.
This procedure extrapolates $\bar\beta_0$ from the 4-body (four-fermion)
phase-space into the $(4+n)$-body phase-space (with the additional $n$-photons).
It is generally not unique and has been defined in a slightly different way in 
{\tt KoralW-1.42} and in {\tt YFSWW3-1.16}. 
This ``extrapolation/reduction procedure'' is marked 
in Eqs.\ (\ref{master})--(\ref{Master}) by superscript ${\cal R}$.
We have now modified it in {\tt KoralW-1.51} to coincide (optionally) with
that of {\tt YFSWW3-1.16}. 
In this way, we removed the last source of discrepancy between the $d\sigma^R$
as implemented in the two programs.

After defining precisely the fully exclusive
differential common reference distribution $d\sigma^R$ 
and implementing the corresponding MC weight in both programs, {\tt KoralW} and {\tt YFSWW3},
the next important step is to check {\em numerically} 
that the integrated cross section and one-dimensional distributions,
such as
the distributions of the $W$ invariant mass, of the $W$ production angle 
and of the photon energy
are (within statistical errors) {\em identically the same},
for the MC weights of $d\sigma^R$.
Such tests were performed very extensively and they have shown the full 
agreement of the distributions and the cross sections from the two programs,
see Section \ref{sec:numeric}.

\subsubsection{Definitions of the correcting weights}
Having completed all the above preparatory steps,
we can now precisely define the actual correction weights used in 
the CMC {\tt KoralW$\&$YFSWW3} and other similar possible scenarios.
Equations (\ref{y/k}) and (\ref{k/y}) can be rewritten as 
\begin{align}
  d\sigma_{Y+\delta K}=&(1+\delta^R_{4f}+\delta^R_{NL})\; d\sigma^R,
  \label{+cc03}\\
  d\sigma_{Y+\delta K}=& \bigg(1+\frac{\delta^R_{4f}}{1+\delta^R_{NL}}\bigg)\;
     d\sigma_{Y_a}^{{\cal O}(\alpha)+ISR_{23}+Cc},
  \label{y+k}\\
  d\sigma_{K+\delta Y}=&
  \bigg(1+\frac{\delta^R_{NL}}{1+\delta^R_{4f}}\bigg)\;
     d\sigma_{K}^{4f+ISR_{123}+Cc},
  \label{k+y}
\end{align}
where
\begin{align}
  \delta^R_{4f} =& \frac{d\sigma_{K}^{4f+ISR_{123}+Cc}}{ d\sigma^R} -1,
  \label{eq-cor4f}\\
  \delta^R_{NL} =& 
  \frac{d\sigma_{Y_a}^{{\cal O}(\alpha)+ISR_{23}+Cc} }{d\sigma^R} -1.
  \label{eq-corO1}
\end{align}
Let us remind the reader that we have chosen
\begin{equation}
  d \sigma^R= d\sigma_{Y_a}^{\rm CC03+ISR_{123}+Cc}\equiv d\sigma_{K}^{\rm CC03+ISR_{123}+Cc},
  \label{eq-dsigma-R}
\end{equation}
identically the same for {\tt KoralW} and {\tt YFSWW3}.
In the MC realization, the bracket factor on the r.h.s. of
Eqs.~(\ref{+cc03})--(\ref{k+y}) 
represents the correction weight in a given reweighting procedure;
for a more detailed discussion see the following Section~\ref{sec:technics}.

The great practical advantage of introducing $d\sigma^R$ is now seen in the 
above definitions of the corrections $\delta^R_{4f}$ and $\delta^R_{NL}$.
They are expressed in Eqs.~(\ref{eq-cor4f}) and (\ref{eq-corO1})
as the MC-generator-independent ratios of $d\sigma$'s;
however, inside a given MC generator they will be calculated
as a ratio of the generator-dependent MC weights
(they are the only available objects there).
The generator dependence of the MC weights (from $d\sigma_{Primary}$)
cancels out in the ratio of the MC weights.

There is also another, multiplicative, way of combining the 
${\cal O}(\alpha)$ and four-fermion corrections
\begin{align}
  d\sigma_{Y*\delta K}=& (1+\delta^R_{4f}) (1+\delta^R_{NL})\; d\sigma^R,
  \label{*cc03}\\
               =& (1+\delta^R_{4f})\; d\sigma_{Y_a}^{{\cal O}(\alpha)+ISR_{23}+Cc},
  \label{y*k}\\
  d\sigma_{K*\delta Y}=& (1+\delta^R_{NL})\; d\sigma_{K}^{4f+ISR_{123}+Cc}.
  \label{k*y}
\end{align}
One can see immediately that it differs from the additive scheme 
of Eqs.~(\ref{y+k}) and (\ref{k+y}) by the term
$\delta^R_{4f} \delta^R_{NL}$,
which is definitively a part of the higher-order corrections 
to the background (non-CC03) graphs. 
As this latter correction has not been
calculated so far, one does not know how close the above 
$\delta^R_{4f} \delta^R_{NL}$ term is to the actual correction.
It can, at best, be treated as a rough indication of the order of 
magnitude of the true correction.

The same general scheme can be applied to reweighting due to any
kind of available corrections.
For example to correct the events generated
with some old versions of {\tt KoralW} one has to define $d\sigma^{R_1}$
equal to the old setup $d\sigma^{old}_K$ , 
calculate the appropriate $\delta^{R_1}_{4f}$ correction
and construct the new distribution $d\sigma_{K}$
\begin{equation}
  \label{k/k}
  \begin{split}
    d\sigma_{K}=&(1+\delta^{R_1}_{4f})\; d\sigma^{R_1}
    =(1+\delta^{R_1}_{4f})\; d\sigma_{K}^{old}
    \\
    \delta^{R_1}_{4f}=&\frac{d\sigma_{K}^{new}}{d\sigma^{R_1}} -1
    =\frac{d\sigma_{K}}{d\sigma_{K}^{old}} -1
  \end{split}
\end{equation}

\subsubsection{Approximated $\delta^R_{NL}$}
In the actual implementation, {\tt YFSWW3} provides 
only an approximate version of the \Order{\alpha^1} correction weight:
\begin{equation}
  \delta^R_{NL-} =
  \frac{d\sigma_{Y^-_a}^{{\cal O}(\alpha)+ISR_{23}+Cc} }{d\sigma^R} -1,
  \label{eq:deltaNLapprox}
\end{equation}
where the differential distribution of {\tt YFSWW3}
is a simplified version of
$d\sigma_{Y_a}^{{\cal O}(\alpha)+ISR_{23}+Cc}$ of Eq.~(\ref{Master}),
in which the sum over partitions is restricted to one term, in which
all photons belong to the ISR
\begin{equation}
  d\sigma_{Y^-_a}^{{\cal O}(\alpha)+ISR_{23}+Cc}=
  d\sigma_{Y_a}^{{\cal O}(\alpha)+ISR_{23}+Cc}\bigg|_{\wp=(I,I,I,...,I)},
\end{equation}
with the partition weight set to 1: $p_{(I,I,I,...,I)}=1$.
As we shall see in the numerical tests
presented in the following section, the above approximation is good enough
for all practical LEP2 applications; its precision is better than $0.1\%$.

\subsubsection{Final discussion}
It is interesting and instructive to look also at the actual form of the $\delta$-corrections
in terms of the $\bar\beta$-series and ultimately the Feynman graphs for
our definition of $d\sigma^R=d\sigma^{\rm CC03+ISR_{123}+Cc}$.
For the background-graphs correction we have
\begin{equation}
\begin{aligned}
\ & 1+\delta^R_{4f} =
\\
& = \frac{         \bbeta^{(3)}_{0,ISR}(\{p,q\}^{\Rcal})
        +\sum\limits_{i=1}^n
        {\bbeta^{(3)}_{1,ISR}( \{p,q\}^{\Rcal}, k_i) \over \tilde{S}_I(k_i)}
        +\sum\limits_{ i>j }^n
        {\bbeta^{(3)}_{2,ISR}( \{p,q\}^{\Rcal}, k_i, k_j)
                             \over \tilde{S}_I(k_i)\tilde{S}_I(k_j)}
        +\sum\limits_{ i>j>l }^n
        {\bbeta^{(3)}_{3,ISR}( \{p,q\}^{\Rcal}, k_i, k_j, k_l)
                \over \tilde{S}_I(k_i) \tilde{S}_I(k_j) \tilde{S}_I(k_l) }
       }
       {  \Bigl[  \bbeta^{(3)}_{0,ISR}(\{p,q\}^{\Rcal})
        +\sum\limits_{i=1}^n
        {\bbeta^{(3)}_{1,ISR}( \{p,q\}^{\Rcal}, k_i) \over \tilde{S}_I(k_i)}
        +\sum\limits_{ i>j }^n
        {\bbeta^{(3)}_{2,ISR}( \{p,q\}^{\Rcal}, k_i, k_j)
                             \over \tilde{S}_I(k_i)\tilde{S}_I(k_j)}
        +\sum\limits_{ i>j>l }^n
        {\bbeta^{(3)}_{3,ISR}( \{p,q\}^{\Rcal}, k_i, k_j, k_l)
                \over \tilde{S}_I(k_i) \tilde{S}_I(k_j) \tilde{S}_I(k_l) }
          \Bigr]_{\rm CC03}
       }
\\ 
& = \frac{\vert M_{}^{4f}(\{p,q\}^{\Rcal})\vert^2}
         {\vert M_{}^{\rm CC03}(\{p,q\}^{\Rcal})\vert^2}.
\end{aligned}
\label{nohot}
\end{equation}
The last equation follows from the fact that the second- and third-order 
LL expressions for the ISR $\bbeta$-functions are exactly the same in both
$d\sigma^R=d\sigma_{K}^{\rm CC03+ISR_{123}+Cc}$ and $d\sigma_{K}^{4f+ISR_{123}+Cc}$,
and
also because we have used the same extrapolation/reduction procedures
in both distributions (note that the extrapolation procedure is fixed by
the requirement that $d\sigma^R$ is identical in {\tt KoralW} and {\tt
YFSWW3}). 
The factorization property of the LL ansatz used in the construction of 
the ISR $\bbeta_i$ series,
cf.\ Ref.\ \cite{koralw:1998}, is, of course, essential.
Both distributions in the numerator and denominator of eq.~(\ref{nohot})
are defined in {\tt KoralW}, that is the appropriate MC weights exist for them.

The case of $\delta^R_{NL-}$, provided by {\tt YFSWW3}, 
can be analysed as follows:
\begin{equation}
\begin{aligned}
1+&\delta^R_{NL-} =
    e^{Y'(p_1,p_2,Q_1,Q_2,k_{\epsilon}) -Y(p_1,p_2,k_{\epsilon})}
    \prod_{i=1}^n  
    \frac{\tilde{S}(p_1,p_2,Q_1,Q_2,k_i)}
         {\tilde{S}(p_1,p_2,k_i)}
\\ \times &
\left({\vbox to 1.1cm {}} 
   1\:\:+\:\:
  \left(1 + \delta_C\,\right)
  \left(1 + \delta_{An}^{\rm TGC}\,\right) \right.
\\&
\left.
\frac
{
  \bbeta^{(1)}_{0}(\{p,Q,q\}^{\Rcal}) -
  \bbeta^{(1)}_{0,ISR}\left(\{p,q\}^{\Rcal}\right)
  +\sum\limits_{i=1}^{n}
  \frac{\bbeta^{(1)}_{1}\left(\{p,Q,q,k_i\}^{\Rcal}
        \right)}{\tilde{S}_I(\{p,Q,q,k_i\}^{\Rcal})}
  -\sum\limits_{i=1}^{n}
  \frac{\bbeta^{(1)}_{1,ISR}\left(\{p,q\}^{\Rcal},
        k_i^I\right)}{\tilde{S}_I(k_i)}
}
{ 
  \Bigl[  \bbeta^{(3)}_0(\{p,q\}^{\Rcal})
  +\sum\limits_{i=1}^n
  {\bbeta^{(3)}_{1}( \{p,q\}^{\Rcal}, k_i) \over \tilde{S}_I(k_i)}
  +\sum\limits_{ i>j }^n
  {\bbeta^{(3)}_{2}( \{p,q\}^{\Rcal}, k_i, k_j)
    \over \tilde{S}_I(k_i)\tilde{S}_I(k_j)}
  +\sum\limits_{ i>j>l }^n
  {\bbeta^{(3)}_{3}( \{p,q\}^{\Rcal}, k_i, k_j, k_l)
    \over \tilde{S}_I(k_i) \tilde{S}_I(k_j) \tilde{S}_I(k_l) }
  \Bigr]_{\rm CC03}
}
\right)
\\ 
=&  \frac{\vert M_{}^{{\cal O}(\alpha)}(\{p,q\}^{\Rcal})\vert^2 
           + \hbox{h.o.t.}}
         {\vert M_{}^{\rm CC03}(\{p,q\}^{\Rcal})\vert^2 + \hbox{h.o.t.}}
\label{hot}
\end{aligned}
\end{equation}
The additional higher-order terms (h.o.t.) in the denominator 
include the \Order{\alpha^1} effects due to the ISR.
The numerator is the \Order{\alpha^1} matrix element squared
for the one-photon ISR$+$WSR.
In the presence of exponentiation, this is true modulo
\Order{\alpha^2} terms, owing to the specific definition
of the $\bbeta_1$ in the YFS exponentiation, 
which admits the presence of the virtual
\Order{\alpha^1} corrections, even for a hard photon.
For two and more hard photons the biggest correction in the h.o.t. will be
of \Order{\alpha^2 \ln(s/m_e^2)}.
Note that there is {\em no} sum over partitions in the above
equation. If, however, we used the exact $\delta^R_{NL}$ instead of the
approximate $\delta^R_{NL-}$ the sum over partitions would reappear and
the above formula (and a relation to the
Feynman graphs) would become even more complicated.

\subsection{Technical aspects}
\label{sec:technics}

In the following section 
we shall discuss more practical aspects of combining the
two programs into one. In particular we shall present, in detail, the
idea of the ``Concurrent Monte Carlo'' realized via the FIFO mechanism.

\subsubsection{Merge for inclusive distributions (histograms)}

The  first, most natural way is to add the four-fermion and 
${\cal O}(\alpha)$ corrections at the level of
the {\em inclusive} observables, that is for integrated  cross sections
and inclusive one- or two-dimensional distributions (histograms).
In this case both sides of Eqs.~(\ref{y/k}) and (\ref{k/y}) are integrated 
either completely
or almost completely, for instance leaving one or two variables
unintegrated.  
The inclusive distributions or the integrated cross sections entering 
Eqs.~(\ref{y/k}) and (\ref{k/y}) are calculated separately from the independent 
runs of the two programs
and combined according to Eqs.~(\ref{y/k}) and (\ref{k/y}).
The whole procedure is most convenient for the variable-weight events,
mainly because one may use the differences of the weights.

The application of this inclusive correcting scheme is quite straightforward.
Suppose that we want to calculate the $W$-mass distribution.
Using the first scheme $Y+\delta K$ of Eq.~(\ref{y/k}), we proceed as follows:
\begin{enumerate}
\item
  Make the properly normalized histogram of
  $d\sigma_{Y_a}^{{\cal O}(\alpha)+ISR_{23}+Cc}/d M_W$ 
  with a sufficiently long run of the {\tt YFSWW3} MC
  with the variable- or constant-weight events.
\item
  Make the properly normalized histogram of
  $\Delta d\sigma_{K}/d M_W$ with {\tt KoralW}.
  This can be done by running {\tt KoralW} once (twice) with 
  the variable-weight (constant-weight) events.
\item
  Add the two histograms.
\end{enumerate}
Using the second scheme $K+\delta Y$ of Eq.~(\ref{k/y}), we proceed in the
analogous way:
\begin{enumerate}
\item
  Make the properly normalized histogram of
  $d\sigma_{K}^{4f+ISR_{123}+Cc}/d M_W$ 
  with the run of the {\tt KoralW}
  with the variable- or constant-weight events.
\item
  Make the properly normalized histogram of
  $\Delta d\sigma_{Y_a}/d M_W$ with the help of {\tt YFSWW3}.
  This can be done by running {\tt YFSWW3} once (twice) with 
  the variable-weight (constant-weight) events.
\item
  Add the two histograms.
\end{enumerate}
Both schemes must give
the same results,
but in practice it is reasonable to use the one that will require, in
the given circumstances, the smaller $\Delta d\sigma$
correction and  less fluctuations in the correcting weight.
For example, in channels such as $u\bar d \mu\bar\nu_\mu$,
the four-fermion correction turns out to be negligible for energies away
from the $W$-threshold; $\Delta d\sigma_{K}$ can therefore be completely
neglected \cite{yfsww3:2000a}. 

The above inclusive approach has serious limitations.
The most important is that it is not fully exclusive (event per event).
Consequently it cannot be used in a data analysis 
that takes the detector simulation properly into account.
On the other hand, this approach can be useful for all kinds of theoretical 
studies.
Note that this method was used in ref.~\cite{koralw:1997}
for combining the results of {\tt KoralW} and {\tt grc4f}.

\subsubsection{Merge for fully exclusive distributions (event per event)}
\label{sec:merg-exclu}

The inclusive method of the previous subsection was not {\em event per event}
because the MC events of {\tt KoralW} and {\tt YFSWW3} were
at different random points of the full phase space.
Consequently, the MC weights of the events from these two generators
were also not calculated at the same points of the phase space; 
therefore, we could not take their ratios (apply correcting weights).

In this section, we shall describe another, more sophisticated, method
of combining fully exclusive differential distributions of {\tt KoralW} 
and {\tt YFSWW3}, in which both programs work with {\em the same} events, 
that is the same random points in the $(4+n)$-dimensional phase space.
One of the MC programs, the master MC,
generates the event and the other one, the slave MC, instead of generating
its own event, reads the event of the master MC and calculates the 
correction weight
due to some missing effect, for instance the ${\cal O}(\alpha)$ correction
or the correction due to the missing background diagrams.
The transfer of the events from one program to the other can be organized
with the help of the external device (disk or tape) or ``in flight''.
In the first case, the master MC writes the events on the disk 
and the slave MC reprocesses the events read from the disk.
Writing the events on the disk is done routinely in the data analysis anyway.
For the description of the second, ``in-flight'' method, 
see the next subsection.

The above procedure of cross-correcting the distributions/events from
one MC program with the help of the second MC is not the only one, and
is not
the simplest one either.
A similar procedure is possible, even with a single MC,
when the events are stored on the external device
and later on, in a separate run, they are corrected by {\em the same} 
MC program with the correcting weight due to the change of the input parameters 
(typically the $W$-mass).
This is very useful for the data analysis, 
where the constant-weight MC events from a physics MC event generator,
like {\tt KoralW} or  {\tt YFSWW3}, are processed through the detector 
simulation. 
In this way the CPU-time-consuming reprocessing of the detector simulation 
is avoided.
It also allows fitting the SM parameters to experimental data 
(typically the $W$-mass)
with full control of the effects due to the detector acceptance.
One has to remember, however, 
that the above procedure provides us essentially with the variable-weight events.
If the correction weight is fluctuating very mildly, then this is not 
a problem; otherwise, 
the strongly fluctuating correction weight would lead to higher statistical 
errors in the calculated observables, and would inhibit the optional
transformation of variable-weight events into constant-weight events.

As seen from the above discussion, it is essential that the MC event generators
can write/read the events into/from disk files.
The appropriate tools for reading the events from the disk and calculating
the correction weight due to the change of the input parameters and the
change of the scattering matrix element have been introduced in the 
version {\tt 1.51} of {\tt KoralW} program presented in this paper,
as described in the next section, and
in the {\tt YFSWW3} program as well%
\footnote{The capabilities of writing/reading the events and calculating
  the correction weight were already present in the previous unpublished
  versions of the programs used by the LEP experiments for the $WW$-physics 
  data analysis.},
see Ref.~\cite{yfsww3:2001}.
Let us now describe, in more detail, the procedures
of cross-correcting the distributions/events from
one MC program with the help of the second.
There are several scenarios for such a procedure
following Eqs.~(\ref{+cc03})--(\ref{k*y}) --
they differ in the choice of the master/slave MC ({\tt KoralW} 
or {\tt YFSWW3})
and in the way the correction weight is constructed 
(additive or multiplicative).

Let us look closer at the ``symmetric'' scheme of Eq.~(\ref{+cc03}).
It consists of the following steps:
\begin{enumerate}
  \item
    Generate the variable- or constant-weight events according to the 
    fully exclusive
    distribution $d\sigma^R=d\sigma^{\rm CC03+ISR_{123}+Cc}$, using 
    either {\tt KoralW} or {\tt YFSWW3}, and store all events on the disk.
  \item
    Calculate the correction $\delta^R_{4f}$ of Eq.~(\ref{eq-cor4f}) using 
    {\tt KoralW} and write it in the disk record for each event.
  \item
    Calculate the correction
    $\delta^R_{NL}$ of  Eq.~(\ref{eq-corO1}) using {\tt YFSWW3}, 
    construct the correction weight 
    $w_{corr}=1+\delta^R_{4f}+\delta^R_{NL}$,
    and write it in the disk-record of each event.
    (In fact, we have to use the substitute $\delta^R_{NL-}$ of 
              Eq.~(\ref{eq:deltaNLapprox}), since the
    $\delta^R_{NL}$ is not provided by {\tt YFSWW3} version 1.16).
  \item
    Optionally reject events according to $w_{corr}$.
\end{enumerate}
Of course, it is not really necessary
to reprocess the MC events twice (total of three runs) and one may follow 
the simpler ``asymmetric'' procedure $Y+\delta K$ 
in which {\tt YFSWW3} is the master MC:
\begin{enumerate}
  \item
    Generate the variable- or constant-weight events according to the 
    fully exclusive distribution $d\sigma_{Y_a}^{{\cal O}(\alpha)+ISR_{23}+Cc}$
    using {\tt YFSWW3}.
    Store each event on the disk, together with the value of $\delta^R_{NL}$.
  \item
    In the second run, for each event use {\tt KoralW} 
    in order to calculate the correction weight, the bracketed expression 
    in Eq.~(\ref{y+k}).
    Optionally, reject the events according to the correction weight.
\end{enumerate}
The analogous scenario $K+\delta Y$ in which {\tt KoralW} is the master MC 
looks as follows:
\begin{enumerate}
  \item
    Generate the variable- or constant-weight events according to the 
    fully exclusive
    distribution $d\sigma_{K}^{4f+ISR_{123}+Cc}$ using {\tt KoralW}.
    Store each event on the disk, together with the value of $\delta^R_{4f}$.
  \item
    In the second run, for each event use {\tt YFSWW3}
    in order to calculate the correction weight, the bracketed expression 
    in Eq.~(\ref{k+y}).
    Optionally, reject the events according to the correction weight.
\end{enumerate}
Alternatively, one may follow the multiplicative schemes
of Eqs.~(\ref{*cc03})--(\ref{k*y}).

As indicated, the above methods provide, 
in principle, a sample of constant-weight events.
However in practice, this is not
convenient, as one would have to deal with millions of 
variable-weight events stored together with their weights, 
and then to be reprocessed again to include the second correction, and finally
to undergo the final rejection. 
Moreover, for the moment only {\tt KoralW} is
capable of continuing the event construction (rejection, decay libraries, 
etc.) based on the input events stored on the external device. 
{\tt YFSWW3} version {\tt 1.16} can only provide the correction weight.
Consequently, the only practical option to provide constant-weight events
is the one in which {\tt KoralW} is the master MC.

The use of the variable-weight events from the master MC is, 
however, possible in the scenario
in which the master and the slave programs exchange the events
in-flight, without recording them in the (large) disk files, 
see the following subsection.

\subsubsection{Concurrent exclusive merge: CMC {\tt KoralW$\&$YFSWW3}}

One of the most important results of the modifications of 
{\tt KoralW} version {\tt 1.51} is the possibility of a direct inclusion
of the ${\cal O}(\alpha)$ NL corrections calculated by the {\tt YFSWW3-1.16}
program into the process of the event generation by {\tt KoralW}.
This is not done, however, by simply compiling and linking them together
into a single executable, because
the source codes of {\tt YFSWW3} and {\tt KoralW} share a number of subroutines,
common blocks and libraries with the same names, but with different
contents.
It is therefore practically impossible (without major rewriting of 
the two source codes)
to merge the two programs into one at the level of the FORTRAN source code%
\footnote{
The example of such a successful, albeit difficult to use, procedure of linking
together a number of different subroutines with identical names was
realized in {\tt KoralW} in an early version of the implementation of
the four-fermion matrix element generated by the GRACE system separately,
channel by channel, for all CC-type final states.
}.
Such a merge might also obstruct a further independent development of 
the programs.

Once both programs have acquired capabilities of exchanging the events
and reprocessing them using the disk file as an intermediate medium,
we have noticed that one can follow another approach, avoiding major
rewriting of both programs,
based on the UNIX/Linux standard facility called 
the ``names pipes'' or the FIFO mechanism%
\footnote{We would like to thank Piotr Golonka for useful discussions
  on that point.}.
It allows two {\em independent} processes to communicate ``in real time''
through the ``named pipe'', into which one process
writes and from which the other reads, in turn. 

\begin{figure}[!ht]
\centering
\setlength{\unitlength}{0.1mm}
\epsfig{file=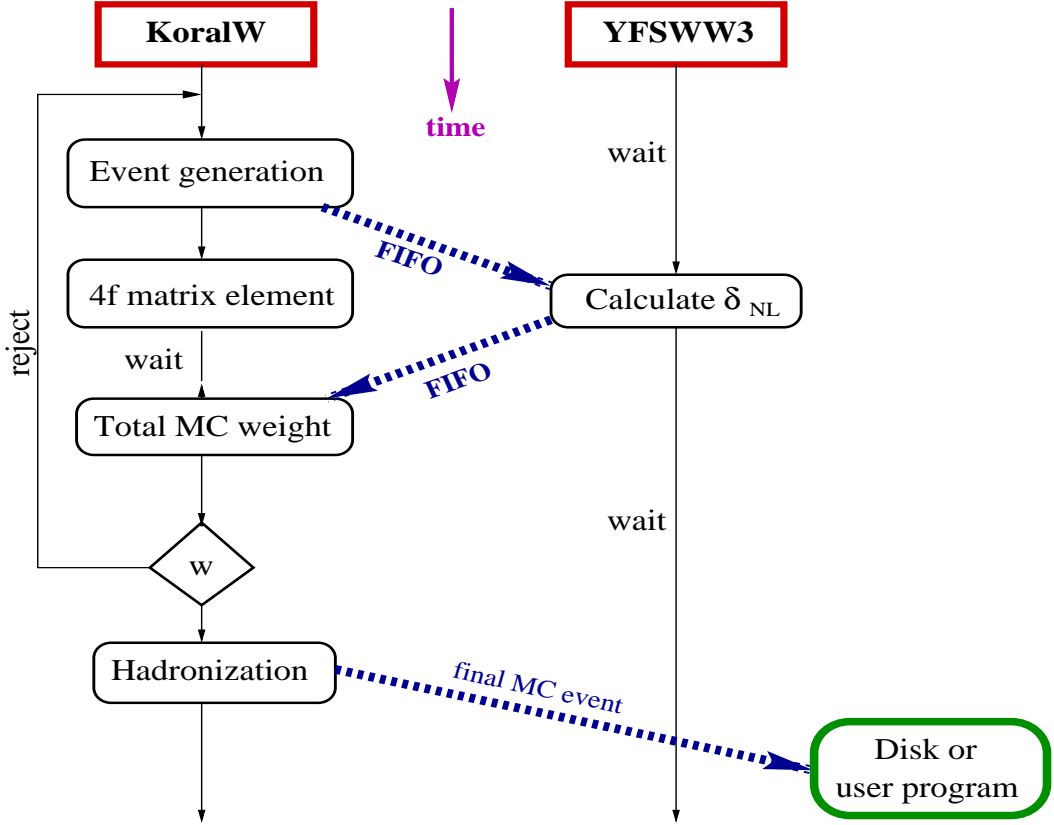,width=140mm,height=110mm}
\caption{\small\sf
  The time-flow chart for the present CMC {\tt KoralW$\&$YFSWW3}.
  The line marked FIFO denotes the transfer of an event record from one
  process to 
  another, using the named-pipes mechanism of UNIX.
}
\label{fig:flow1}
\end{figure}

In our case, the scheme is such
that {\tt KoralW} generates an event and writes its four-momenta into
the FIFO special file. 
These four-vectors are then read by {\tt YFSWW3}, which is running in
parallel (concurrently) and calculates the correcting 
${\cal O}(\alpha)$ NL weight,
or more precisely the $\delta^R_{NL}$ quantity, 
and then writes it to another FIFO special file. 
In the final step, {\tt KoralW} reads this weight from the FIFO special file,
includes it in the total weight and finishes the event construction. 
This Concurrent Monte Carlo {\tt KoralW$\&$YFSWW3} scheme is
schematically depicted in Fig.\ \ref{fig:flow1}.
As a result, {\tt KoralW} can now, for example, perform the rejection 
and provide the {\em constant-weight} events 
with both the four-fermion background 
correction $\delta^R_{4f}$ and the ${\cal O}(\alpha)$ NL
correction $\delta^R_{NL}$ 
to the $W$-pair production and decay process taken into account.
Another application of such a scheme
is the reweighting of events with both the
four-fermion and ${\cal O}(\alpha)$ corrections included simultaneously 
in order to perform ``Monte Carlo fits''.

To summarize it shortly:
from the point of view of the user, the CMC program {\tt KoralW$\&$YFSWW3}  
works as a single Monte Carlo generator with all its normal features. 
This might be, to our knowledge, the first practical working implementation
of such a scheme to the problem of the MC event generation in particle physics.

Remarkably, the efficiency of  the CMC {\tt KoralW$\&$YFSWW3}
in generating the {\em constant-weight} events according to the
four-fermion background and the ${\cal O}(\alpha)$ NL corrected distributions
is not bad.
In principle
one could worry that an additional {\tt YFSWW3} weight of the order of a few tens 
would translate into a similar increase of the maximal weight for rejection 
and, consequently, into a similar increase of the CPU time needed for
the event generation. 
Surprisingly, however, this is not true! In the additive mode
of combining {\tt KoralW} and {\tt YFSWW3}
the increase of the maximal weight that we recorded on the $2\times10^8$
sample of variable-weight events was $1\to 5$, for example. Overweighted
events were of two topologies. 
The first topology 
was strongly $WW$-like with multiple, i.e.\ at least triple, soft
bremsstrahlung and led to overweights below 2 (in an extreme
case with seven photons, we recorded an overweight of 4). The other
topology was very far from $WW$-like 
(small values of the $W^-$, $W^+$ invariant
masses) and gave the highest over-weights ($\sim 5$). 
In the latter cases the use of the on-shell ${\cal O}(\alpha)$ NL 
library is less
justified anyway and these few events can simply be discarded, instead of
increasing the maximal weight for the rejection%
\footnote{
  In the {\tt KoralW} we set (over)conservatively the maximal weight
  of the rejection above the {\em highest} possible generated weight.
  This strategy can easily be relaxed,
  by requesting that the events with weights over the maximal weight
  contribute to the total cross-section less than a certain, predefined, 
  small number. This would
  lower substantially the maximal weight for the rejection.
  }.
Why was the overweighting so small? Because in {\tt KoralW} to cover 
the four-fermion-dominated configurations 
the maximal weights are adjusted higher than 
what is needed by the $WW$-like configurations. 
This extra safety margin turns out
to be sufficient to absorb fluctuations of the additional $WW$-like 
${\cal O}(\alpha)$ NL weights. The situation is different
for the multiplicative prescription. Here, the big four-fermion weight
is multiplied
by the ${\cal O}(\alpha)$ NL correction, on occasion leading to
overweights of the expected order of a few tens. This happens, typically,
for a very hard bremsstrahlung photon, with energy of the order of 40~GeV.
However, as we have explained earlier, the ${\cal O}(\alpha)$ correction 
for the configurations far away
from being $WW$-like is less justified and the overweights provide
merely another reason for using the additive scheme rather than
the multiplicative one.

Let us add a few additional comments on the {\tt KoralW$\&$YFSWW3} CMC:
\begin{enumerate}
\item
  The corrections can be combined in an additive or multiplicative way, 
  according to the discussion from the previous subsections.
\item
  One may also wonder about other possible source of efficiency loss 
  caused by running two programs at the same
  time on a single processor. 
  Our experience shows, however, that there is very little loss
  due to an alternate suspension of the programs, performed
  automatically by the operating system (in the UNIX kernel).
\item
  The big advantage of the FIFO special files is that they are in practice 
  done by the operating system ``in flight'', in virtual memory. 
  Consequently, there is no need for a huge amount of disk space to 
  store all the millions of generated events needed for generating
  the sample of the constant-weight events.  
\item
  In order to introduce the ``named pipes'', there is {\em no need to
    modify} in the FORTRAN source code of either of the
  programs. The only change is the replacement of regular
  input/output files by the special FIFO files, done at the level of the 
  operating
  system and not in the FORTRAN source code. We give an example of such
  a procedure 
  in the demo run of the CMC {\tt KoralW$\&$YFSWW3}, 
  which can be invoked through: {\tt make KandY} 
  ({\tt KandY} stands for ``{\tt K} and {\tt Y}''). 
  It creates two special FIFO files 
  {\tt 4vect.data.special} and {\tt wtext.data.special} that will serve 
  for transmitting the four-momenta and the weights between the programs. 
  These special FIFO files are 
  then linked to the appropriate input/output file names for {\tt KoralW} and 
  {\tt YFSWW3} in their local working directories. 
\item
  Finally, the two programs are executed with the {\em common default data cards}!
  They are modified later on 
  by the user cards located in the local working directory.
\end{enumerate}

The proposed scheme of CMC 
{\tt KoralW} $\to$ {\tt YFSWW3} $\to$ {\tt KoralW} is
not the only possibility of realizing the  four-fermion background and 
${\cal O}(\alpha)$ NL corrected events. 
Alternatively, one can construct the CMC the other way around:
{\tt YFSWW3} $\to$ {\tt KoralW} $\to$ {\tt YFSWW3} 
(symbolically {\tt YFSWW3$\&$KoralW}). 
In general, this direction
has a serious drawback -- the four-fermion correction weight from 
{\tt KoralW} can be very high, especially for the final states with
electrons, spoiling the convergence of the series. On the contrary, the
{\tt YFSWW3} weight for the ${\cal O}(\alpha)$ NL 
correction is well behaved and
for LEP2 energies it seems not to exceed a few tens. 
However, if the phase space for generation were restricted to $WW$-like
configurations the situation would be reversed -- the four-fermion
correction weight would be smaller than the ${\cal O}(\alpha)$ NL one
and the scheme {\tt YFSWW3$\&$KoralW} would be a good choice.

We conclude this section with two technical remarks on the FIFO
mechanism. They may prove useful in practical implementations of FIFOs.
\begin{enumerate}
\item
The FIFO special files must be created (by the command
``{\tt mkfifo }{\it file\_name}'')
on a {\em local} disk of a
computer on which the program will be executed. This means in particular
that it cannot be created directly on the AFS file
system or even on a mounted remote file system. The good
location is, for example, the {\tt /tmp} directory. After creation, however,
the FIFO special file can be symbolically linked to any location in
the mounted or AFS file systems.
The FIFO special file itself requires only 4kB of disk space.
\item
In certain configurations of the data transfer through named pipes, it may
be at some point necessary to flush the output buffers of the programs.
The syntax of the FORTRAN command for that operation is
{\tt call flush(}{\it unit\_number}{\tt )}.
\end{enumerate} 

\subsubsection{Possible future development}
Our present modest application of the parallel processing to MC event generation
is schematically depicted in Fig.~\ref{fig:flow1}.
Note that on a dual-processor machine, at a certain moment {\tt KoralW}
is calculating the
four-fermion matrix element while {\tt YFSWW3} is {\em simultaneously} 
calculating the electroweak \Order{\alpha} corrections.
In the present form, however, {\tt KoralW$\&$YFSWW3} does not really
allow us to profit fully
from running on the multiprocessor machine.
It is mainly because 
the calculation of the \Order{\alpha} NL correction takes on average
longer than the calculation of the four-fermion correction; as a
result the 
{\tt KoralW} process is sometimes waiting for
{\tt YFSWW3} before the final reprocessing of the event.
Also hadronization by {\tt JETSET}, which takes a substantial amount of
CPU time and is independent from {\tt KoralW} and {\tt YFSWW3},
is not delegated to a separate process, but included as a part of {\tt KoralW}.

\begin{figure}[!ht]
\centering
\setlength{\unitlength}{0.1mm}
\epsfig{file=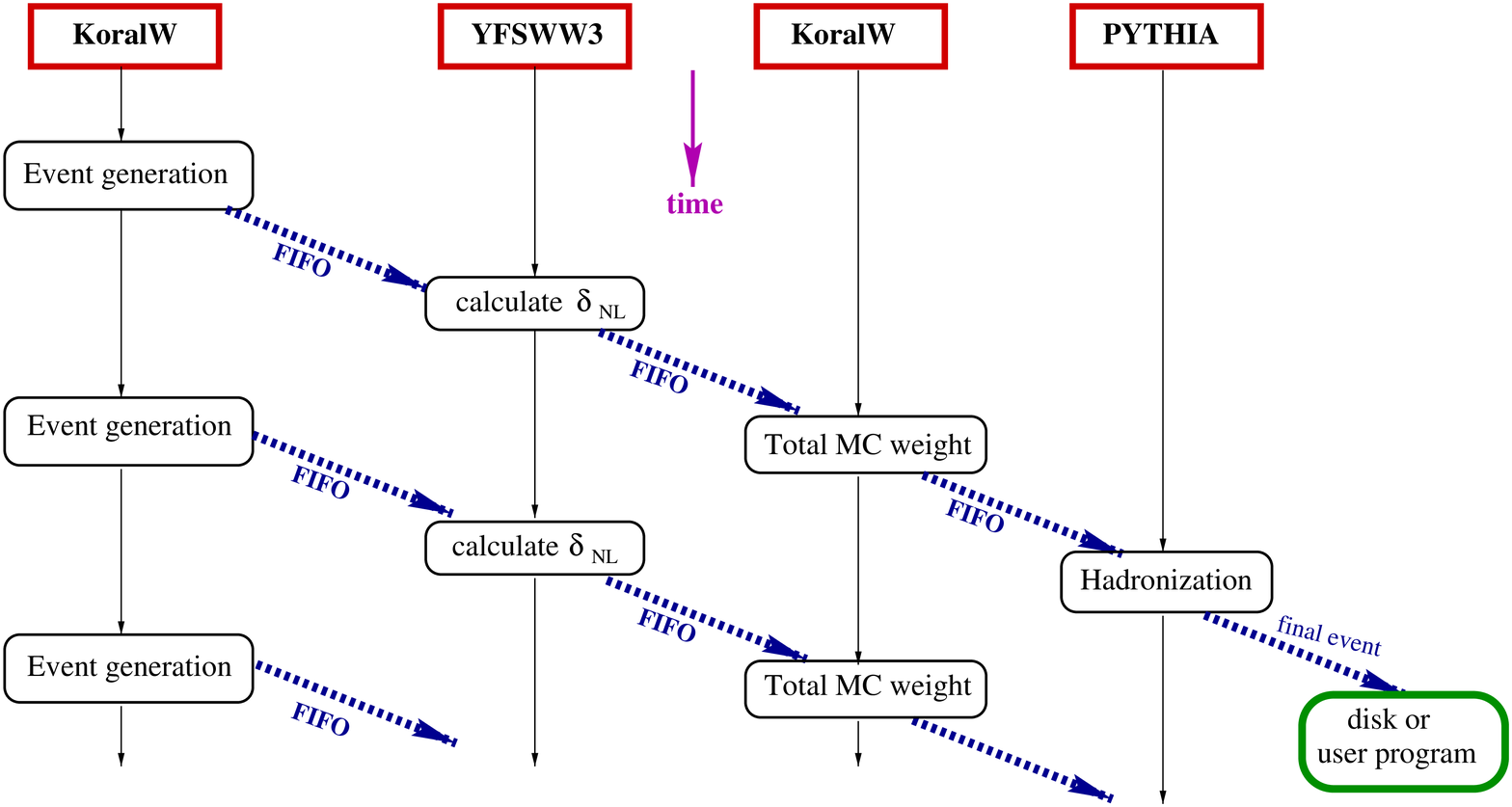,width=160mm,height=100mm}
\caption{\small\sf
  The time-flow chart for the possible future concurrent cascade-type 
  arrangement of {\tt KoralW},
  {\tt YFSWW3} and a hadronization package such as {\tt PYTHIA}.
}
\label{fig:flow2}
\end{figure}
In Fig.~\ref{fig:flow2} we show another possible future 
concurrent arrangement, which would
work more efficiently on the multiprocessor installation.
Here, four independent processes work  ``in the cascade'',
in such a way that when the last process ({\tt PYTHIA}) finishes
to hadronize an event, 
the first process may have already started to construct the next event.
A similar solution was proposed in Ref.~\cite{1989:eloisatron},
however limited to a solution in which MC generators communicate through
disk files (database)%
\footnote{To our knowledge it was not realized in some widely used practical application.}.
Of course, there are many other possible
variants of such a scheme -- the best one should be adapted
to a particular MC generation problem and to available hardware.

The important advantage of such a concurrent arrangement is that
it provides ``encapsulation'' for the ``dusted deck programs'',
without the need of laboriously translating them to object-oriented C++ or 
another OO programming language.
It also allows an easy combination of programs written in 
different programming languages.
This may prove to be in practice a rather effective solution, 
before eventual emergence of the next generation of event generators,
written 
from scratch in the OO environment.

\section{Numerical tests}
\label{sec:numeric}

Although the basic idea of reweighting MC events, 
produced either by the same MC event generator or the other one,
is relatively simple,
its actual implementation has to be tested very carefully.
Such tests are the principal aim of the present section.
We shall concentrate
on the reweighting of MC events produced by {\tt KoralW}
using the correction weight produced by {\tt YFSWW3},
that is on the ``asymmetric'' procedure $K+\delta Y$
described in Subsection \ref{sec:merg-exclu}.
Before the programming tools for such a scenario can be fully trusted,
we have to perform certain important introductory numerical tests:
\begin{description}
\item[(A)] 
  We are going to check numerically that the common reference
  differential distributions $d\sigma^R$ 
  generated by {\tt KoralW} and {\tt YFSWW3} are the same.
\item[(B)]
  For the distribution $d\sigma^{Test}=d\sigma^{ {\cal O}(\alpha)+ISR_{23}+Cc}$
  we are going to test the reweighting tools of {\tt KoralW} and {\tt YFSWW3}:
  \begin{description}
    \item[(B.1)] Self-test of {\tt YFSWW3}: we shall compare results
      from the standard run of {\tt YFSWW3} and from the run $Y+\delta Y$
      in which $d\sigma^{Test}$ is obtained by reweighting events
      generated according to $d\sigma^R$.
    \item[(B.2)] Calibration of $Y+\delta K$ using {\tt YFSWW3}: we shall compare
      results of $Y+\delta K$ (in which events generated according
      to $d\sigma^R$ by {\tt KoralW} are reweighted with the help of {\tt YFSWW3})
      with the direct results of {\tt YFSWW3} and with the results of the
      $Y+\delta Y$ scheme. 
  \end{description}
\end{description}
In both kinds of tests, numerical results for the cross sections
and the key distributions,
for instance the distributions of the $W$ mass, $W$ scattering angle and photon energy,
should agree within statistical errors.
We shall use the approximate version of the correcting weight 
($\delta_{NL-}$ from {\tt YFSWW3}).
Test (B) will provide the measure of the quality of the approximation.

In all numerical tests in this section,
the input parameter set-up and
the definitions of event acceptances are that of Ref.~\cite{4f-LEP2YR:2000},
unless stated otherwise.
For the convenience of the reader, let us recall briefly these acceptance
conditions. 

For the distributions we used the following cuts/acceptances:
\begin{enumerate}
\item
  We required that the polar angle of any charged final-state fermion
  with respect to
  the beams be $\theta_{f_{ch}}>10^{\circ}$.
\item
  All photons within a cone of $5^{\circ}$ around the beams were treated
  as invisible, i.e. they were disregarded in the calculation of any observable.
\item
  The invariant mass of a visible photon with each 
  charged final-state fermion, $M_{f_{ch}}$, was calculated, 
  and the minimum value 
  $M^{min}_{f_{ch}}$ was found. If $M^{min}_{f_{ch}}< M_{rec}$ 
  or if the photon energy $E_{\gamma}<1\,$GeV,
  the photon was combined with the corresponding fermion, 
  i.e. the photon four-momentum was added to the fermion four-momentum
  and the photon was discarded. This was repeated for all visible photons. 

  In our numerical tests we used two values of the recombination cut: 
  $$
  M_{rec} = \left\{
  \begin{tabular}{l}
   \: 5\,{\rm GeV:} \hspace{1cm} {\rm CAL05}, \\
   25\,{\rm GeV:} \hspace{1cm} {\rm CAL25}.
  \end{tabular}
  \right.  
  $$
  Let us remark that we have changed here the labeling of these
  recombination cuts from the slightly misleading BARE and CALO
   names used in
  Ref.~\cite{4f-LEP2YR:2000}. This change allows us to reserve the BARE
  name for a truly bare setup (without any recombination).
\end{enumerate}

The integrated cross sections presented here were obtained without any
cuts (labeled in the tables ``NO CUTS'') or with the cut No.~1 from the above 
acceptance conditions (labeled in the tables ``WITH CUTS'').

\subsection{Integrated cross sections}

\begin{table*}[!th]
\centering
\begin{tabular}{||c|c||c|c||c|c||c||}
\hline\hline
\multicolumn{7}{||c||}{ NO CUTS}\\
\hline
Description &Program &$\sigma^{\rm CC03}$[fb]  &$\sigma^{R}$[fb] {\vbox to 0.4cm {}}
            &${\bar \delta}^{\rm CC03}_{4f}[\%]$ &$\bar{\delta}^{R}_{4f}[\%]$ &$\bar{\delta}^{R}_{NL}[\%]$ \\
\hline\hline
            &YFSWW3   &$219.793\,(16)$ &$204.198\,(09)$  &---      &---    &$-1.92\,(4)$\\
$\nu_{\mu}\mu^+\tau^-\bar{\nu}_{\tau}$
            &KoralW   &$219.766\,(26)$ &$204.178\,(21)$  &$0.041$&$0.044$ &--- \\
\cline{2-4}
200 GeV      &(Y$-$K)/Y&$0.01\,(1)\% $ &$0.01\,(1)\% $    &---      & ---     & --- \\
\hline\hline
            &YFSWW3   &$659.69\,(5)$  &$635.81\,(3)$     &---      & ---   &$-1.99\,(4)$\\
$u\bar{d}\mu^-\bar{\nu}_{\mu}$
            &KoralW   &$659.59\,(8)$  &$635.69\,(7)$     &$ 0.073$ &$ 0.073$ & --- \\
\cline{2-4}
200 GeV      &(Y$-$K)/Y&$ 0.02\,(1)\% $ &$ 0.02\,(1)\%$   &--- & --- & --- \\
\hline\hline
            &YFSWW3   &$1978.37\,(14)$ &$1978.00\,(09)$  &--- & --- & $-2.06\,(4)$ \\
$u\bar{d} s\bar{c} $
            &KoralW   &$1977.89\,(25)$ &$1977.64\,(21)$  &$ 0.060$ &$ 0.061$ & --- \\
\cline{2-4}
200 GeV      &(Y$-$K)/Y&$ 0.02\,(1)\% $ &$ 0.02 \,(1)\%$  &--- & --- & --- \\
\hline\hline
\multicolumn{7}{||c||}{ WITH CUTS}\\
\hline
            & YFSWW3  &$210.938\,(16)$ &$196.205\,(09)$  &--- & --- & $-1.93\,(4)$ \\
$\nu_{\mu}\mu^+\tau^-\bar{\nu}_{\tau}$
            &KoralW   &$210.911\,(26)$ &$196.174\,(21)$  &$0.041$ &$0.044$ & --- \\
\cline{2-4}
200 GeV      &(Y$-$K)/Y&$ 0.01\,(1)\% $ &$0.02\,(1)\%$    &--- & --- & --- \\
\hline\hline
            & YFSWW3  & $627.22\,(5)$  &$605.18\,(3)$    &--- & --- &$-2.00\,(4)$ \\
$u\bar{d}\mu^-\bar{\nu}_{\mu}$
            &  KoralW & $627.13\,(8)$  &$605.03 \,(7)$   &$0.074$ &$0.074$ & --- \\
\cline{2-4}
200 GeV      &(Y$-$K)/Y&$ 0.01\,(1)\% $ &$ 0.02 \,(1)\%$  &--- & --- & --- \\
\hline\hline
            & YFSWW3 & $1863.60\,(15)$ &$1865.00\,(09)$  &--- & --- &$-2.06 \,(4)$ \\
$u\bar{d} s\bar{c} $
            &KoralW  & $1863.07\,(25)$ &$1864.62\,(21)$  &$0.065$ &$0.064$ & --- \\
\cline{2-4}
200 GeV      &(Y$-$K)/Y  &$ 0.03\,(2)\%$&$ 0.02\,(1)\%$   &--- & --- & --- \\
\hline\hline
\end{tabular}
\caption{\sf
  The numerical check of equality of the integrated reference cross sections 
  $\sigma^R$ from {\tt YFSWW3} and {\tt KoralW}. 
  We also include $\sigma^{\rm CC03}$ in which the ISR is switched off 
  (CC03 Born).
  All results are at $\sqrt{s} = 200$~GeV, with and without cuts.
  Corrections due to the background diagrams and the missing \Order{\alpha} 
  are also indicated; see the text for more explanation.
  The statistical errors corresponding to the last digits 
  are indicated in parentheses.
}
\label{tab:YR-xtot-withcuts-200}
\end{table*}
The first test of type (A) is presented in
Table~\ref{tab:YR-xtot-withcuts-200}, 
where we check whether the integrated reference cross section $\sigma^R$
defined in the previous section is numerically the same
when calculated by {\tt KoralW} and {\tt YFSWW3}.
It is done for three examples of the CC11 class process 
(see Ref.~\cite{4f-LEP2YR:2000} for its definition)
at the center-of-mass system (CMS) energy $\sqrt{s}=200$~GeV.
In addition to the standard $\sigma^R$ of the previous section,
we also show results for $\sigma^{\rm CC03}$, which is a variant of 
$\sigma^R$ in which the ISR is switched off.
As already indicated,
it is the so-called CC03 process (in the 't~Hooft--Feynman gauge).
As we see in Table~\ref{tab:YR-xtot-withcuts-200},
the relative differences of the reference cross sections 
$\sigma^R$ and $\sigma^{\rm CC03}$ of {\tt YFSWW3} and {\tt KoralW}
are below $3\times 10^{-4}$.
This test indicates that the reference differential cross section $d\sigma^R$
is implemented correctly through the corresponding MC weight 
in both  {\tt KoralW} and {\tt YFSWW3}.
In Table~\ref{tab:YR-xtot-withcuts-200}, we also indicate the size
of the correction $\bar{\delta}^{R}_{4f}$ due to the background diagrams and
$\bar{\delta}^{R}_{NL}$ due to the missing \Order{\alpha}, which are
defined as follows:
\begin{equation}
\begin{aligned}
\bar{\delta}^{R}_{4f} & = \frac{\sigma_{K}^{4f+{\rm ISR_{123}+Cc}} 
                  - \sigma^{\rm CC03+ISR_{123}+Cc}}{\sigma^{\rm CC03}},\\
\bar{\delta}^{\rm CC03}_{4f} & = \frac{\sigma_{K}^{4f} 
                  - \sigma^{\rm CC03}}{\sigma^{\rm CC03}},\\
\bar{\delta}^{R}_{NL} & = \frac{\sigma_{Y_a}^{\rm {\cal O}(\alpha)+ISR_{23}+Cc}
                  - \sigma^{\rm CC03+ISR_{123}+Cc}}{\sigma^{\rm CC03}}.
\end{aligned}
\label{deltas:tab}
\end{equation}
%
Note that in Table~\ref{tab:YR-xtot-withcuts-200}
the final-state QCD correction is excluded%
\footnote{ This helps the direct comparison
  between $\bar{\delta}^{R}_{4f}$ and $\bar{\delta}^{\rm CC03}_{4f}$.
  See also Ref.~\cite{4f-LEP2YR:2000} for a description of the QCD correction.}
from $\bar{\delta}^{R}_{4f}$ (i.e.\ $\bar{\delta}^{R}_{4f}$ is
divided by $(1+\alpha_S/\pi)$).
Note also that the Coulomb correction is taken here without screening.

\begin{table*}[!th]
\centering
\begin{tabular}{||c|c||c|c||c|c||c||}
\hline\hline
\multicolumn{7}{||c||}{ NO CUTS}\\
\hline
Description &Program &$\sigma^{\rm CC03}$[fb]  &$\sigma_{R}$[fb] {\vbox to 0.4cm {}}
            &$\bar{\delta}^{\rm CC03}_{4f}[\%]$ &$\bar{\delta}^{R}_{4f}[\%]$ &$\bar{\delta}^{R}_{NL}[\%]$ \\
\hline\hline
          &YFSWW3 & $156.670 \,(16)$ &$122.832 \,(08)$ &--- & --- & $-1.41 \, (4)$ \\
$u\bar{d}\mu^-\bar{\nu}_{\mu}$
          &KoralW & $156.601 \,(24)$ &$122.836 \,(11) $ &$ 0.29 $ &$ 0.25 $ & --- \\
\cline{2-4}
161 GeV    &(Y$-$K)/Y &$ 0.04 \,(2)\% $ &$ 0.00 \,(1)\% $ &--- & --- & --- \\
\hline\hline
\multicolumn{7}{||c||}{ WITH CUTS}\\
\hline
          & YFSWW3 & $151.158 \,(16)$ &$118.482 \,(08)$ &--- & --- &$-1.41 \, (4)$ \\
$u\bar{d}\mu^-\bar{\nu}_{\mu}$
          & KoralW & $151.089 \,(24)$ &$118.485 \,(11)$ &$ 0.29 $ &$ 0.25 $ & --- \\
\cline{2-4}
161 GeV    &(Y$-$K)/Y &$ 0.05 \,(2)\% $ &$ 0.00 \,(1)\% $ &--- & --- & --- \\
\hline\hline
\multicolumn{7}{||c||}{ NO CUTS}\\
\hline
                  & YFSWW3 & $261.368 \,(23)$ &$292.029 \,(18)$ &--- & --- &$-4.95 \, (4)$ \\
$u\bar{d}\mu^-\bar{\nu}_{\mu}$
                  & KoralW & $261.348\,(17)$ &$291.979\,(19) $ &$-0.51 $ &$-0.51 $ & --- \\
\cline{2-4}
 500 GeV           &(Y$-$K)/Y &$ 0.01 \,(1)\% $ &$ 0.02 \,(1)\% $ &--- & --- & --- \\
\hline
{\tt YFSWW3}-like & KoralW & $261.348\,(17)$ &$291.980\,(22) $ &$-0.51 $ &$-0.51 $ & --- \\
\cline{2-4}
extrapol.         &(Y$-$K)/Y &$ 0.01 \,(1)\% $ &$ 0.02 \,(1)\% $ &--- & --- & --- \\
\hline\hline
\multicolumn{7}{||c||}{ WITH CUTS}\\
\hline
                  & YFSWW3 & $181.505 \,(22)$ &$209.449 \,(17)$ &--- & --- &$-6.34 \, (4)$ \\
$u\bar{d}\mu^-\bar{\nu}_{\mu}$
                  &KoralW & $181.480 \,(17)$ &$209.592 \,(18) $ &$-0.69 $ &$-0.69  $ & --- \\
\cline{2-4}
500 GeV            &(Y$-$K)/Y &$ 0.01 \,(1)\% $ &$-0.07 \,(1)\% $ &--- & --- & --- \\
\hline
{\tt YFSWW3}-like &KoralW & $181.480 \,(17)$ &$209.426 \,(21) $ &$-0.69 $ &$-0.69  $ & --- \\
\cline{2-4}
extrapol.         &(Y$-$K)/Y &$ 0.01 \,(1)\% $ &$ 0.01 \,(1)\% $ &--- & --- & --- \\
\hline\hline
\end{tabular}
\caption{\sf
  The numerical check of equality of the integrated reference cross sections 
  $\sigma^R$ from {\tt YFSWW3} and {\tt KoralW},
  as in Table~\protect\ref{tab:YR-xtot-withcuts-200},
  but for two other energies: 161~GeV and 500~GeV.
}
\label{tab:YR-xtot-withcuts-161-500}
\end{table*}
In Table~\ref{tab:YR-xtot-withcuts-161-500}, we
present the analogous (A)-type test for the integrated cross sections
as in Table~\ref{tab:YR-xtot-withcuts-200} but
for two other CMS energies: close to the $WW$-threshold, 161~GeV,
and at 500~GeV, within the range of the future Linear Collider.
As we see, the result of the test is again positive.
The integrated reference cross section $\sigma^R$ is again the same from both
{\tt YFSWW3} and {\tt KoralW} to within $5\times 10^{-4}$.
In Table \ref{tab:YR-xtot-withcuts-161-500}, we also show the effect
of switching from the extrapolation/reduction procedure of {\tt KoralW}
to that of {\tt YFSWW3}
(which is the source of the largest discrepancy of $-0.07\%$).
One can verify that the size of this effect is at the sub-per mille level
and appears only in the case with imposed cuts. For the case without
any cuts there is no difference in the total cross section, as
expected. Note that Table \ref{tab:YR-xtot-withcuts-161-500} shows
the case of $\sqrt{s}=500$~GeV. At $\sqrt{s}=200$~GeV, there is no difference
between these two extrapolation procedures.


\begin{table*}[!th]
\centering
\begin{tabular}{||c|l||c|c||c|c||}
\hline\hline
  \multicolumn{2}{||c||}{ Type of calculation} 
& \multicolumn{2}{c||}{ NO CUTS} 
& \multicolumn{2}{c||}{ WITH CUTS} \\
\hline
 MC & Formula  &$\sigma_i$ [fb] &$\frac{\sigma_i}{\sigma_1} -1$ 
                     &$\sigma_i$ [fb] &$\frac{\sigma_i}{\sigma_1} -1$\\
\hline\hline
$Y$           &$\sigma_1=\sigma_{Y_a}^{{\cal O}(\alpha)+ISR_{23}+Cc}$
    &$623.07\,(14)$  &---                &$593.03\,(15)$ &--- \\
$Y+\delta Y$  &$\sigma_2=\int d\sigma_{Y_a}^R [1+\delta^R_{NL-}]_Y   $
    &$622.59\,(14)$  &$-0.08(3)\%$        &$592.57\,(14)$ &$-0.08(3)\%$ \\ 
$K+\delta Y$    &$\sigma_3=\int d\sigma_{K}^R   [1+\delta^R_{NL-}]_Y $
  & $622.68\,(17)$   &$-0.06(4)\%$        &$592.58\,(17)$ &$-0.06(4)\%$ \\
$K+\delta Y$    &$\sigma_4=\int d\sigma_{K} \left[1+\frac{\delta^R_{NL-}}{1+\delta^R_{4f}}\right] $
  & $623.24\,(6)$   &$+0.03(2)\%$        &$593.14\,(6)$ &$+0.02(3)\%$ \\
$K+\delta Y$    &$\sigma_5=\sigma_3$({\small $4f$ presampler})
  & $622.74\,(6)$   &$-0.05(2)\%$        &$592.66\,(6)$ &$-0.06(3)\%$ \\
\hline
\hline
\end{tabular}
\caption{\sf
  The self-test of {\tt YFSWW3} ($Y+\delta Y$) and the
  test of CMC {\tt KoralW$\&$YFSWW3} ($K+\delta Y$)
  for the \Order{\alpha}-corrected integrated cross section 
  $\sigma^{{\cal O}(\alpha)+ISR_{23}+Cc}$
  (no background diagrams).
  All results are at $\sqrt{s} = 200$~GeV, for the 
  $u\bar{d}\mu^-\bar{\nu}_{\mu}$ final state,
  with and without cuts.
  The statistical errors corresponding to the last digits are
  indicated in parentheses.
}
\label{tab:kandy-200}
\end{table*}
In Table~\ref{tab:kandy-200} (test (B)), we show in the {\em first line}
the standard ``best'' result of {\tt YFSWW3}.
In the {\em second line} we present the
self-test of {\tt YFSWW3} of the type $Y+\delta Y$
in which events are primarily generated with {\tt YFSWW3} according to the
reference distribution $d\sigma^R=d\sigma^{CC03+ISR_{123}+Cc}$ 
and are later reweighted using the weight
$[1+\delta^R_{NL-}]_Y$ calculated also by {\tt YFSWW3}.
In the {\em third line} we show the test of the CMC {\tt KoralW$\&$YFSWW3} of the type $K+\delta Y$,
in which events are generated using {\tt KoralW} according to $d\sigma^R$
and are reweighted using the same weight $[1+\delta^R_{NL-}]_Y$
provided by {\tt YFSWW3}.
The $\sigma_2$ and $\sigma_3$ represent 
 the same quantity generated in two different ways, and indeed the numerical
agreement between them is well within the statistical errors.
The agreement of the latter two results with the first one is within $0.08\%$.
It is sufficient for the purpose of LEP2.
Note that we do not expect perfect agreement, owing to the use of 
the approximate $\delta^R_{NL-}$.
The above discrepancy is also much smaller than the size of the NL correction
itself, which is often up to $2\%$.
In the {\em fourth line} we show the result of the CMC {\tt KoralW$\&$YFSWW3}
in the full operational mode.
Since the four-fermion correction is rather small we have checked, using differences
of the MC weights,
that the four-fermion correction contribution in $\sigma_4$ is $0.4985(17)\,$fb,
that is $0.0728(2)\%$ in units
of the CC03 cross section (adjusting also for the QCD factor
$(1+\alpha_S/\pi)$) for the ``No-Cuts'' case, and $0.4822(17)\,$fb, i.e.
$0.0741(3)\%$ for the ``With-Cuts'' case.
This is fully compatible with the results 
shown in Table~\ref{tab:YR-xtot-withcuts-200}.
Finally, in the {\em fifth line} we show again calculation of the cross section
equal to $\sigma_3$, obtained, however, 
as a by-product of the previous CMC {\tt KoralW$\&$YFSWW3} run
in the full operational mode, with different arrangement of the MC weights.
As compared with the original $\sigma_3$ calculation, {\tt KoralW} is now run in the CCall mode
instead of the CC03.
The equality $\sigma_3=\sigma_5$ (within the statistical errors),
provides yet another consistency check of the CMC {\tt KoralW$\&$YFSWW3}.

\subsection{One-dimensional distributions}

\begin{figure}[!ht]
\centering
\setlength{\unitlength}{0.1mm}
\begin{picture}(1600,1480)
\put( 450,1420){\makebox(0,0)[cb]{\bf CAL05} }
\put(1250,1420){\makebox(0,0)[cb]{\bf CAL05} }
\put( 850,1410){\makebox(0,0)[cb]{\large $u \bar{d} \mu^- \bar{\nu}_{\mu}$} }
\put(  0,700){\makebox(0,0)[lb]{
\epsfig{file=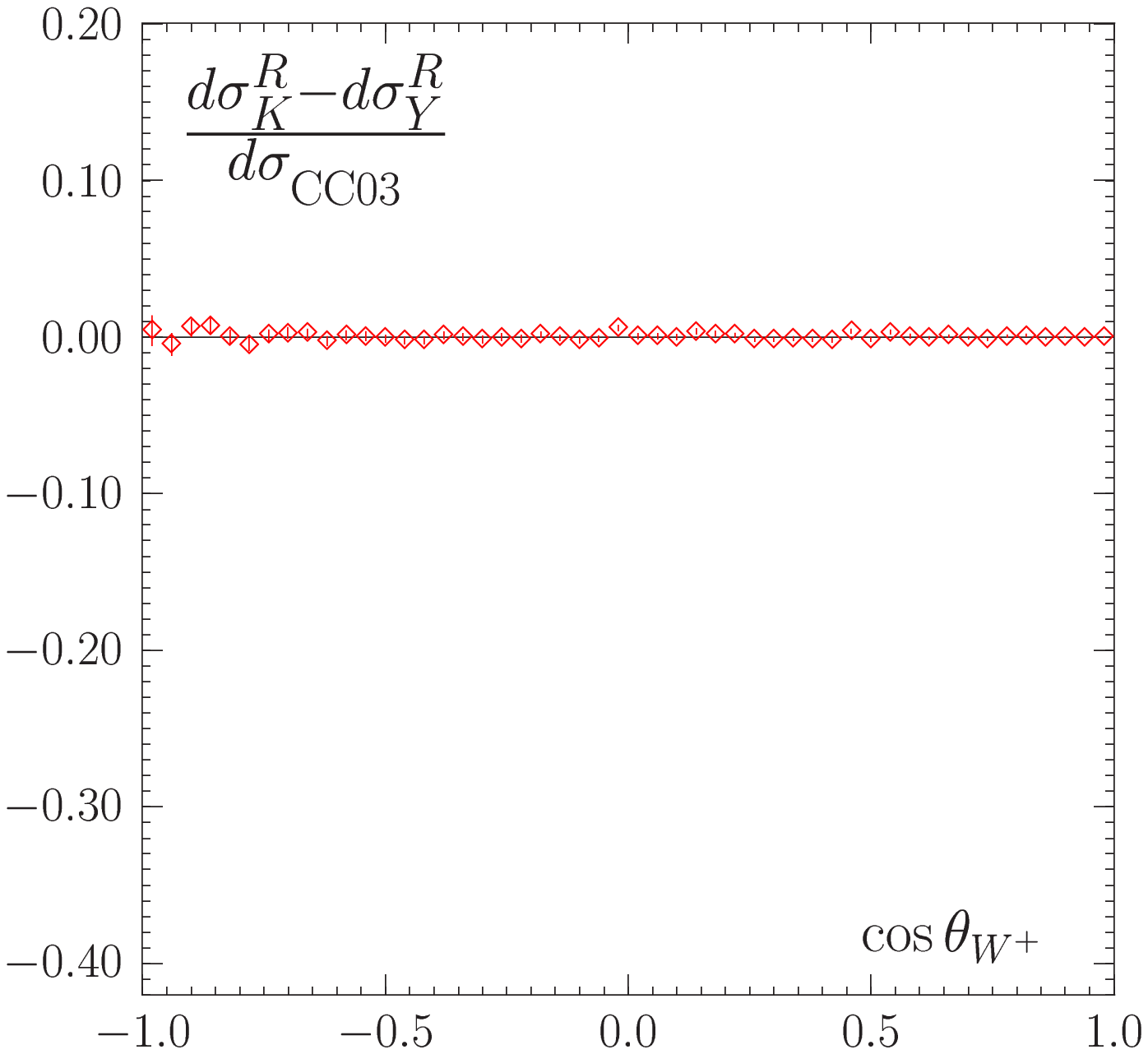, width=80mm,height=70mm}
}}
\put(  0,  0){\makebox(0,0)[lb]{
\epsfig{file=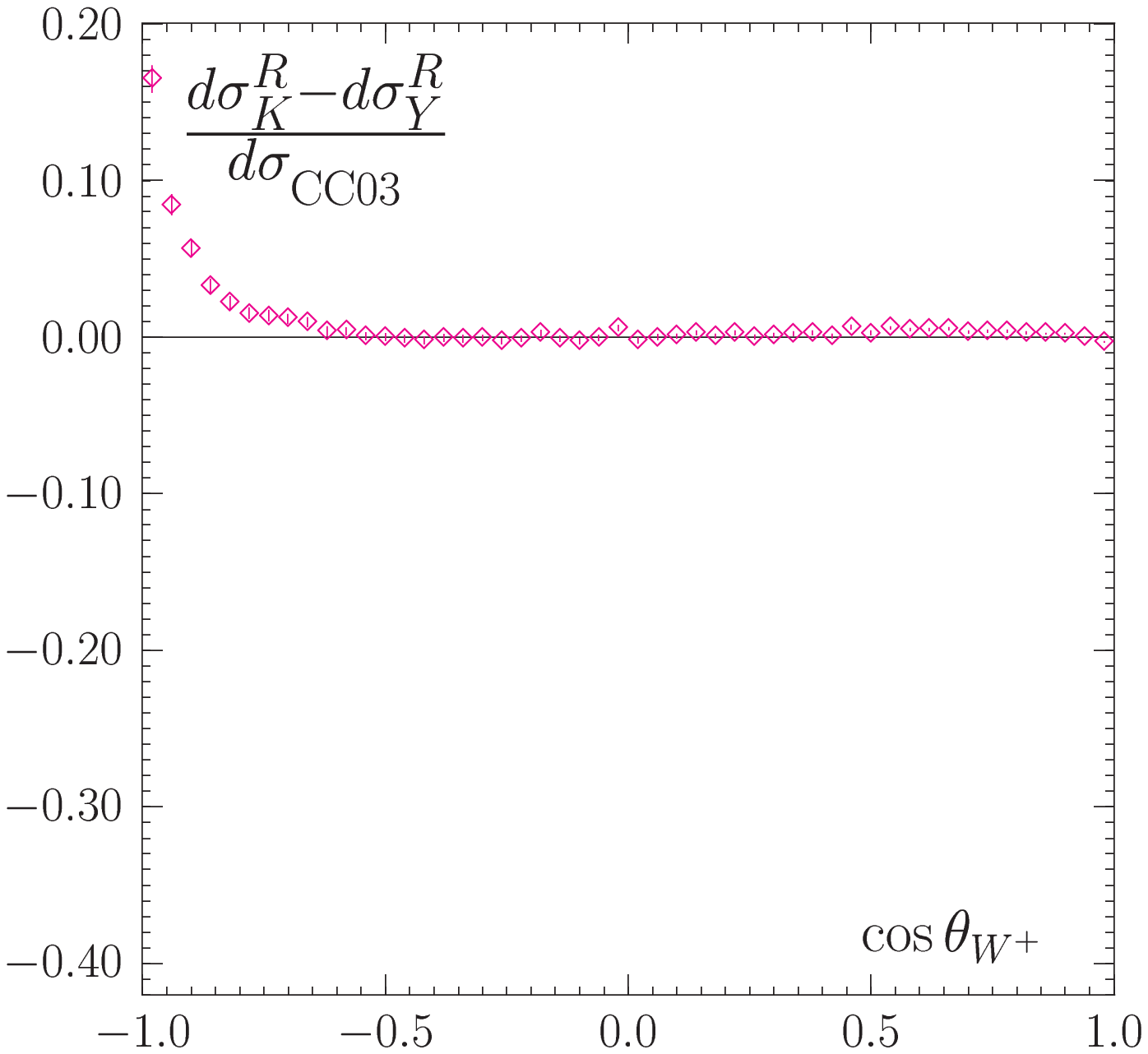,     width=80mm,height=70mm}
}}
\put(800,700){\makebox(0,0)[lb]{
\epsfig{file=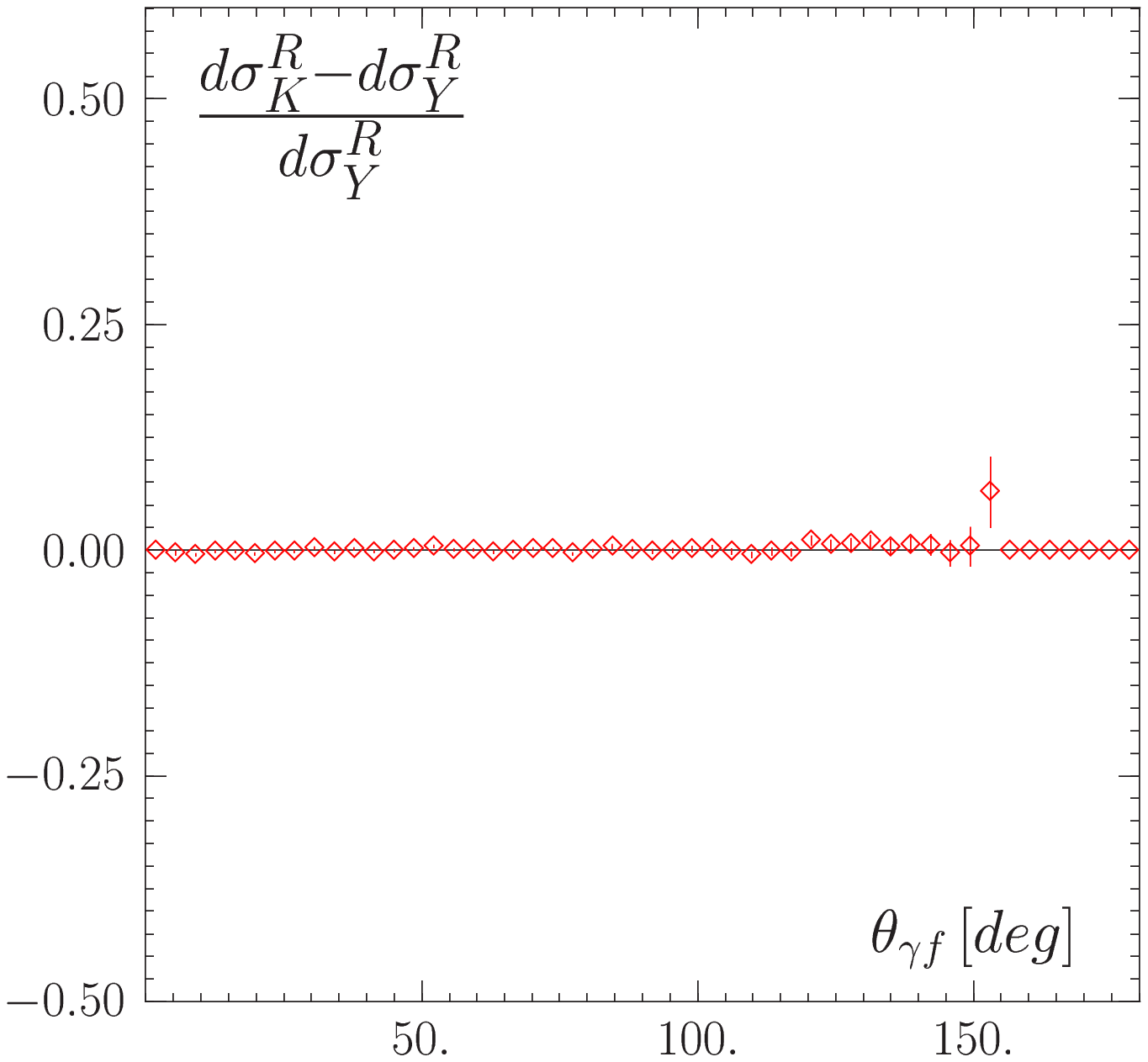, width=80mm,height=70mm}
}}
\put(800,  0){\makebox(0,0)[lb]{
\epsfig{file=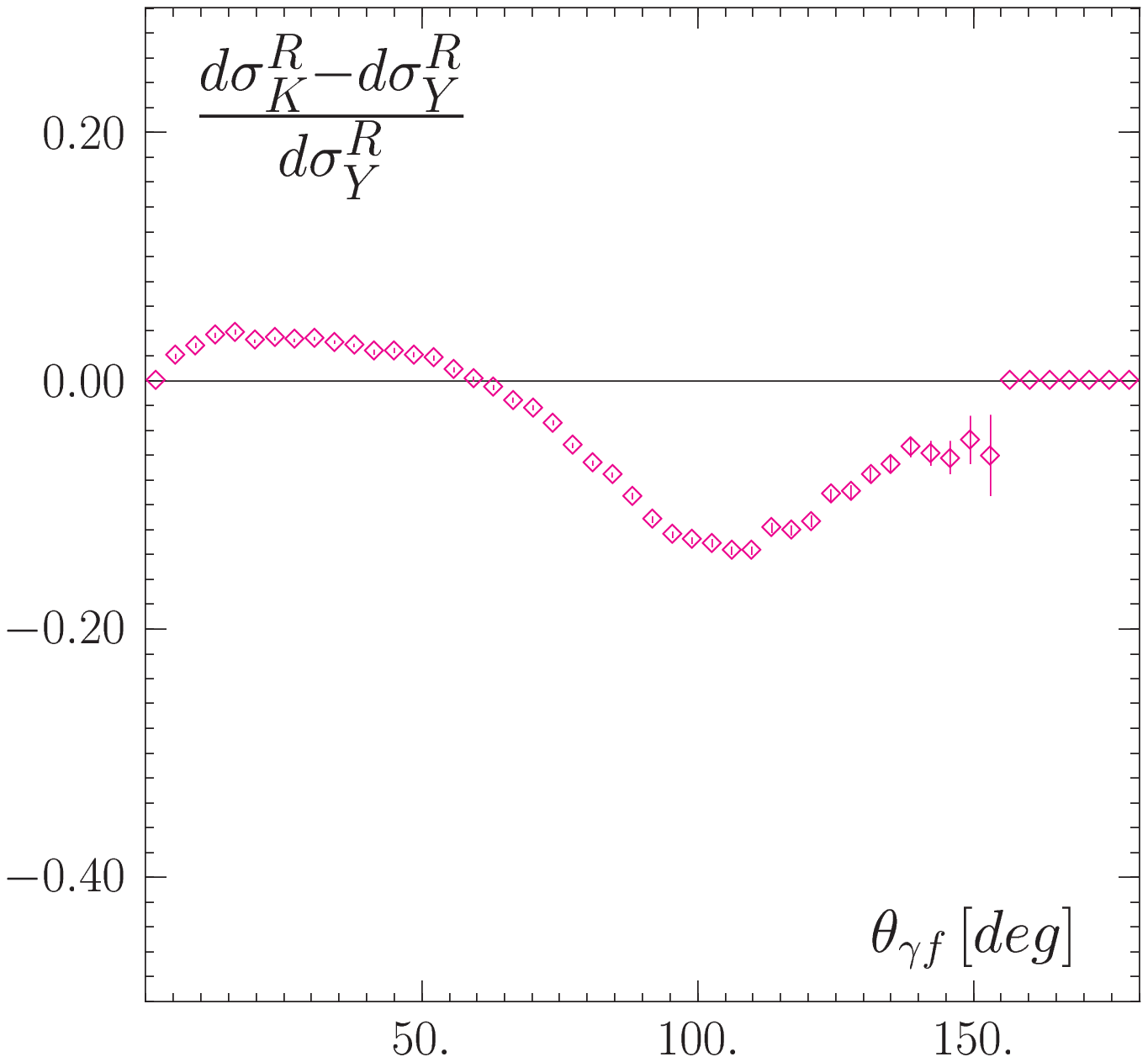,     width=80mm,height=70mm}
}}
\end{picture}
\caption{\small\sf
  The comparison of the reduction procedures in {\tt KoralW} and  {\tt YFSWW3}.
  On the lower plots they are different, i.e.\ in {\tt KoralW} we keep 
  that of 1.41,
  while in the upper plot they are the same, i.e.\ in {\tt KoralW}
  it is adjusted to be the same as in {\tt YFSWW3}.
  Plotted are the distributions of the $W^+$ polar angle w.r.t. the $e^+$ beam
  and of the angle between the hardest photon with respect to    
  the nearest final-state charged fermion at $\sqrt{s}=500$~GeV.
}
\label{fig:udmv-thgf}    
\end{figure}

The example of a purely technical test is also presented in
Fig.~\ref{fig:udmv-thgf}, 
where we check that the $d\sigma^R$ distribution of the polar angle of the 
$W$ and of the photon angle with respect to the final charged fermion 
is the same,
after adjusting the reduction/extrapolation procedure to be the same in
{\tt KoralW} as in {\tt YFSWW3}, while it was not the same for the original
reduction/extrapolation procedure of {\tt KoralW} version {\tt 1.41}.

\begin{figure}[!ht]
\centering
\setlength{\unitlength}{0.1mm}
\begin{picture}(1600,1480)
\put( 450,1430){\makebox(0,0)[cb]{\bf CAL05} }
\put(1250,1430){\makebox(0,0)[cb]{\bf CAL25} }
\put( 850,1430){\makebox(0,0)[cb]{\large $\nu_{\mu} \mu^+ \tau^- \bar{\nu}_{\tau}$} }
\put(  0,700){\makebox(0,0)[lb]{
\epsfig{file=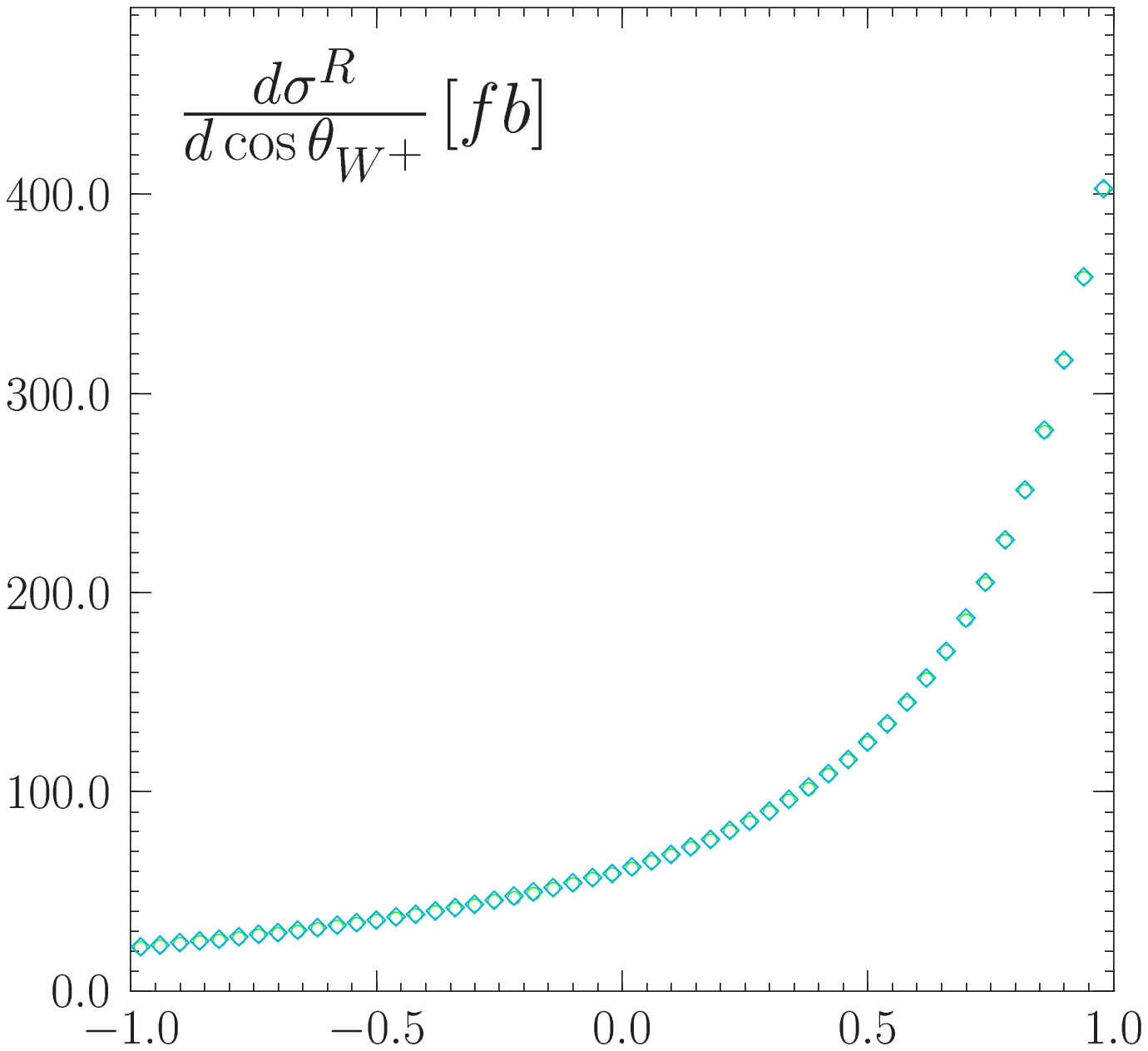            ,width=80mm,height=70mm}
}}
\put(  0,  0){\makebox(0,0)[lb]{
\epsfig{file=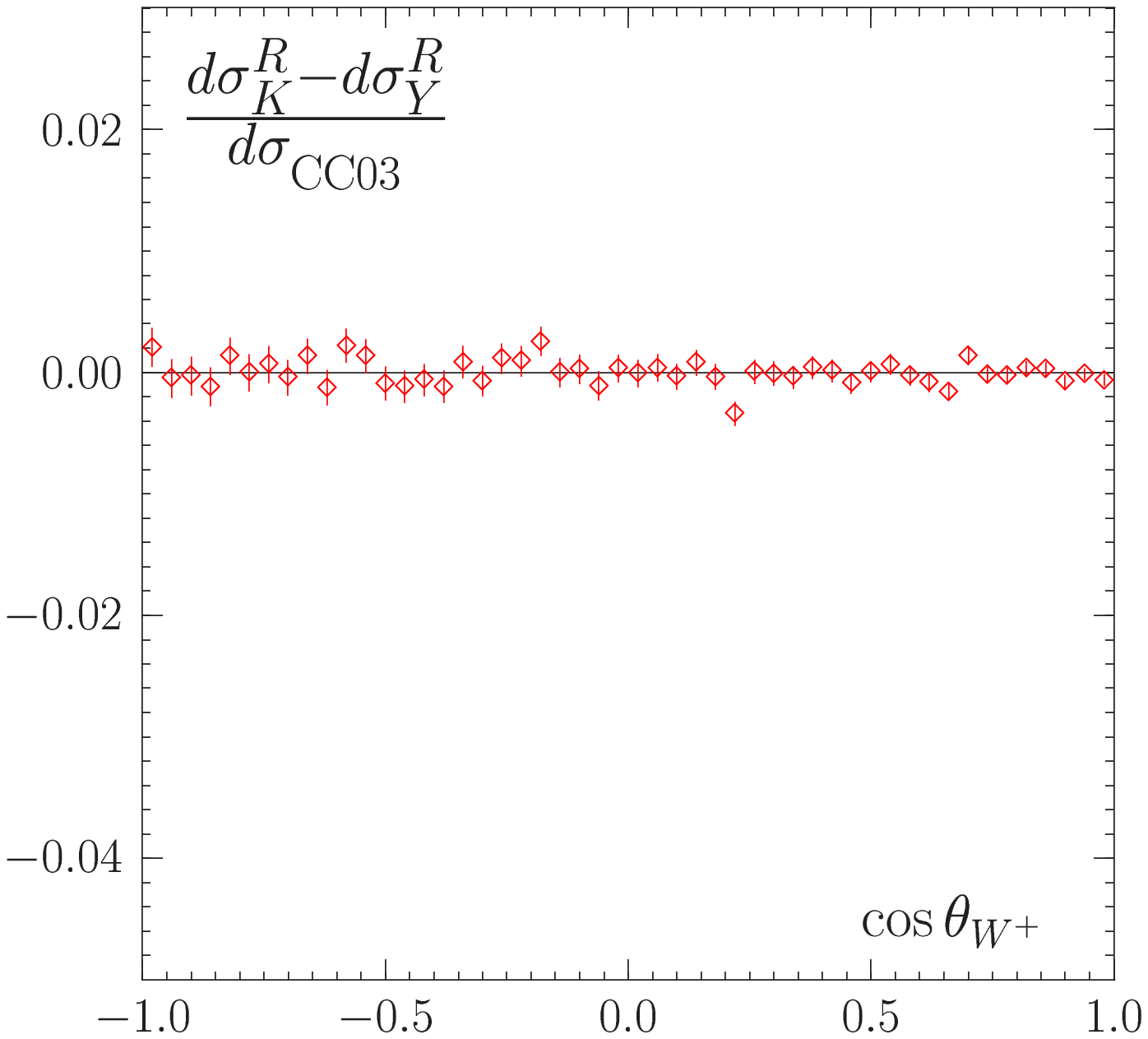        ,width=80mm,height=70mm}
}}
\put(800,700){\makebox(0,0)[lb]{
\epsfig{file=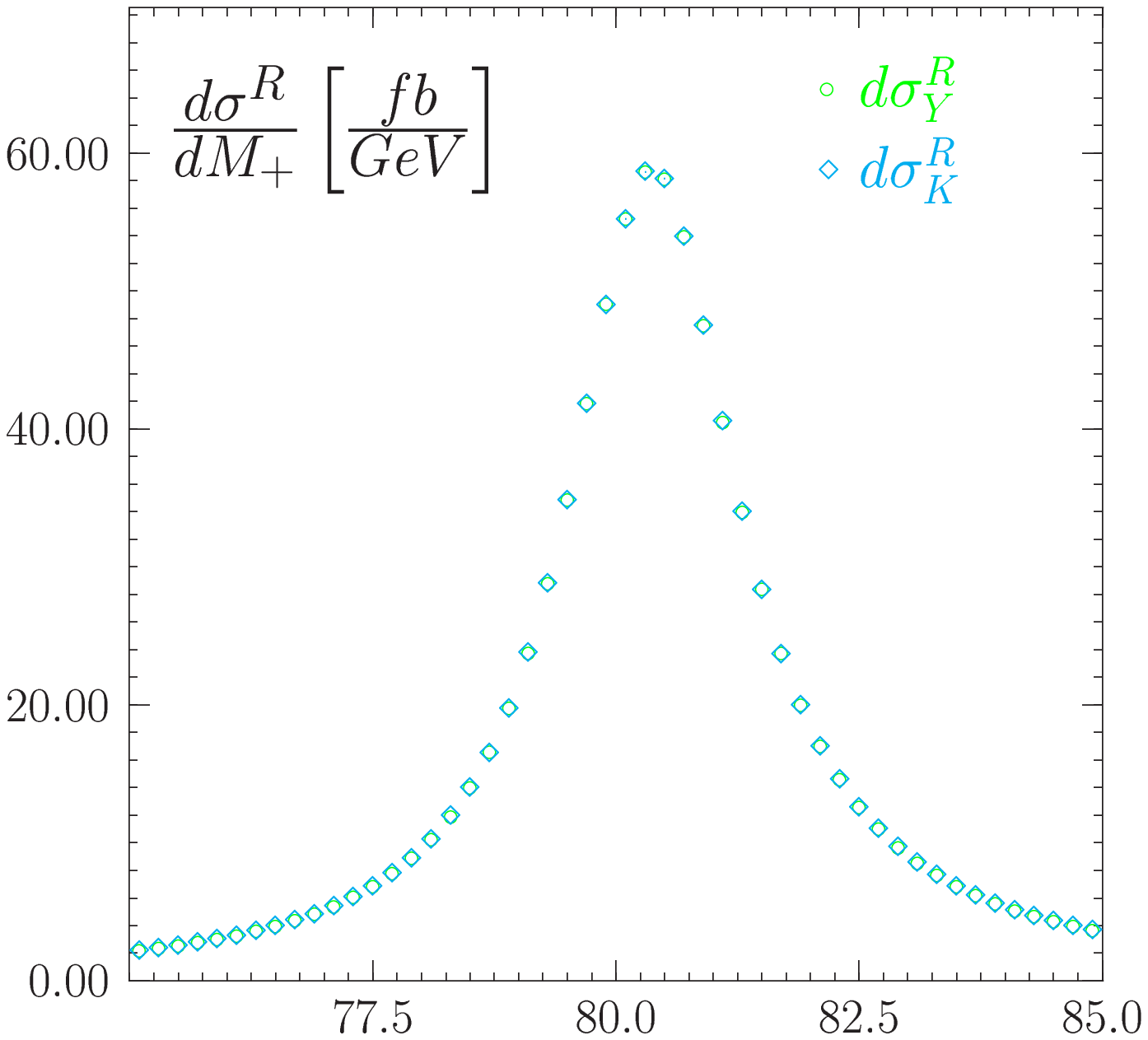            ,width=80mm,height=70mm}
}}
\put(800,  0){\makebox(0,0)[lb]{
\epsfig{file=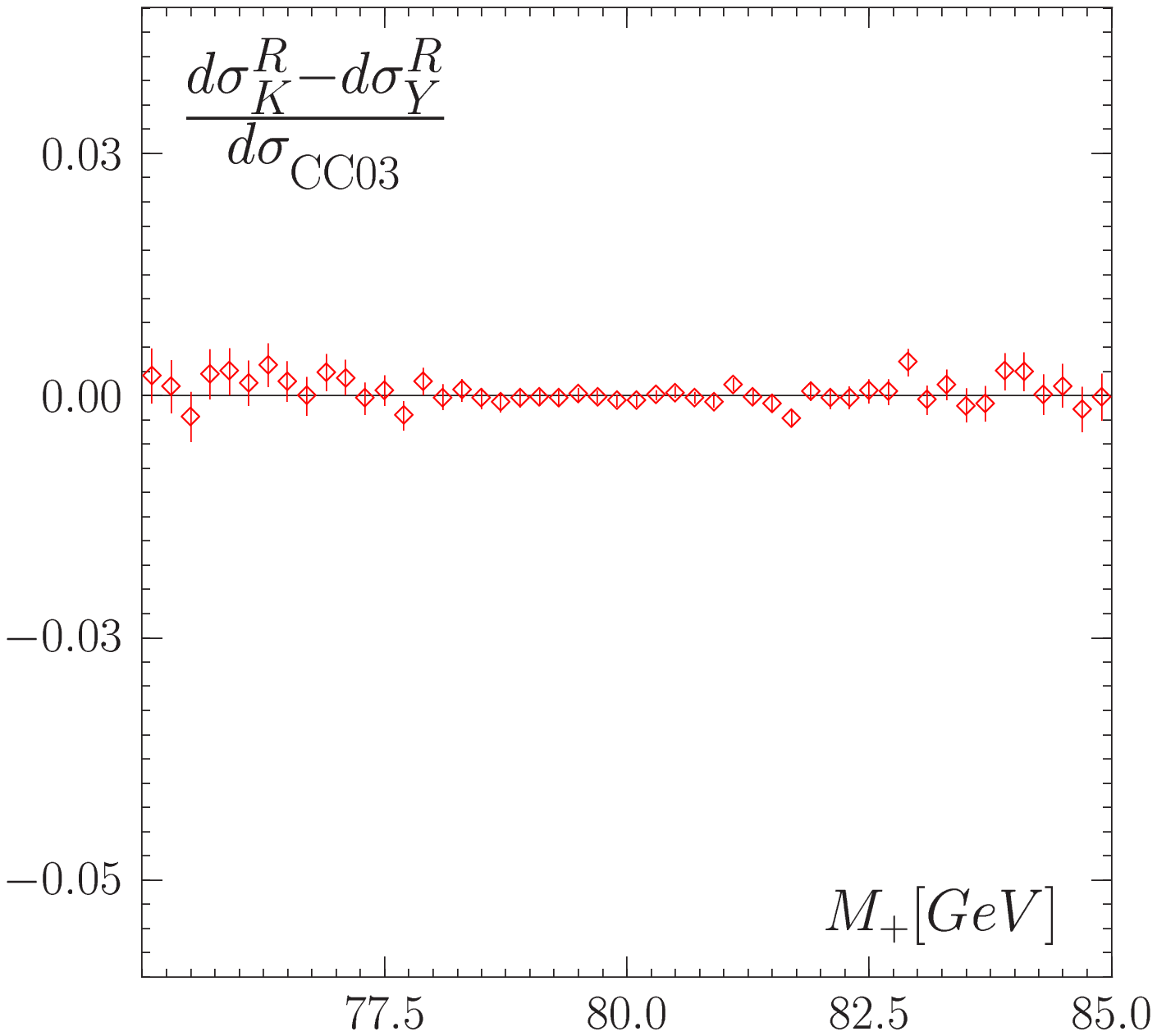        ,width=80mm,height=70mm}
}}
\end{picture}
\caption{\small\sf
  The technical test of the universality of $d\sigma_{R}$ from {\tt YFSWW3} and {\tt KoralW}.
  Plotted are the distributions of the $W^+$ mass and of the $W^+$  polar angle
  w.r.t. the $e^+$ beam at $\sqrt{s}=200$~GeV.
}
\label{fig:vmtv-MWp}     
\end{figure}

\begin{figure}[!ht]
\centering
\setlength{\unitlength}{0.1mm}
\begin{picture}(1600,1480)
\put( 450,1430){\makebox(0,0)[cb]{\bf CAL05} }
\put(1250,1430){\makebox(0,0)[cb]{\bf CAL25} }
\put( 850,1430){\makebox(0,0)[cb]{\large $u \bar{d} s \bar{c}$} }
\put(  0,700){\makebox(0,0)[lb]{
\epsfig{file=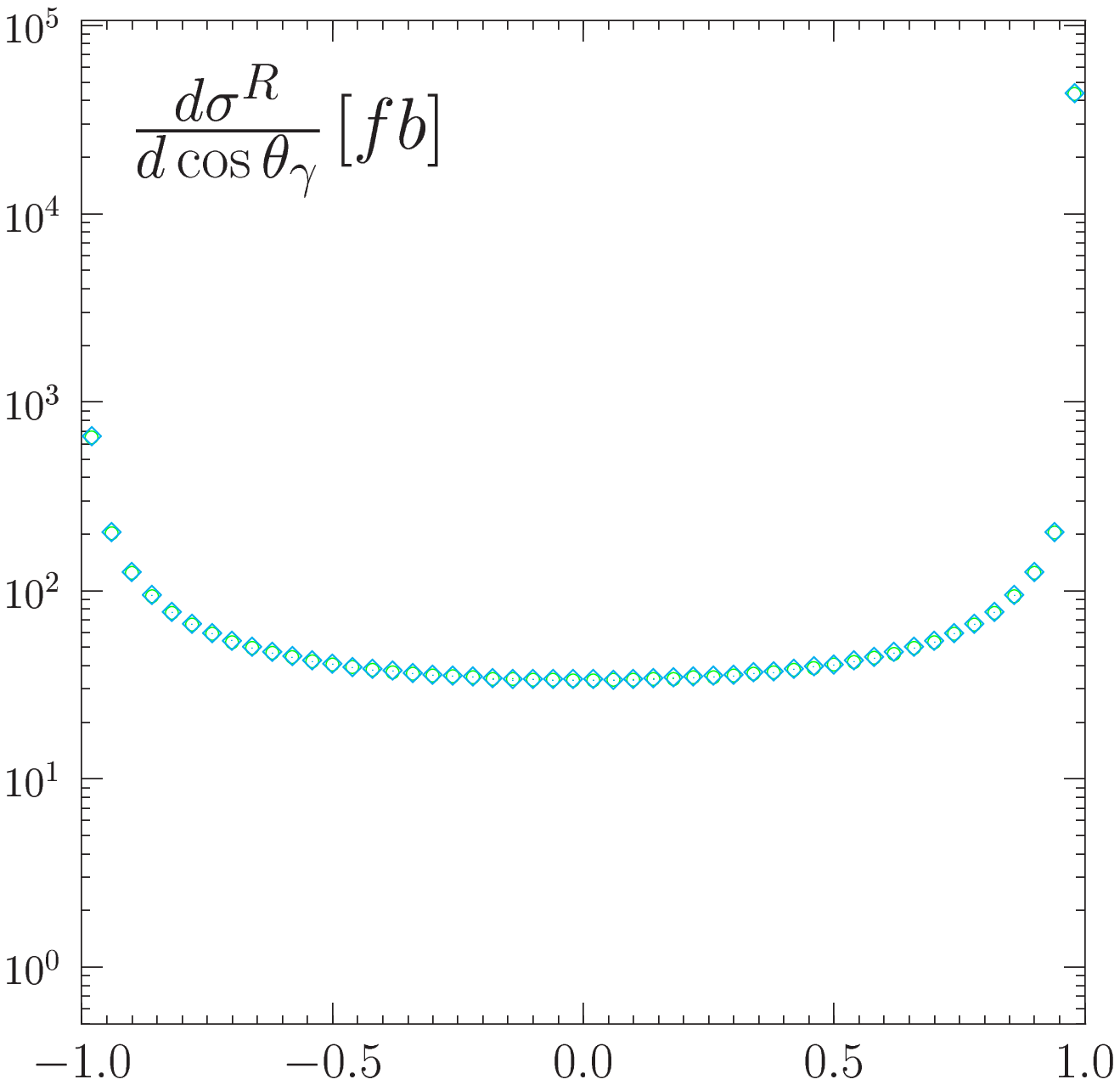          ,width=80mm,height=70mm}
}}
\put(  0,  0){\makebox(0,0)[lb]{
\epsfig{file=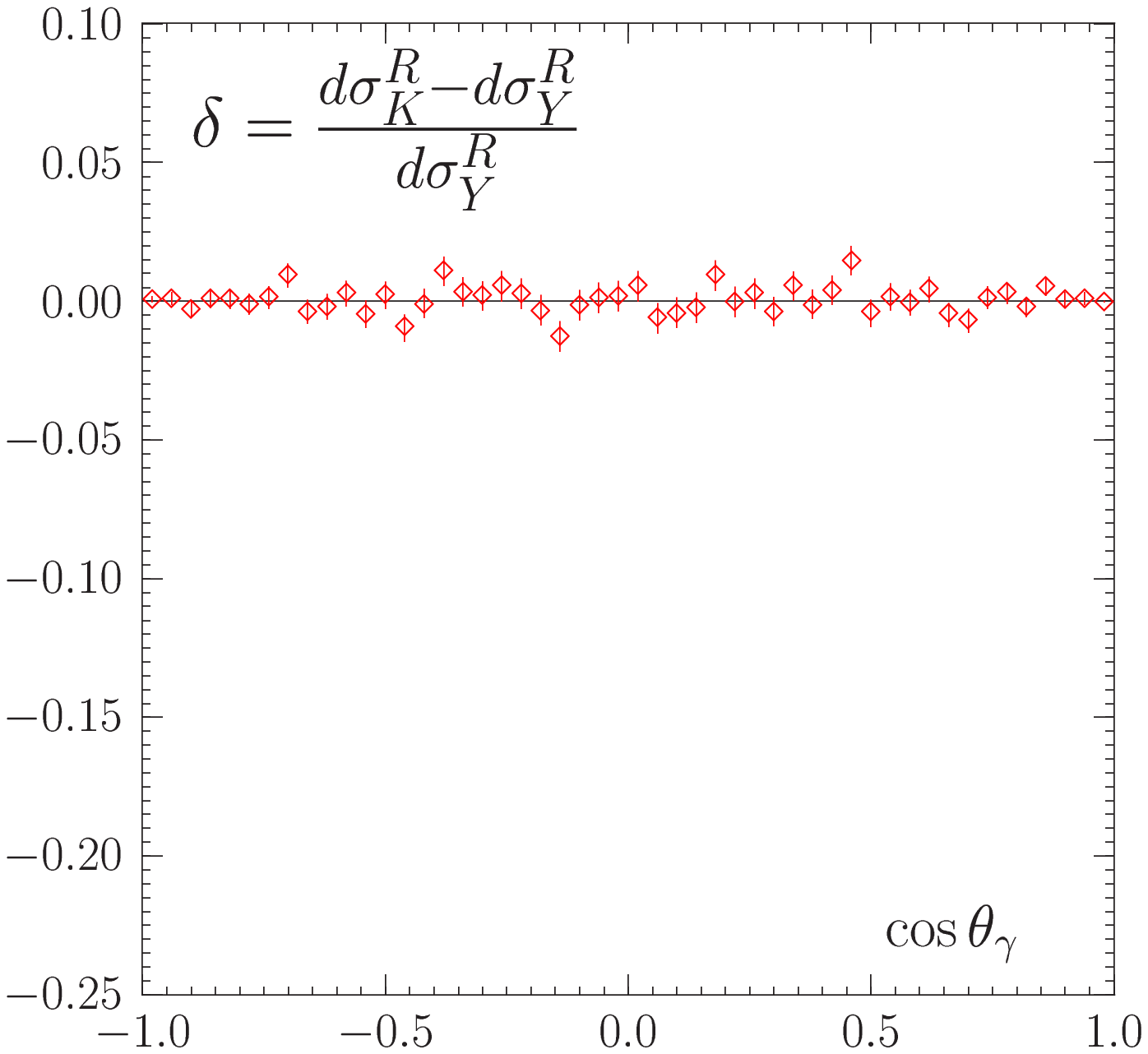      ,width=80mm,height=70mm}
}}
\put(800,700){\makebox(0,0)[lb]{
\epsfig{file=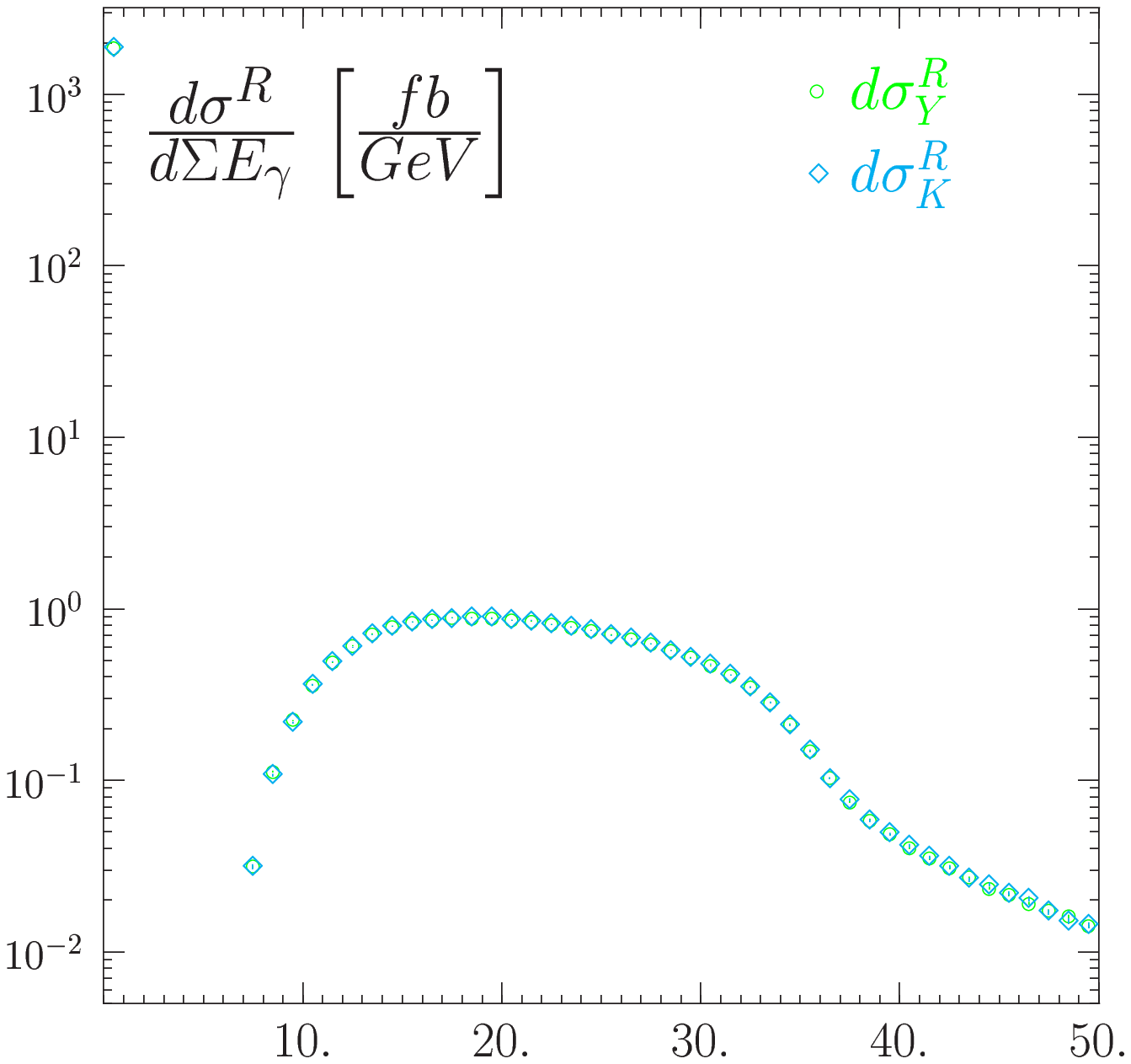          ,width=80mm,height=70mm}
}}
\put(800,  0){\makebox(0,0)[lb]{
\epsfig{file=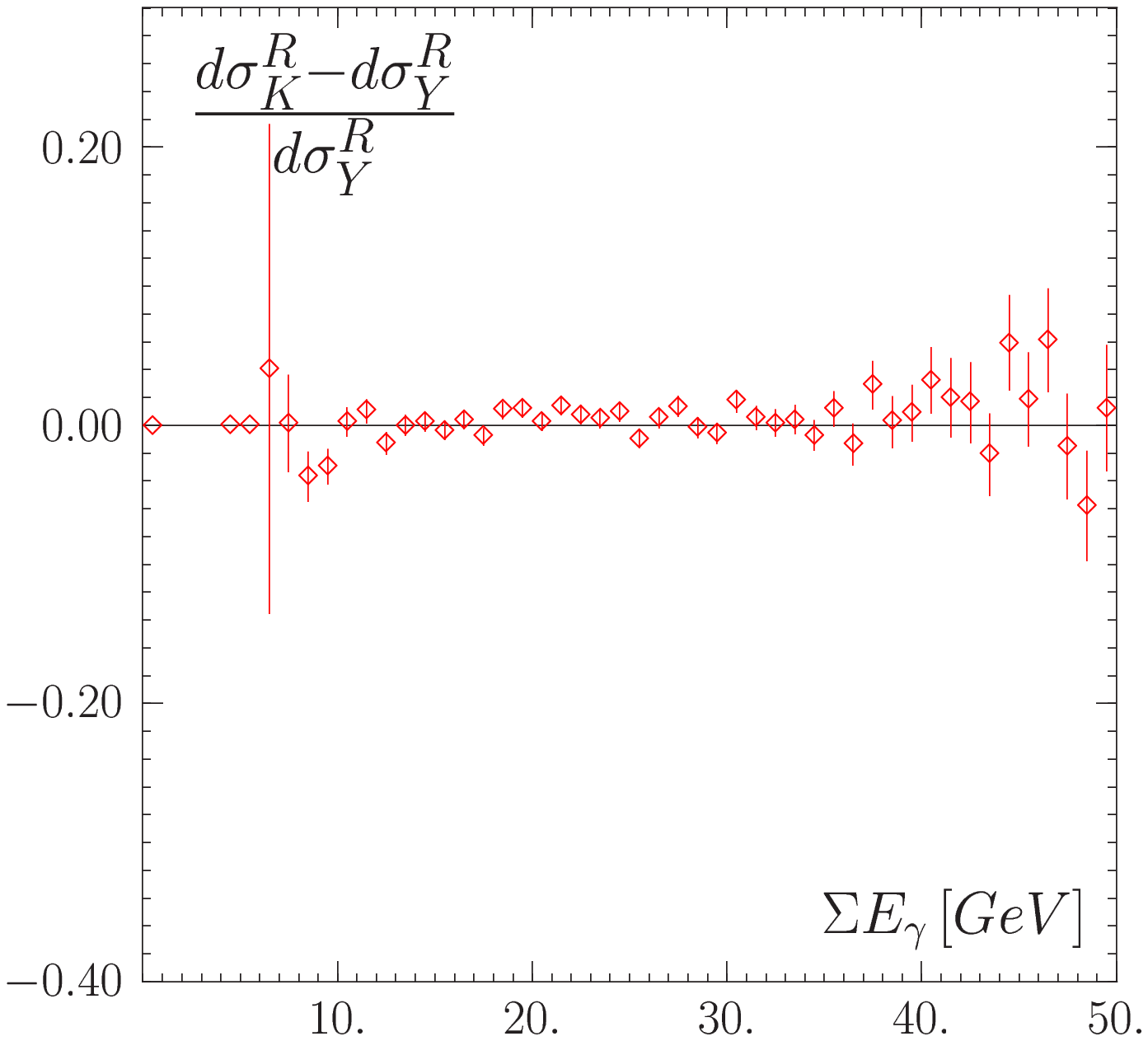      ,width=80mm,height=70mm}
}}
\end{picture}
\caption{\small\sf
  The technical test of the universality of $d\sigma_{R}$ from {\tt YFSWW3} and {\tt KoralW}.
  Plotted are the distributions of the angle of the photon w.r.t. the $e^+$ beam
  and the sum of the photon energy
  at $\sqrt{s}=200$~GeV.
}
\label{fig:udsc-Egsum}   
\end{figure}

Another technical test is presented in Figs.\ 
\ref{fig:vmtv-MWp} and \ref{fig:udsc-Egsum}
(test (A))
where we check, for various distributions, that the reference distribution
$d\sigma^R$ is identically implemented in {\tt KoralW} and {\tt YFSWW3}.
The distributions of the $W$ mass and angle, of the photon energy and angle
are the same within the statistical errors.
We conclude that $d\sigma^R$ is implemented correctly, 
at the precision level relevant to LEP2.

\begin{figure}[!ht]
\centering
\setlength{\unitlength}{0.1mm}
\begin{picture}(1600,1480)
\put( 450,1430){\makebox(0,0)[cb]{\bf CAL05} }
\put(1250,1430){\makebox(0,0)[cb]{\bf CAL25} }
\put( 850,1430){\makebox(0,0)[cb]{\large $u \bar{d} \mu^- \bar{\nu}_{\mu}$} }
\put(  0,700){\makebox(0,0)[lb]{
\epsfig{file=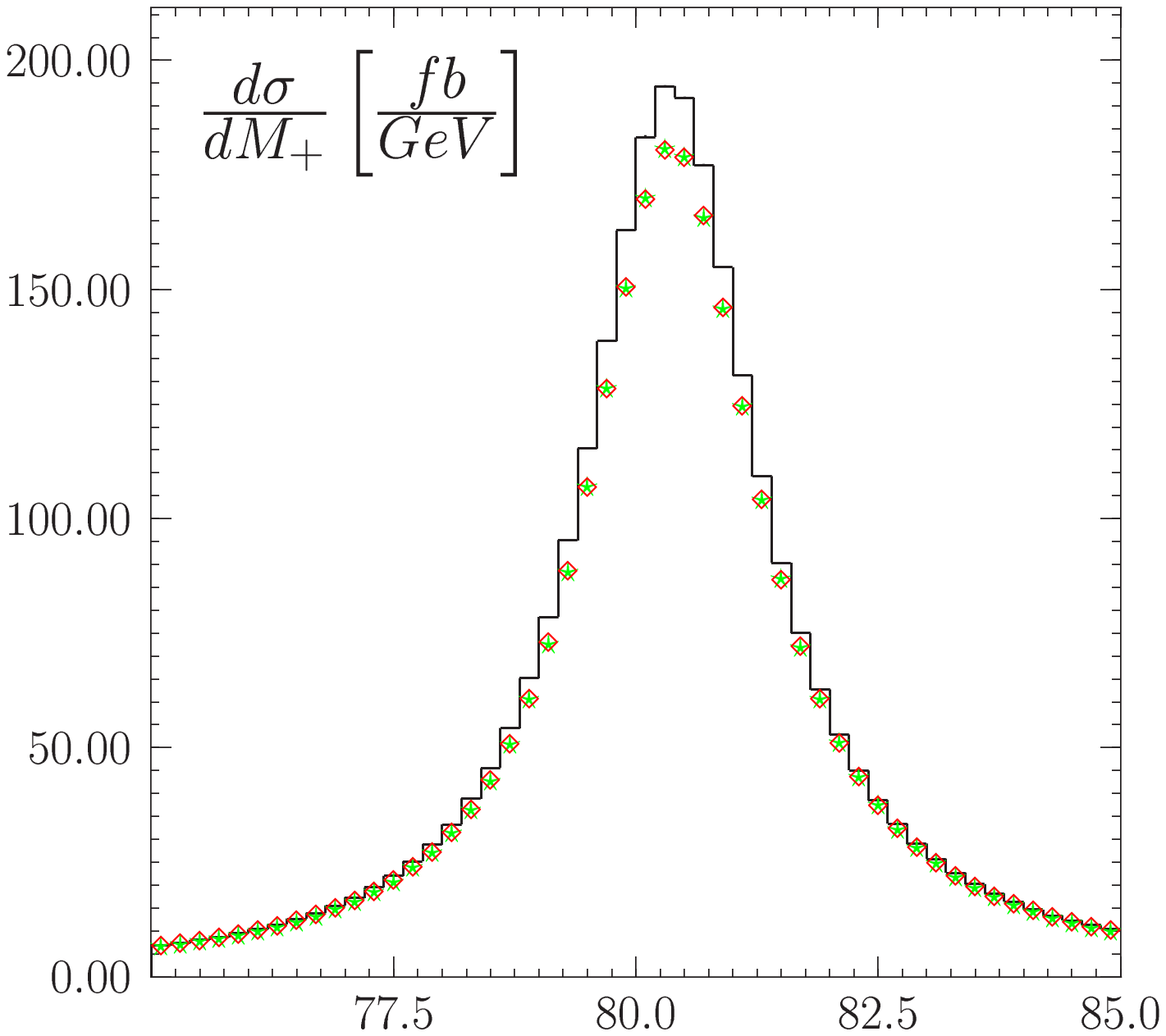            ,width=80mm,height=70mm}
}}
\put(  0,  0){\makebox(0,0)[lb]{
\epsfig{file=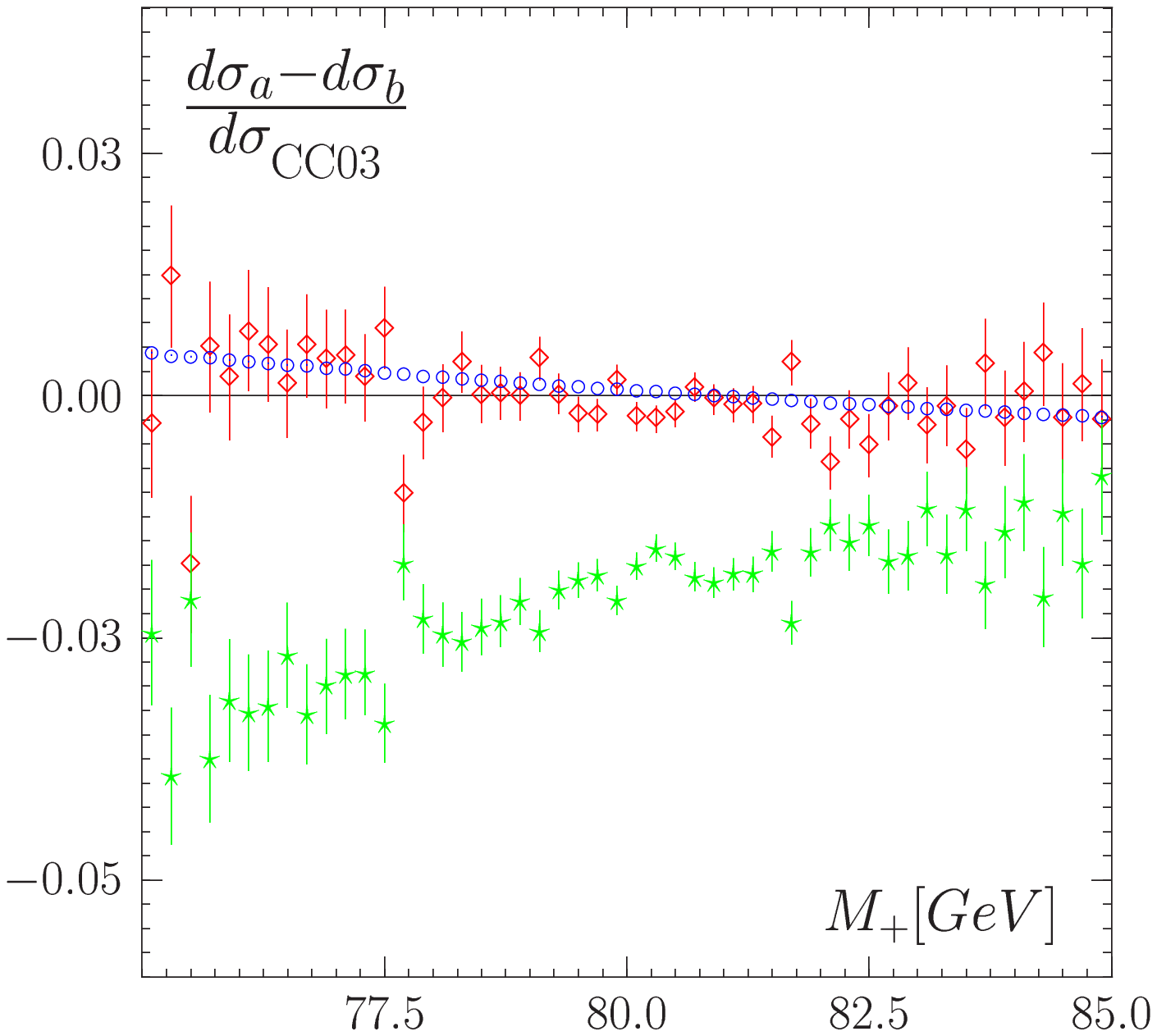        ,width=80mm,height=70mm}
}}
\put(800,700){\makebox(0,0)[lb]{
\epsfig{file=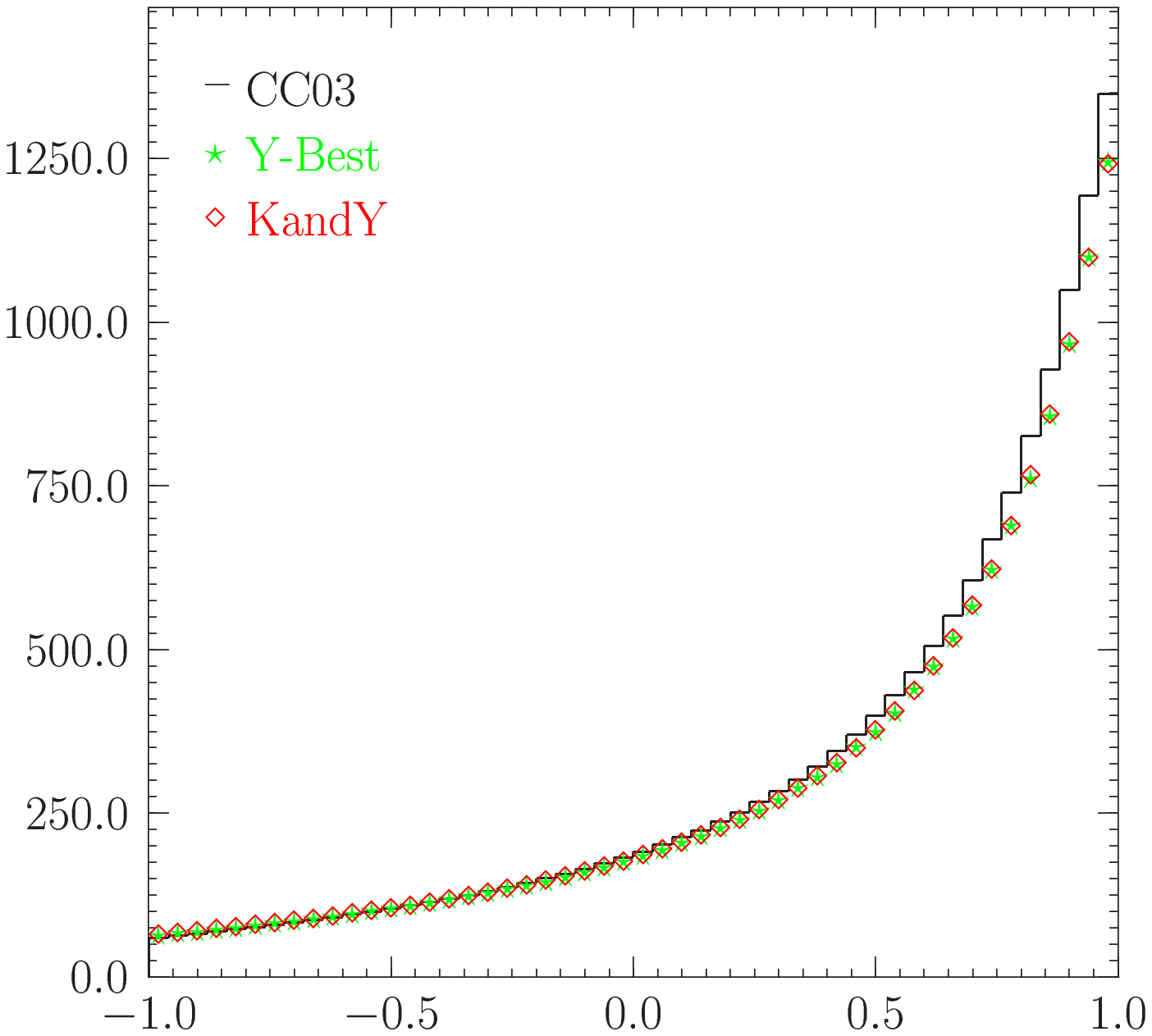        ,width=80mm,height=70mm}
\put(-300, 620){\makebox(0,0)[l]{\large                          
$\frac{d\sigma}{d \cos\theta_{W^+} }\, [fb] $ }}                
}}
\put(800,  0){\makebox(0,0)[lb]{
\epsfig{file=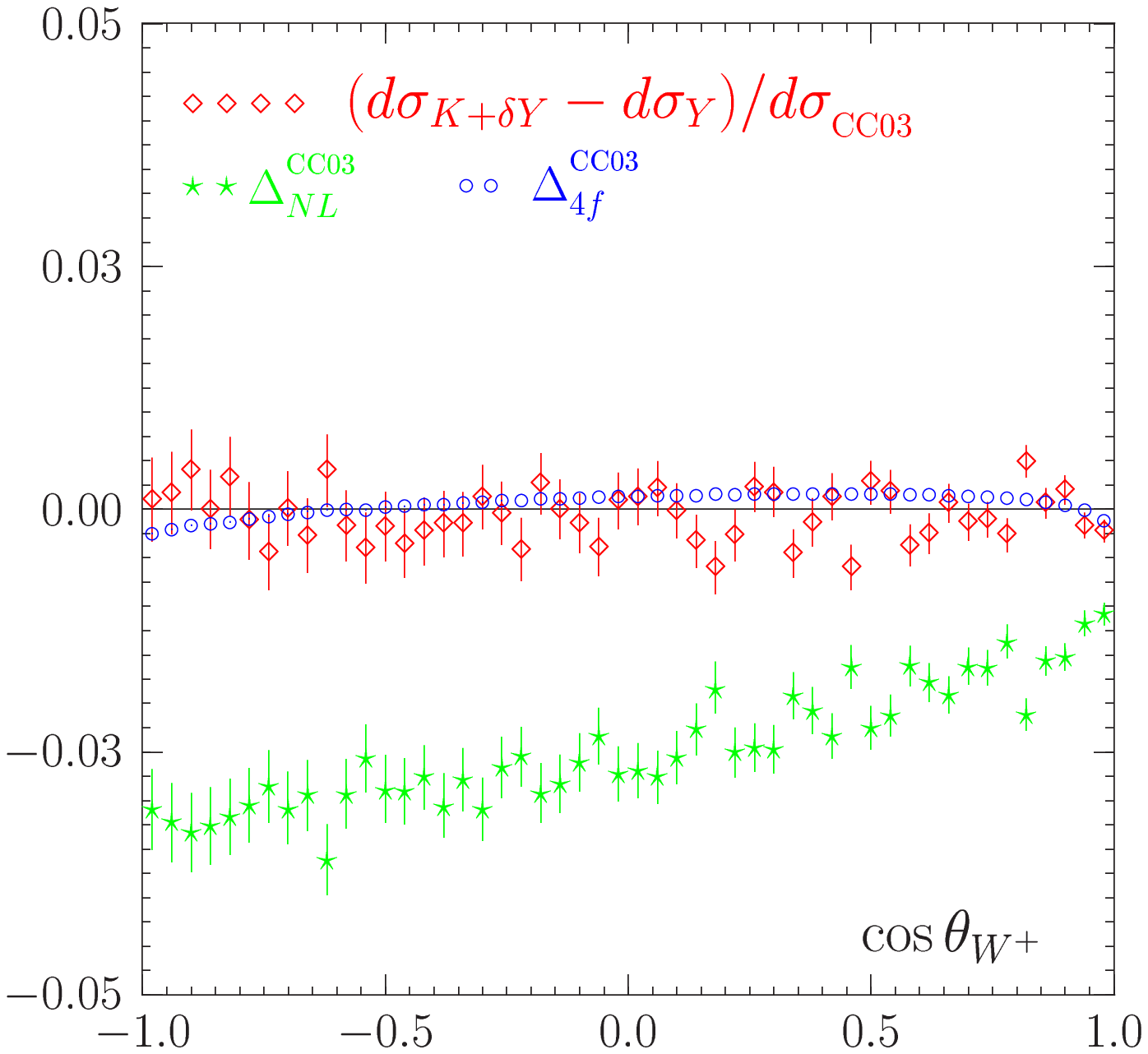    ,width=80mm,height=70mm}
}}
\end{picture}
\caption{\small\sf
  The calibration of the CMC {\tt KoralW$\&$YFSWW3} (labeled ``KandY'') 
  using {\tt YFSWW3}
  for $d\sigma^{\rm {\cal O}(\alpha)+ISR_{23}+Cc}$ at $\sqrt{s}=200$~GeV.
  Plotted are the distributions of the $W^+$ invariant mass       
  and of the $W^+$ polar angle w.r.t. the $e^+$ beam
  and their relative differences (diamonds on the lower pictures).
  We also include the plots for
  the relative size of the $4f$ background corrections (dots)
  and of the ${\cal O}(\alpha)$ NL (stars).
}
\label{fig:kandy-udmv-MWp}     
\end{figure}

\begin{figure}[!ht]
\centering
\setlength{\unitlength}{0.1mm}
\begin{picture}(1600,1480)
\put( 450,1430){\makebox(0,0)[cb]{\bf CAL05} }
\put(1250,1430){\makebox(0,0)[cb]{\bf CAL25} }
\put( 850,1430){\makebox(0,0)[cb]{\large $u \bar{d} \mu^- \bar{\nu}_{\mu}$} }
\put(  0,700){\makebox(0,0)[lb]{
\epsfig{file=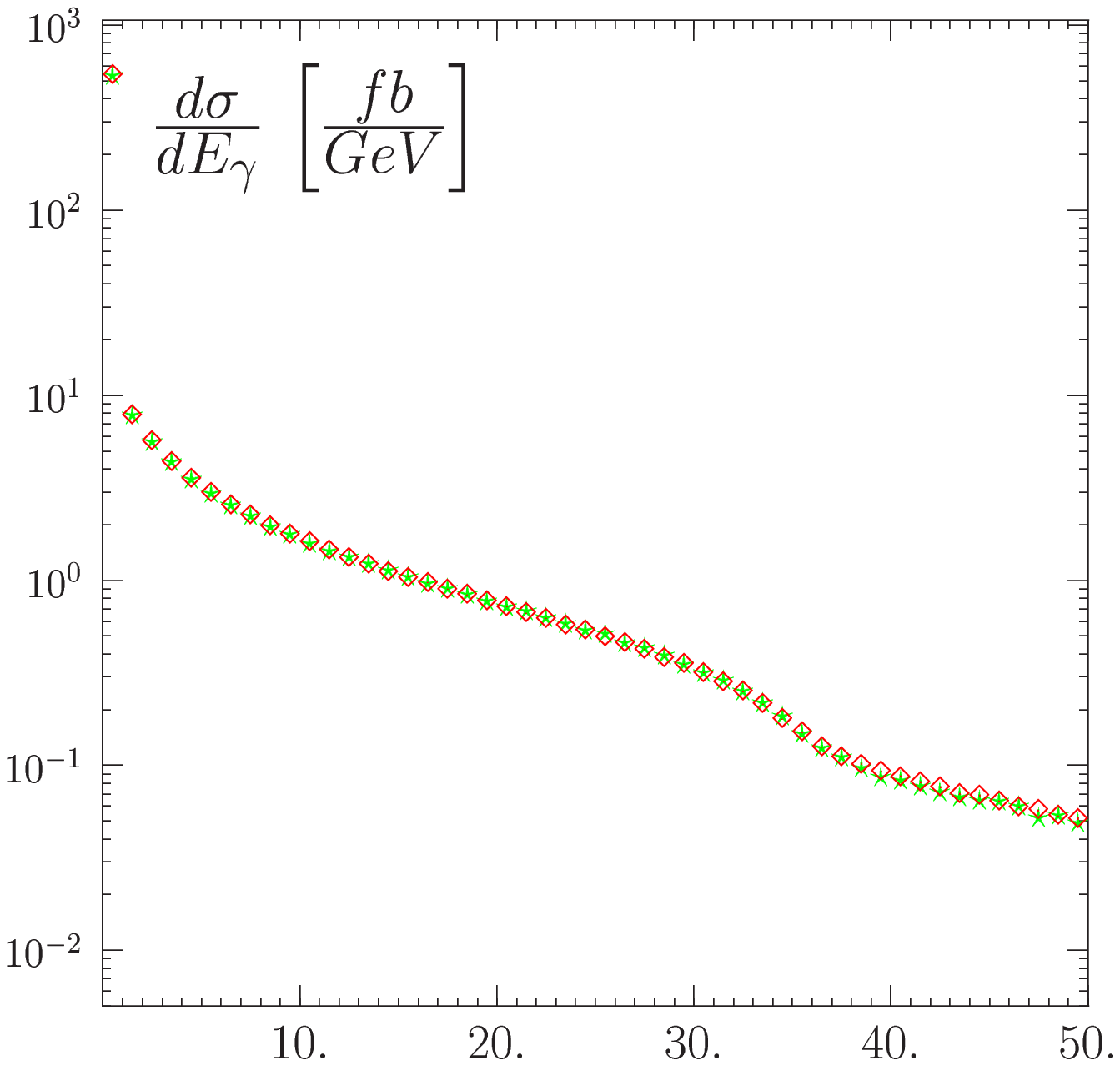           ,width=80mm,height=70mm}
}}
\put(  0,  0){\makebox(0,0)[lb]{
\epsfig{file=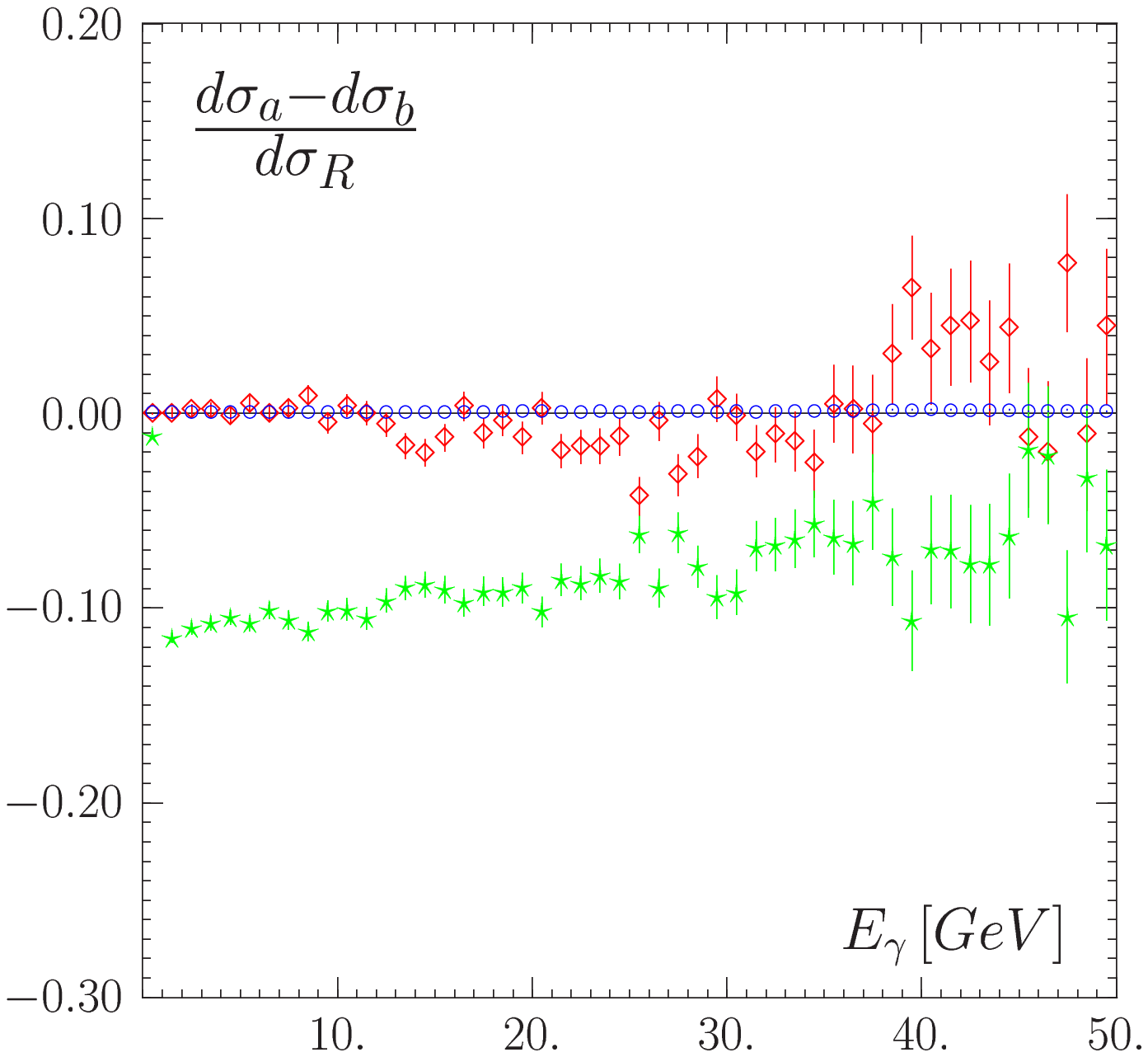       ,width=80mm,height=70mm}
}}
\put(800,700){\makebox(0,0)[lb]{
\epsfig{file=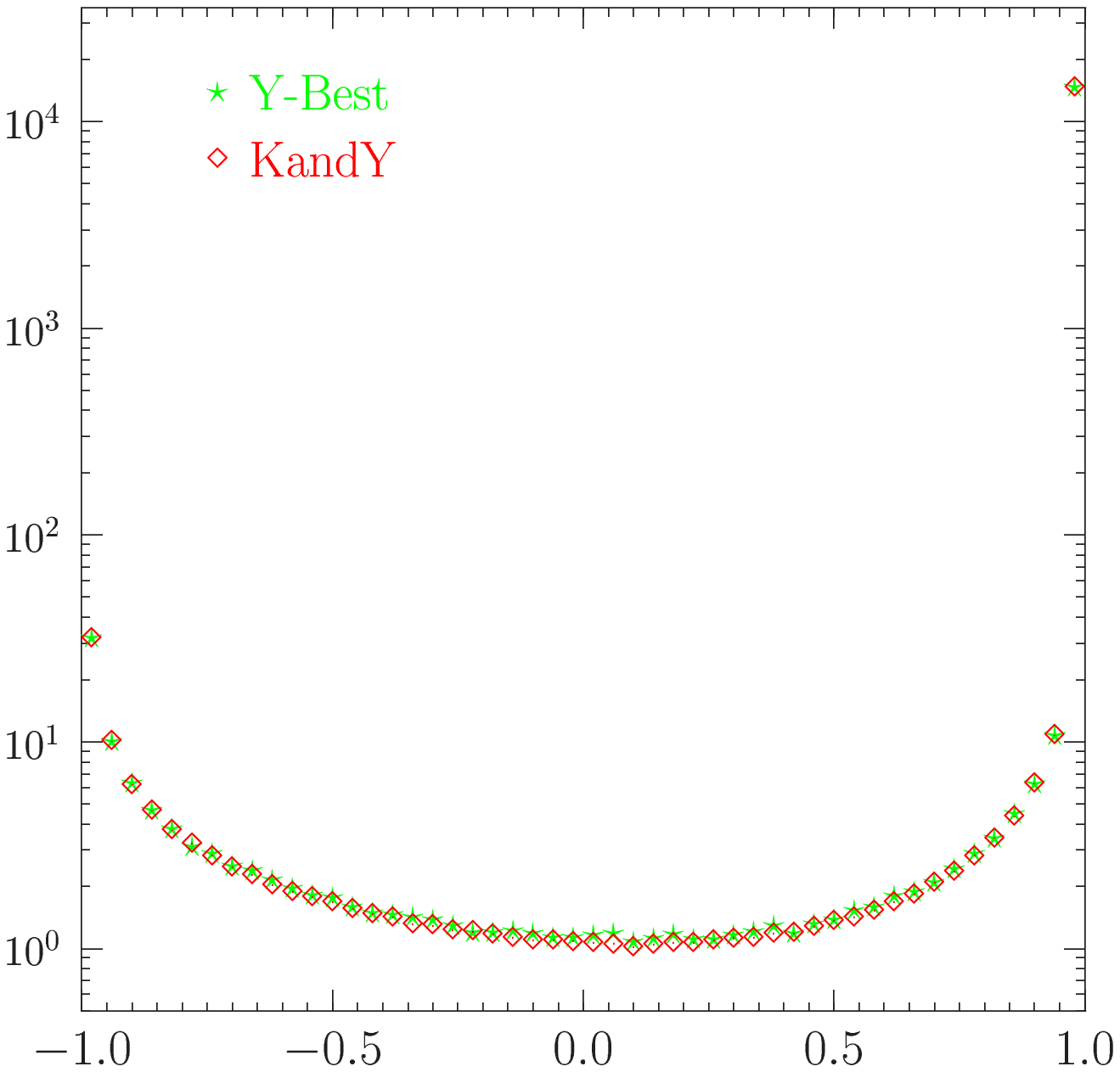       ,width=80mm,height=70mm}
\put(-300, 620){\makebox(0,0)[l]{\large                          
$\frac{d\sigma}{d \cos\theta_{\gamma} }\, [fb] $ }}             
}}
\put(800,  0){\makebox(0,0)[lb]{
\epsfig{file=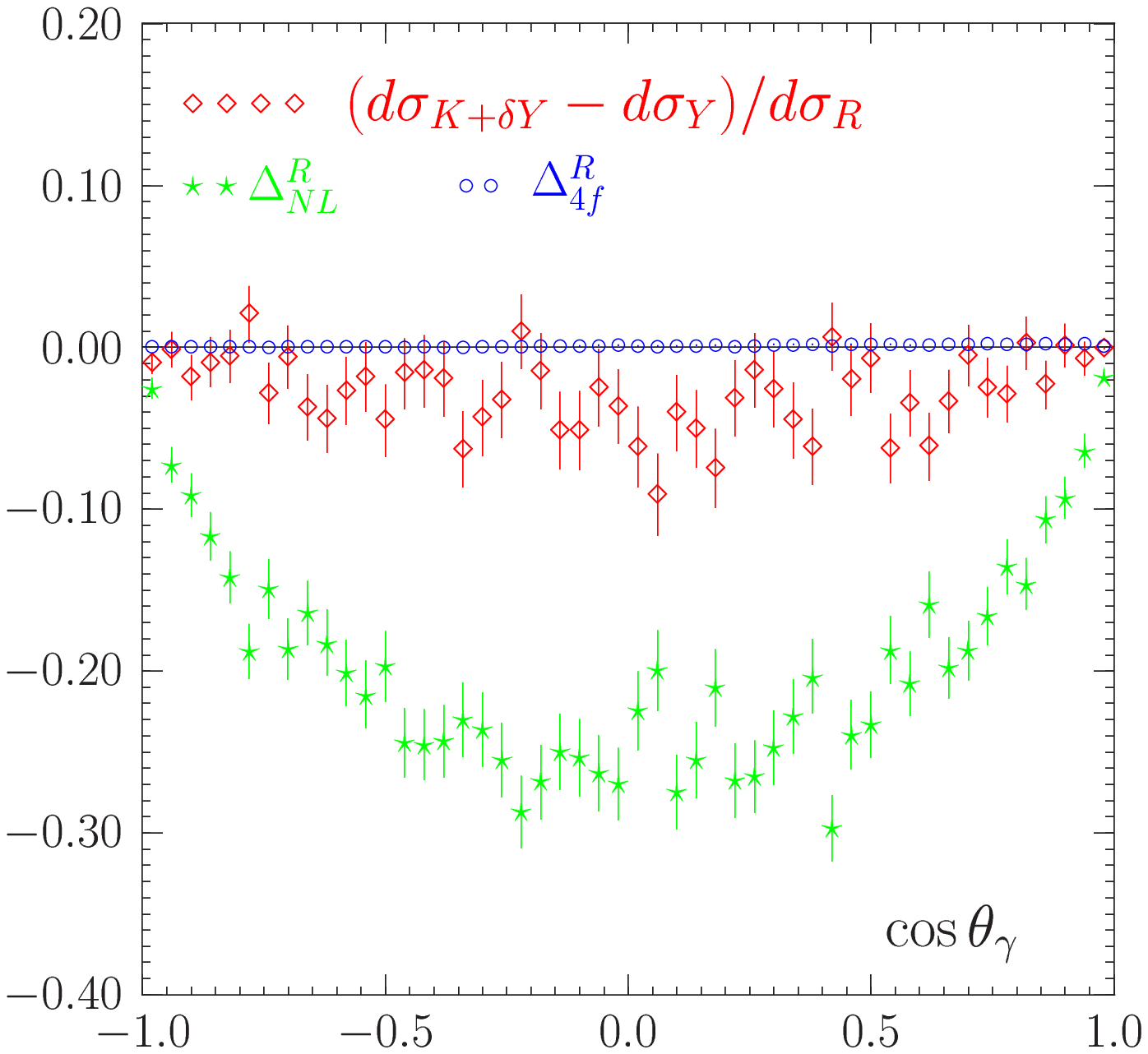   ,width=80mm,height=70mm}
}}
\end{picture}
\caption{\small\sf
  The calibration of the CMC {\tt KoralW$\&$YFSWW3} (labeled ``KandY'') 
  using {\tt YFSWW3}
  for $d\sigma^{\rm {\cal O}(\alpha)+ISR_{23}+Cc}$ at $\sqrt{s}=200$~GeV.
  Plotted are the distributions of the hardest photon energy 
  and of the photon angle w.r.t. the $e^+$ beam,
  and their relative differences (diamonds on the lower pictures).
  We also include the plots for
  the relative size of the $4f$ background corrections (dots)
  and of the ${\cal O}(\alpha)$ NL (stars).
}
\label{fig:kandy-udmv-Egam}    
\end{figure}

Finally, in Figs.~\ref{fig:kandy-udmv-MWp} and \ref{fig:kandy-udmv-Egam},
we calibrate the CMC {\tt KoralW$\&$YFSWW3} using {\tt YFSWW3}
for the distribution $d\sigma^{\rm {\cal O}(\alpha)+ISR_{23}+Cc}$
(test (B)).
Again, for the distributions of the $W$ mass and angle, of the photon energy 
and angle, we see the satisfactory agreement between the results
of the CMC {\tt KoralW$\&$YFSWW3} and of {\tt YFSWW3},
with the exception of the angular distribution of
the hardest photon, where a small deviation shows up for large angles.
We attribute it to the use of the approximate character of $\delta^R_{NL-}$
provided by {\tt YFSWW3}.
In the last two figures, we also indicate the size of the NL and four-fermion 
corrections.
For a more exhaustive discussion of the complete ``best'' results of the
CMC {\tt KoralW$\&$YFSWW3} we refer the reader to other works~\cite{yfsww3:2000a}.

\section{Details of modifications of {\tt KoralW} version {\tt 1.51}}

In this section we describe in detail all the modifications introduced
in the version {\tt 1.51} of {\tt KoralW} with respect to the previous version 
{\tt 1.41}.

In the first subsection we present the changes that 
allow the use of {\tt KoralW} together with
{\tt YFSWW3} in the form of Concurrent
Monte Carlo {\tt KoralW$\&$YFSWW3}.
%
The key modification necessary for this scheme is the
option of reprocessing by {\tt
KoralW} the events stored on some external device, 
generated earlier by {\tt KoralW} or {\tt YFSWW3}. 
In addition, a new optional ``extrapolation procedure'',
as in {\tt YFSWW3}, and a new, screened, Coulomb correction have been
implemented in {\tt KoralW} to ensure the full
compatibility of {\tt KoralW} with {\tt YFSWW3}, 
whereas a new optional normalization
scheme allows for cross-checks with other programs, such as {\tt RacoonWW} for
example.

In the next subsection we will describe modifications that will be helpful in  
the application of {\tt KoralW} to study the background to
the two-fermion processes due to the emission
of a secondary fermion pair. The idea is to calculate
separately the contribution from the real fermion-pair emission 
in the complete phase space and the virtual corrections;
see \cite{2f-LEP2YR:2000} for more details.
The corresponding virtual pair
contribution should be calculated by the
{$\cal KK$} Monte Carlo program \cite{Jadach:1999vf} and the cancellation 
between the
leading real and virtual logarithms is done numerically. 
The use of the MC program for calculating these corrections may be 
an interesting alternative to the semi-analytical approach.
More information on 
this approach to the fermion-pair corrections in
the two-fermion processes 
can be found in \cite{2f-LEP2YR:2000}.
The presented modifications of {\tt KoralW} provide a number
of approximate matrix elements (ISNS, FSNS, etc.) and a new
``extrapolation procedure'' oriented toward the $t$-channel-dominated
photonic radiation. 

Finally, in the last subsection 
we shall describe ``miscellaneous'' modifications not related
directly to any of the above subjects.

\subsection{Modifications related to $W$-pair processes 
  and communication with {\tt YFSWW3}}

\subsubsection{Switches activating reading and writing events}
We have added switches to activate and steer the process of
reading events in the form of a list of four-vectors 
from the external file, instead of generating them. 
The four-vectors must be located in the file {\tt 4vect.data.in}
and the exact format is specified in 
the subroutine {\tt reader2} in {\tt korww/karludw.f}. 
The relevant input parameter {\tt i\_disk} is set in the standard way,
as other input parameters; see Table~\ref{tab:5}.
It may be set as follows:
{\tt i\_disk=0} (default), standard MC generation of four-vectors (no reading events);
{\tt i\_disk=1}, internal tests on the four-fermion presampler;
{\tt i\_disk=2,3,4}, new settings for reading the four-vectors for FIFO.
When using FIFO
the user of the program should adjust the style formats
for the reading of four-momenta from the storage file by setting
the values 2, 3 or 4 to the variable {\tt i\_disk}.
Let us describe this organization in more detail:
\begin{description}
\item[{\tt i\_disk}=0:] 
  standard MC generation of four-vectors (no reading).
\item[{\tt i\_disk}=1:] 
  internal tests of the four-fermion presampler (special tests).
\item[{\tt i\_disk}=2 (reading from the disk file):] 
  an event is read from an external ASCII file in a format close to
  the PDG common block; four-momenta of all four final fermions,
  the number of photons and their four-momenta (if present) --
  let us call the above a ``PDG event record'';
  for the actual formats see the subroutine {\tt reader2}.
\item[{\tt i\_disk}=3 (for FIFO):]
  in this case a normal event starts with 
  a line with any single character ({\tt character*1}) other than ``E''
  followed by
  a ``PDG event record'' in the format of the {\tt i\_disk}=2 case;
  if the first line contains the character
  ``E'' then the reading program exits immediately%
  \footnote{ Such an elaborate organization is convenient (albeit not necessary)
   while communicating with another generator through FIFO.
   The normal end-of-file does not work any more for the FIFO
mechanism and the slave program would not terminate
automatically when the master program finishes. Alternatively, this
termination can be done directly by the operating system.}.
\item[{\tt i\_disk}=4 (for FIFO):]
  in this case a normal event starts with 
  a line with any single character ({\tt character*1}) other than
``R'' and ``E'' 
followed by
  the ``PDG event record'' and followed by the MC weight;
  alternatively, the event may start with the marker ``R'' in the first line,
  followed by the (almost) empty PDG record and the weight%
  \footnote{This kind of event record serves the purpose of determining
    the normalization of the integrated cross section,
    using the total number of events (rejected and accepted)
    and the value of the stored weight.};
  finally (alternatively),
  the line with marker ``E'' terminates the reading process immediately.
\end{description}
There is a corresponding switch {\tt i\_writ\_4v=2,3,4}
which, upon activation, causes {\tt KoralW} to write the file 
{\tt 4vect.data.out} in the formats corresponding 
exactly to the ones specified for the above settings of {\tt i\_disk}.

It must be kept in mind that in the mode {\tt i\_disk=2,3,4},
{\tt KoralW} will {\em not} calculate
certain parts of the differential distributions $d\sigma_K$.
In this case only the ratios of the four-fermion matrix element weights 
({\tt wtset(1-4,6-9)}) are meaningful; for example, we have 
{\tt wtset(i\_prwt)/wtset(10-i\_prwt)}$=1+\delta^R_{4f}$, 
see eq.~(\ref{eq-cor4f}). 

Note that {\tt i\_disk=2,3,4} works for the ISR as well as for the CC03.
An inclusive mixture of the final states with different flavor composition
is also allowed.

It is also sometimes useful to switch off the printouts for weights over {\tt wtmax}
by setting {\tt i\_prnt=0} while reading four-vectors from the file
(as {\tt wtmax} makes no sense in this context).

\subsubsection{Switches activating reading and writing weights}
There are two additional keys, which are steering
the process of writing and reading Monte Carlo weights from the external files
{\tt i\_writ\_wt} and {\tt i\_read\_wt}.  
The {\tt i\_writ\_wt} key activates writing weights into the file 
{\tt wtext.data.out} in the
subroutine {\tt writer\_wt} in {\tt korww/karludw.f}:
\begin{description}
\item[{\tt i\_writ\_wt}=1:] 
  Only one external weight
  {\tt wtext = wtset(i\_prwt)/wtset(10-i\_prwt)} is written. 
  The input variable {\tt i\_prwt} (see Table~\ref{tab:5}) defines
  the perturbative order of the ISR for the best (principal) weight.
\item[{\tt i\_writ\_wt}=3:] 
  The first nine entries from the {\tt wtset} matrix are written. 
\end{description}
The  {\tt i\_read\_wt} key has a twofold function:
it controls reading weights 
from the external device  (from the file {\tt wtext.data.in}) 
and it optionally activates special tests of this procedure
(of no interest for the user).
For positive values, various modes of
reading weights from the file are invoked:
\begin{description}
\item[{\tt i\_read\_wt}=1:]
  One external weight {\tt wtEXT} is read and 
  combined in an additive way 
  with the entries 1--9 of the {\tt wtset} matrix:\\
  {\tt wtset(i)=wtset(i)+wtset(10-i\_prwt)*(wtEXT - 1)}.
\item[{\tt i\_read\_wt}=2:]
  One external weight {\tt wtEXT} is read and combined in a
  multiplicative way with the entries 1--9 of the 
  {\tt wtset} matrix:
  {\tt wtset(i)=wtset(i)*wtEXT}.
\item[{\tt i\_read\_wt}=3:]
  Nine weights are read and stored as entries 1--9 of the {\tt wtset}
  matrix, overwriting the old values.
\item[{\tt i\_read\_wt}=4:]
  Nine weights are read and stored as entries 91--99 of the {\tt wtset} 
  matrix.
\end{description}
For negative values of {\tt i\_read\_wt} 
certain tests are performed on the weights from the
disk. 
\begin{description}
\item[{\tt i\_read\_wt}=--3:]
  The {\tt wtset} entries 1--9 are read and compared
  with the ones calculated directly by {\tt KoralW}. 
\item[{\tt i\_read\_wt}=--2:]
  A single {\tt wtEXT} is read and compared with the 
  four-fermion matrix element
  calculated directly by {\tt KoralW} (this is useful for checking 
  the conventions of the matrix element).
\item[{\tt i\_read\_wt}=--1:]
  A single {\tt wtEXT} is read and compared with the CC03 matrix element
  calculated directly by {\tt KoralW} (this is useful for checking the 
  conventions of the matrix element).
\end{description}
User must be aware of the high
sensitivity of the four-fermion matrix element
with respect to small changes of the
four-vectors. For example, in order to reproduce the original {\tt wtset}
weights when recalculated by {\tt KoralW} in the {\tt i\_disk > 1}
mode, an additional kinematical tune-up 
(the Lorentz boost from the LAB frame to the effective CMS$_{eff}$) 
has been permanently introduced in version {\tt 1.51}
(marked  with the special string {\tt <<<< tune-up >>>>} in the source file {\tt karlud.f}).
If four-momenta are generated by {\tt YFSWW3}, then {\tt KeyISR=2} should be
chosen (see below for more details on the new meaning of {\tt KeyISR=2})
to ensure the compatibility of the ISR ``extrapolation procedures''.

\subsubsection{New MC weights related to CC03}
Already in the previous version, {\tt 1.42}, of {\tt KoralW} 
the CC03 matrix element
was always calculated, even in the CCall mode ({\tt Key4f=1}).
In the present version, the MC weight corresponding
of the CC03 matrix element is now always provided (also for {\tt Key4f=1})
by {\tt KoralW} as {\tt wtset(10-i)},
where {\tt i=4} corresponds to the standard (best)
\Order{\alpha^3} exponentiated LL variant of the ISR distribution ($d\sigma^R$).
The other variants \Order{\alpha^{i-1}} $i=1,2,3$ are also provided for each event.
For {\tt Key4f=1} the other {\tt wtset(i)} provide, as before,
the values of the MC weights corresponding to the CCall matrix element.
The above arrangement provides an
{\em event-per-event} access to the difference CCall$-$CC03:
\begin{equation}
{\tt  wtdiff = wtcrud *(wtset(i) - wtset(10-i))},
\end{equation}
as well as the elements for the construction of the 
four-fermion correction weight
\begin{equation}
  [1+\delta_{4f}^R]_K
  = \frac{d\sigma_{K}^{4f+ISR_{1...(i-1)}+Cc}}{ d\sigma^R}
  = \frac{\vert M_{}^{4f}(\{p,q\}^{\Rcal})\vert^2}
         {\vert M_{}^{\rm CC03}(\{p,q\}^{\Rcal})\vert^2}
  = \frac{\tt wtset(i)}{\tt wtset(10-i)},
\end{equation}
see Eqs.~(\ref{eq-cor4f}) and (\ref{nohot}),
necessary for reweighting the
{\tt YFSWW3} events.

\subsubsection{New extrapolation/reduction procedure}
The next modification concerns the ``extrapolation procedure'', i.e.\ the
calculation of the Born-level four-fermion matrix element in the presence of
multiple photons. The four-fermion matrix element, defined for points
from the four-body phase space, has to be extrapolated
to points from the multibody phase space.
In other words, the multibody phase space
has to be projected onto the four-body phase space. There is a freedom in
defining this procedure and the default ones in {\tt KoralW} and {\tt
  YFSWW3} differ by a finite rotation of the effective CMS frame of the 
final-state fermions with respect to the laboratory frame,
leading to differences in some
photonic distributions (but not in the total
cross-section!). Therefore, the new procedure we introduced in {\tt
KoralW} is
{\em identical} to that in {\tt YFSWW3}. 
It is accessible under the {\tt KeyISR=2}
setting in the input cards. Note that the physical origin of this
freedom is the lack of the complete $4f+n\gamma$ matrix elements. 
They are approximated correctly in both the soft and collinear limits 
(up to $n=3$) but miss some non-leading-logarithmic corrections for transverse photons,
responsible for this ambiguity.

\subsubsection{Screened Coulomb correction}
The screened Coulomb correction as proposed in Ref.\ 
\cite{Chapovsky:1999kv} has been added. 
It can be activated by setting the {\tt KeyCul} input
parameter to {\tt KeyCul=2}. This ansatz is an efficient approximation 
of the non-factorizable corrections. It is also helpful for comparisons, with
{\tt YFSWW3}, with the screened Coulomb correction as the default option.

\subsubsection{New meanings of {\tt KeyBra} switch}
The {\tt KeyBra} switch has changed some of its meaning. Namely, the
setting {\tt KeyBra=1} has been modified. In version {\tt 1.42} it took the
arbitrary values of $W$ decay branching ratios from the input while 
fixing the value of $\alpha_S$ at $0.12$ and recalculating the $W$ width
$\Gamma_W= 3/(2\sqrt{2}\pi) M_W^3 G_\mu (1+(2/3) (\alpha_S/\pi))$.
Such an input set-up could have led to inconsistency
if the branching ratios had been changed
from the supplied default values.

The modified setting {\tt KeyBra=1} 
allows for arbitrary $W$ decay branching
ratios as well as $\alpha_S$ and $\Gamma_W$, taken without any
consistency checks and
modifications from the input cards. This
option changes also the way the CC03 matrix element is normalized.
The new normalization of the Born CC03 matrix element in the {\tt KeyBra=1} 
mode is the following.
Instead of the Br($W\to e\nu_e$) from the input, the
Standard Model value
Br$_{el} = {\alpha_W M_W}/(12\sin^2\theta_W \Gamma_W)$ is taken,
and the normalization of the decay channel $(i,j)$ is set by the factor 
$ \hbox{Br}_i\hbox{Br}_j /\hbox{Br}_{el}^2 $ with respect to the
($e\bar\nu_e,\nu_e\bar e$) decay channel.
%
%
%

These changes were necessary in order to allow for running
{\tt KoralW} with the input parameters as defined by {\tt RacoonWW} in its
comparisons with {\tt YFSWW3} in \cite{4f-LEP2YR:2000}.
In this way, {\tt KoralW} and the CMC {\tt KoralW$\&$YFSWW3} can be
directly used for the comparisons with {\tt RacoonWW}.

The ``old'' setting {\tt KeyBra=1} of the version {\tt 1.42} has been preserved
under the new setting {\tt KeyBra=3} with the only change that now the
$W$ branching ratios are {\em not} taken from the input but are
``hardwired'' into
the source code (to their default values of the version {\tt 1.42}) enforcing 
consistency with the $W$ width and $\alpha_S$, which are also set in the program.

{\tt KeyBra=2} remains unchanged.

\subsubsection{New meanings of the {\tt KeyMix} switch}
The default value of the {\tt KeyMix} 
parameter has been changed from {\tt KeyMix}=0 to
{\tt KeyMix}=1. 
This key chooses the electroweak ``Input Parameter Scheme", and the
accessible settings are
{\tt KeyMix}=0: ``LEP2 Workshop 1995'' scheme and 
{\tt KeyMix}=1: $G_\mu$-scheme. 
In the previous versions we recommended the scheme {\tt KeyMix}=0, 
worked out throughout the 1995 LEP2 Workshop. 
However in {\tt YFSWW3} only the
standard $G_\mu$-scheme is available for the ${\cal O}(\alpha)$ NL
corrections. Therefore, in order to make the two
programs compatible, we changed the default scheme to $G_\mu$ in 
{\tt KoralW} as well. 
It must be stressed that the Born level difference between
the ``LEP2''-scheme and $G_\mu$-scheme is well below
the quoted 2\% physical precision of {\tt KoralW}, and both choices are
equally legitimate. 
This difference itself is due to the ${\cal O}(\alpha)$ NL 
corrections to the Born process missing in {\tt KoralW}. 
Therefore, when the ${\cal O}(\alpha)$ NL corrections are calculated with
the {\tt YFSWW3}, the problem is solved -- now the difference
between any two schemes is due to the second-order corrections and can be
neglected.

\subsection{Modifications related to  two-fermion 
  and $t$-channel-dominated processes}

\subsubsection{New {\tt i\_sw4f} switch for selecting subgroups of
Feynman graphs in the matrix element}
The option of downgrading the complete four-fermion matrix element to
certain sub-classes
of graphs has been implemented for some final states.
It can be activated with the input
parameter {\tt i\_sw4f}. It has been installed with an eye toward
applications of {\tt KoralW} in the calculation of the background to 
the two-fermion
processes due to the emission of an additional fermion pair. Therefore the
implemented options have been motivated by the options available in the
{\tt Gentle} program \cite{Gentle}.
Denoting initial states non-singlets as ISNS and final-state
non-singlets as FSNS
(see also ref.~\cite{4f-LEP2YR:2000})
the following options are available:
\begin{description}
  \item[{\tt i\_sw4f}=--1:] all approximations set in the data parameter
             {\tt /isw4f/} in
  {\tt amp4f.f}; transmitted out with the {\tt wt4f(9)} weight,
  \item[{\tt i\_sw4f}=0:] the CC03 approximation,
  \item[{\tt i\_sw4f}=1:] the complete four-fermion (default),
  \item[{\tt i\_sw4f}=2:] the ISNS$_{\gamma+Z}$ $\tau^+\tau^-$ pair emission in
            the $e^+e^-\to \mu^+\mu^-$ process,
  \item[{\tt i\_sw4f}=3:] the FSNS$_{\gamma+Z}$ $\tau^+\tau^-$ pair emission in
            the $e^+e^-\to \mu^+\mu^-$ process,
  \item[{\tt i\_sw4f}=4:] the ISNS$_\gamma$ +FSNS$_\gamma$ $\tau^+\tau^-$ or 
            $e^+e^-$
            pair emission in the $e^+e^-\to \mu^+\mu^-$ process,
  \item[{\tt i\_sw4f}=5:] the ISNS$_\gamma$ $\tau^+\tau^-$ or $e^+e^-$ pair 
            emission in the $e^+e^-\to \mu^+\mu^-$ process,
  \item[{\tt i\_sw4f}=6:] the FSNS$_{\gamma+Z}$ $\mu^+\mu^-$ pair emission in
            the $e^+e^-\to \tau^+\tau^-$ process.
\end{description}
Further approximations can be added in the subroutine {\tt selgrf}
located in the file\\
{\tt grc4f\_init/selgrf.f}. The {\tt ibackgr}
variable denotes the value of {\tt i\_sw4f}, whereas {\tt nthprc} is
the decay channel number as defined in the {\tt amp4f} routine
(with the variable {\tt idef}).

\subsubsection{New reduction/extrapolation procedure for $t$-channel processes}
Another ``extrapolation procedure'' (accessible with the {\tt KeyISR=3}
setting) has been implemented in 
{\tt KoralW} in order to improve on photonic emission in the case of
the $t$-channel-dominated processes.  The standard option of {\tt KoralW}
(i.e.\ {\tt KeyISR=1}) for matching the ISR QED bremsstrahlung generation
with the generation of
the Born-level hard process is designed in a way that assumes that
the four-fermion process always involves a substantial
contribution from the $s$-channel
interactions. That is the preferred approach in the case of the CC03
$W$-pair production and decay.
However, in such cases as
the generation of the fermion-pair corrections to the two-fermion 
final states, it is no longer the optimal one.
Of course, in such configurations the whole assumption of the separate
treatment of the QED photonic ISR bremsstrahlung and the fermion-pair emission 
is not the best one, even for the lowest-order approximation.  
The standard {\tt KoralW} solution is however much worse, since 
it breaks the principle
of the leading-log treatment.  This leads to some pathologies as
described in the program documentation, which consist of uncontrolled
wash-out of the fermions originally generated close to the beam pipe
toward larger angles and eventually down to the acceptance region.

Fortunately this inconsistency is easy to fix, by following the
logic of the LL approximation in which the photon and 
fermion-pair emissions
should not affect each other.  This means that the bremsstrahlung should
not modify the numerical value of the small $t$-channel transfers with
which the four-fermion process matrix elements vary a lot.

The modification for the algorithm is localized in the subroutine
{\tt from\_cms\_eff} in the file {\tt karludw.f} only. The
kinematical transformation between the laboratory system and the four-fermion
rest frame CMS$_{eff}$ must depend not only on the momenta of the ISR
photons but also on the final-state fermions. The freedom of rotating
the effective CMS$_{eff}$ frame of the final-state
fermions with respect to the laboratory frame is used to ensure that
the smallest scalar product of all eight products
built out of one of the effective beams (in CMS$_{eff}$) and one of the
final state four-momenta will be identical to the same product
built out of the respective laboratory-frame four-momenta for the
same beam and the final-state fermion.

\subsection{Miscellaneous modifications}
\label{misc}

\begin{enumerate}

\item
The semi-analytical part ({\tt
  KorWan}) now supports a non-running $W$ width as well as the running one,
i.e.\ the key {\tt
  KeyWu} is now fully functional in semi-analytical calculations.

\item
Automatic resetting of the low-level cuts to zero in the mode with the CC03
matrix element has been disabled. Now the cuts must be set in the input
cards for both CC03 and four-fermion modes in an identical way.

\item 
The convention of writing four-momenta of the MIX-type CKM-suppressed final
states ($d\bar d c\bar c$, $u\bar u s\bar s$, $u\bar u b\bar b$, 
$c\bar c b\bar b$) in the common blocks 
{\tt momdec} and {\tt cms\_eff\_momdec} 
has been unified in both the CC03 and four-fermion modes to the four-fermion 
convention (as given above in parentheses). This has been done in order
to define uniquely the form of the input in the case of reading 
the four-momenta
from the external file. 
The way the Lund common block is filled and hadronization is done for 
these states has not been changed, i.e.\ it is performed as 
$WW$ in the CC03 mode and as $ZZ$ in the four-fermion mode. 
This change removes the inconsistency of notation present 
in version {\tt 1.42}.

\item
Some of the dip-switches have been replaced by the input parameters and some
new parameters have been added.
See the Appendix for the complete list of the new input parameters in the
{\tt data\_DEFAULTS} file.

\item
 The {\tt PHOTOS} package~\cite{photos:1994}
 was updated with the new version, better adjusted 
 to the present software requirements (Linux, {\tt HEPEVT}
 of the non-standard dimensionality, etc.). The functionality of the
 program was not 
 changed, except for the removal of the security check on photon emission from
 light quarks. Such an option is often in use for some $W$-pair studies, so
 even though from the physics point of view it is, as yet, 
 of not much interest,
 we leave it here to enable tests/comparisons with other
 calculations.
 The radiation from both leptons and quarks is activated by the new setting 
 {\tt ifphot=2} in the input parameters. 
 The old {\tt ifphot=1} setting has not changed its
 meaning of the photon emission from leptons only.
 The appropriate warning message is printed by the
 version of {\tt PHOTOS} included here.

 Note also that in the distribution version of {\tt TAUOLA} the parameters in
 the $\tau$  decay modes are not adjusted to the recent experimental
 data. We recommend the user to replace this version of {\tt TAUOLA}
 \cite{tauola:1992,tauola:1993} with the
 one accepted within her/his own collaboration. The technical update
 explained in Refs.~\cite{golonka:2000,pierzchala:2001} will resolve 
 this inconvenience in
 the future releases of the program.
\end{enumerate}

\section{Installation of {\tt KoralW} version {\tt 1.51}}
The version {\tt 1.51} of {\tt KoralW} is distributed as a stand-alone package
as well as in the form of
an update to the old version {\tt 1.42.3}.

\subsection{Stand-alone version}
If one plans to use {\tt KoralW} together with {\tt YFSWW3}, the
following steps should be observed:
\begin{itemize}
\item
  The main directory of the {\tt YFSWW3} program 
  should be placed next to the {\tt KoralW} main directory.
  (That means {\tt YFSWW3} should be visible from the {\tt KoralW} main 
  directory as {../\tt yfsww3-1.16-export}.)
\item
  Next, one should 
  descend into the directory {\tt demo.yfsww} of {\tt KoralW} 
  and execute the
  {\tt make yfsww3-install} command to perform the installation of the
  {\tt demo.koralw} subdirectory in the {\tt YFSWW3} main directory. The
  name of the {\tt YFSWW3} directory can be specified in the 
  {\tt makefile} file in the {\tt demo.yfsww} directory; the default name is  
  {\tt yfsww3-1.16-export}. This installation ({\tt make yfsww3-install})
  will be automatically performed when a demo run of {\tt KoralW$\&$YFSWW3} 
  is invoked by {\tt make KandY} or {\tt make XKandY}.
\item
  It may sometimes be necessary to set appropriate compiler
  flags, dependent on the operating system (the default is Linux 
   RedHat versions 6.x and 7.x). 
  To do this one must go to
  the main {\tt KoralW} directory, set all relevant options in the {\tt
  makefile} file 
  in the first place and then execute the  command {\tt make makfil}. 
  A similar operation may also be needed for
  the {\tt YFSWW3} program. 
\item
  At this moment the program is ready to use. One can
  descend again to the {\tt demo.yfsww} directory and perform various tests.
\end{itemize}

\subsection{Update of the old version}
The patch source code is located in the {\tt
koralw-1.51.x-export/demo.yfsww} directory.
To perform the update one should follow the following steps:
\begin{itemize}
\item
  Extract the directory {\tt demo.yfsww}
  from the distribution version of {\tt KoralW} 1.51
  and move it to the main directory of {\tt KoralW} 1.42.3, i.e. to 
  {\tt koralw-1.42.3-export}.
\item
  If one plans to use {\tt KoralW} together with {\tt YFSWW3}, the
  main directory of the {\tt YFSWW3} source code (at present {\tt yfsww3-1.16-export}) 
  should be placed next to the directory {\tt koralw-1.42.3-export}. 
\item
  After descending into the directory {\tt demo.yfsww} one should execute 
  the command {\tt make install} to perform the actual installation.
  Upon installation, some of the original files of
  {\tt KoralW} will be modified and, optionally, a new directory 
  {\tt demo.koralw} will be created in the {\tt yfsww3-1.16-export}
  directory (the name of the {\tt YFSWW3} directory can be changed 
  in the {\tt makefile} file in {\tt demo.yfsww} directory, the default is 
  {\tt yfsww3-1.16-export}).
  One can undo the changes by the {\tt make uninstall} command.
\item
  After installation, it is usually necessary to set appropriate compiler
  flags, dependent on the operating system. To do this one must go up to
  the main {\tt KoralW} directory, set the options in the {\tt makefile} file
  and execute the {\tt make makfil} command. 
  A similar operation may also be needed for the {\tt YFSWW3} source code. 
\item
  At this moment the source code is ready to compile and execute. One can
  descend again to the {\tt demo.yfsww} directory and perform various tests.
\end{itemize}

\subsection{Summary of the {\tt make} commands relevant to the installation}

In this subsection we summarize the options,
available in the master {\tt makefile} in the {\tt demo.yfsww}
directory, which are relevant to the installation.

\subsubsection{Stand-alone version}

\begin{itemize}
\item
    {\tt make yfsww3-install} -- installs the {\tt demo.koralw} directory in
    the {\tt YFSWW3} main directory, necessary for the CMC {\tt KoralW$\&$YFSWW3}
\end{itemize}

\subsubsection{Update of the old version}
\begin{itemize}
\item
    {\tt make install} -- creates the version {\tt 1.51} of {\tt KoralW} 
    out of the version {\tt 1.42.3} 
\item
    {\tt make update} -- copies any changes in the source directory {\tt
    src} into the {\tt KoralW-1.51} source code (useful for adding corrections)
\item
    {\tt make uninstall} -- recovers the original {\tt KoralW} version 
    {\tt 1.42.3}
\end{itemize}

\section{Organization of the source code}
The copy of FORTRAN files modified in the version {\tt 1.51} 
along with the tests specific to 
the $W$-physics are located in the
single directory 
{\tt demo.yfsww}. In this directory there are the demo program 
used for all the demo runs and the master {\tt makefile} that
defines the demo runs.
The source code FORTRAN files of the version {\tt 1.51} are located in the 
{\tt demo.yfsww/src} subdirectory. 
If the installation is performed in the form of an update --
these files will replace the corresponding
files of {\tt KoralW 1.42.3}. 
Input/output files for tests are stored in a working
directory ({\tt demo.yfsww/work}). In the following we shall briefly
describe some of the files in the {\tt demo.yfsww} subdirectory.

\begin{itemize}
\item
{\tt KWyfsww.f} is the main program for tests. It does all the
bookkeeping of
cross-sections and histograms. The actual job is done by two
subroutines: {\tt Prod1} for the Born-level comparisons and {\tt Prod3} for
the ISR comparisons.
Both routines use the set of cuts used in the comparisons between 
{\tt YFSWW3} and {\tt RacoonWW} in \cite{4f-LEP2YR:2000}
and plot several distributions for both the
CCall cross-sections and the CCall$-$CC03 difference.
\item
{\tt makefile}
is the master makefile for the installation as well as for the tests.
\item
{\tt user\_selecto.f} -- a dummy file for setting low-level cuts, 
no pre-cuts are set here.
\item
Subdirectory {\tt work}\\
In this subdirectory all the input files for the demo programs
can be found and the
output files are generated. For the exact assignment of the files to 
specific demo runs, we refer to the {\tt makefile} file. Another three 
auxiliary
files, {\tt iniseed} with a seed for the random-number generator and 
the {\tt semaphore} and {\tt semaphore.START} flag-files for restarting
the generation from a given event or to start from scratch, are also
located there.
\item
Subdirectory {\tt work.tmp}\\
Some of the tests will create this clone of the {\tt work} directory in
order to run two copies of {\tt KoralW} at the same time.
\item
Subdirectory {\tt src}\\
This subdirectory contains the FORTRAN source code of the patch for the 
version {\tt 1.51} and
the new, updated {\tt data\_DEFAULTS} file.
\item
Subdirectory {\tt krfarm}\\
Test subdirectory with the setup for running the program on a cluster of
computers. 
\end{itemize}

\section{Test and demonstration programs}
In this section we describe the test and demonstration programs
distributed in version {\tt 1.51}. These demos cover basic demo
programs for checking the correctness of the installation process, 
examples of the procedure of reading events from the disk file, and the
examples of running the Concurrent MC based on the FIFO mechanism. These latter
tests show how to set up the communication for the concurrent runs of {\tt
KoralW} with {\tt YFSWW3} ({\tt KoralW$\&$YFSWW3}) as well as of 
{\tt KoralW} with itself ({\tt KoralW$\&$KoralW}). All tests are
available as options of the master {\tt makefile} in the {\tt
demo.yfsww} directory.

\subsection{Basic demo runs (tests of the installation)}
\begin{itemize}
\item
    {\tt make KWyfsww} -- the Born-level test run with the set-up as 
    used for the 
    comparisons with {\tt YFSWW3} for the $\mu\bar\nu_\mu u\bar d$ final
    state. The 
    output file is compared with the stored benchmark (for Linux).
\item
    {\tt make KWyfswwISR} -- a similar run with the ISR.
    The output file is compared with the stored benchmark (for Linux).
\item
    {\tt make KWdisk.yfsww} -- the comparison of the CC03 matrix element of 
    {\tt KoralW} and {\tt YFSWW3}. The four-vectors generated earlier by 
    {\tt YFSWW3}
    along with the corresponding matrix element values are read from the
    disk and compared with the matrix element recalculated by {\tt
    KoralW} on an event-per-event basis. 
    By manipulating the {\tt KeyISR} switch, one can inspect the
    effect due to the change of the extrapolation procedures in {\tt KoralW}.
\end{itemize}

\subsection{Advanced technical tests (reading from the disk file)}
\begin{itemize}
\item
    {\tt make KWreadWTset} -- a complex test of recalculation of the matrix
    elements from the events stored on the disk. It consists of four sequential
    runs of {\tt KoralW}:

\begin{enumerate}
\item
     {\tt KoralW} generates the four-fermion events and writes them onto
       the disk together with the values of the {\tt wtset} weights.
\item
     {\tt KoralW} reads the generated four-vectors in the 
       CC03 mode and writes
       them back onto the disk in the CC03 mode.  
       This is a purely technical step to assure that both the CC03 and
       four-fermion modes lead to the same results (in the previous
       version, the conventions of writing four-vectors were slightly
       different for these modes, see Section \ref{misc}).
\item
     {\tt KoralW} recalculates {\tt wtset} based on the four-vectors from
       step 1. The program reports only if the relative discrepancy between
       {\tt wtset} on the disk and the recalculated one 
       is greater than $10^{-12}$.
\item
     {\tt KoralW} recalculates {\tt wtset} based on the four-vectors from
       step 2. The program reports only if the relative discrepancy between
       {\tt wtset} on the disk and the recalculated one is 
       greater than $10^{-12}$.
\end{enumerate}
\end{itemize}

\subsection{Demo of the CMC {\tt KoralW$\&$KoralW} with the FIFO mechanism}

\begin{figure}[!ht]
\centering
\setlength{\unitlength}{0.1mm}
\epsfig{file=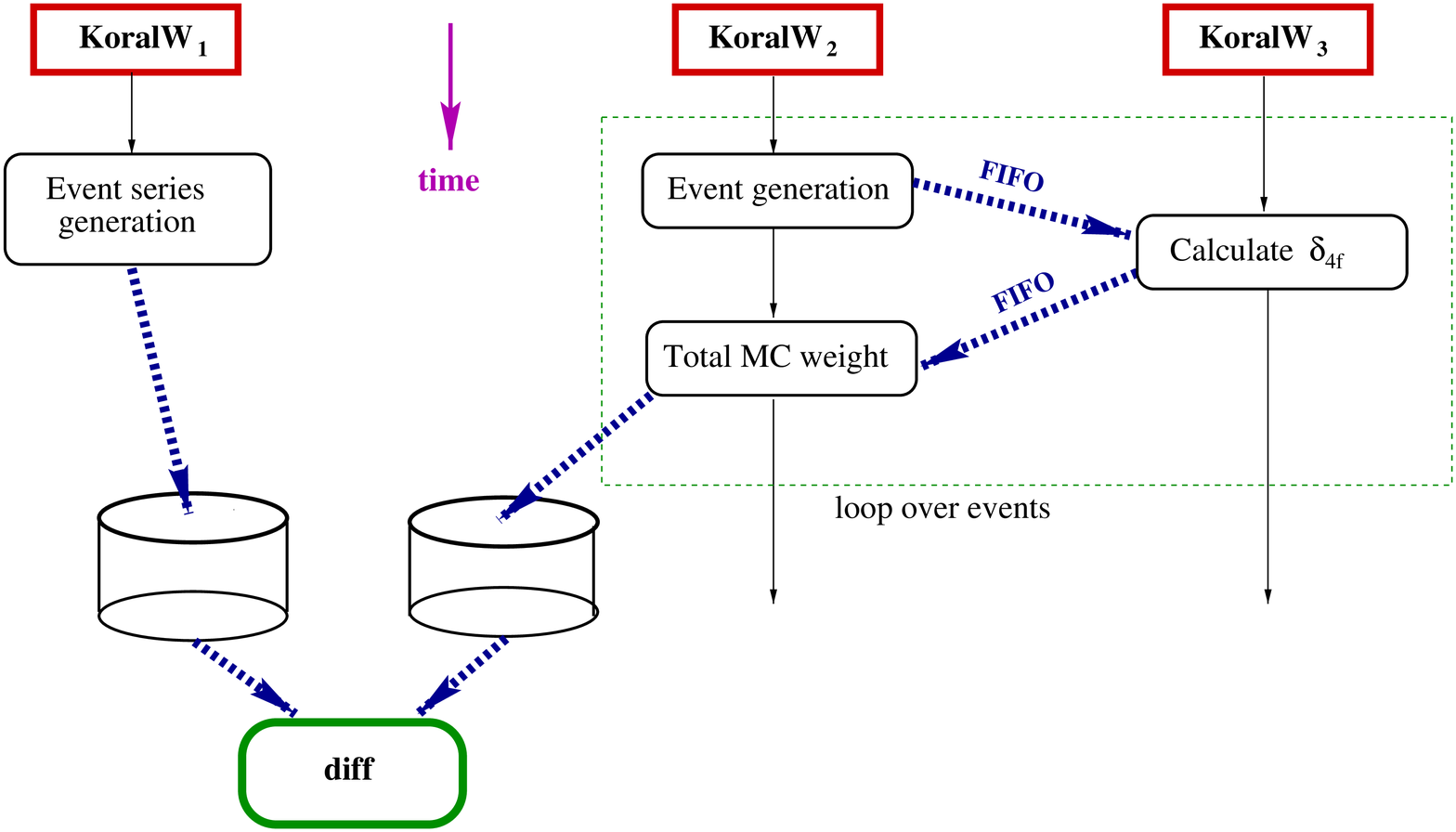,width=160mm,height=100mm}
\caption{\small\sf
  Test {\tt KWspecial} of the FIFO
  usage employing two concurrent processes of {\tt KoralW}. 
  The output is compared (using ``diff'')
  with the output of the standard run of {\tt KoralW}.
}
\label{fig:flow3}
\end{figure}
\begin{figure}[!ht]
\centering
\setlength{\unitlength}{0.1mm}
\epsfig{file=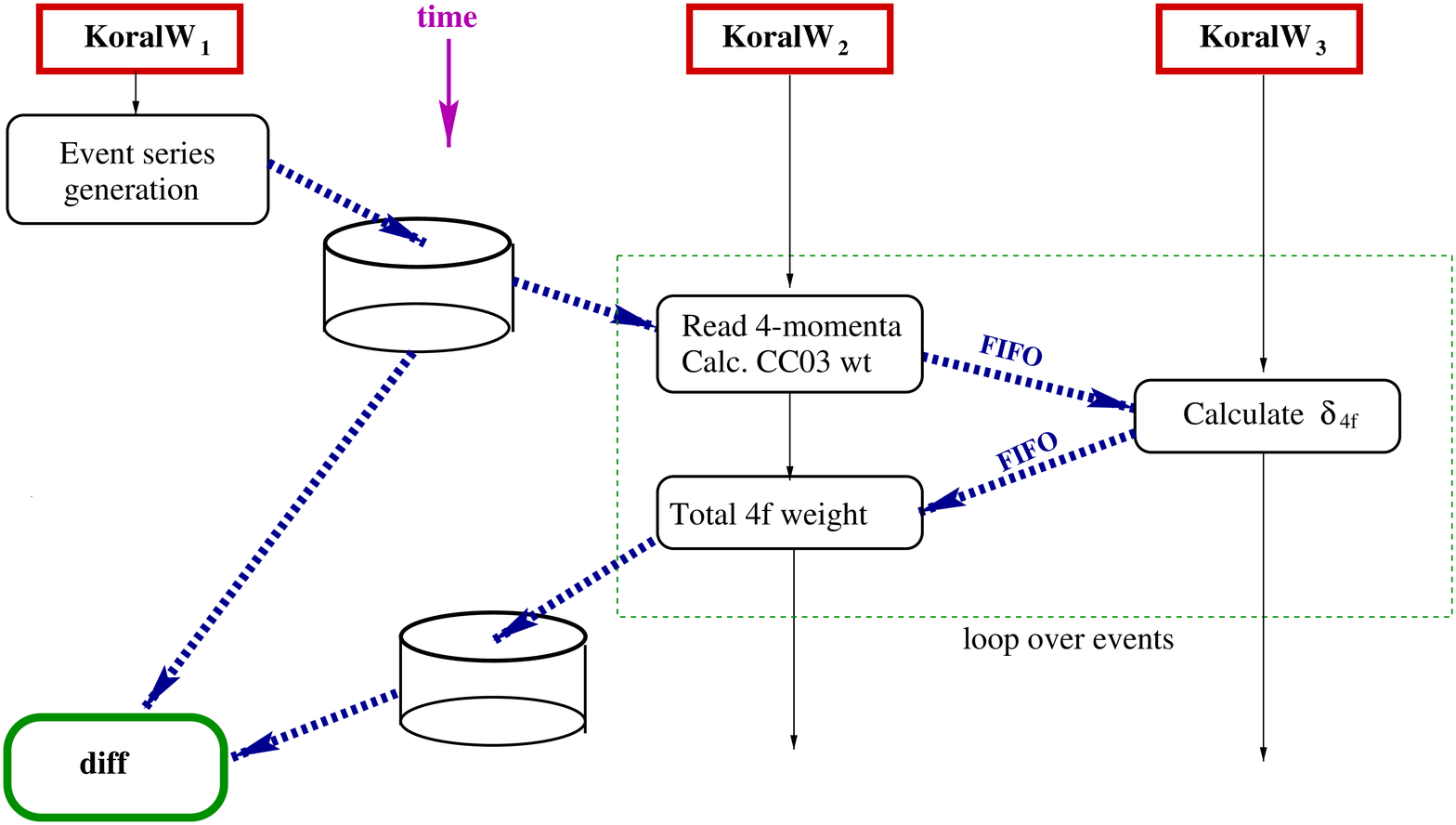,width=160mm,height=100mm}
\caption{\small\sf
  Test {\tt KWspecialDisk} of the FIFO
  usage employing two concurrent processes of {\tt KoralW}. 
  The output is compared (using ``diff'')
  with the output of the standard run of {\tt KoralW}.
}
\label{fig:flow4}
\end{figure}
\begin{itemize}
\item
       {\tt make KWspecial} -- a technical test of the FIFO mechanism
       used in the event reweighting.
       If you work in the X environment, consider {\tt make XKWspecial}
       instead.
       This important testing demo program is schematically depicted in
       Fig.~\ref{fig:flow3}.
       The demo goes as follows:
\begin{enumerate}
\item
       At the beginning, a single run of {\tt KoralW$_1$} in the
       four-fermion mode  
       is performed in order to generate a benchmark output file. 
\item
       Next, the true demo begins. The first, master, {\tt KoralW$_2$}, 
       located
       in the directory {\tt work.tmp}, is started. It generates
       a series of four-fermion events and writes the four-vectors to
       the FIFO special file {\tt 4vect.data.special}.
       After each set of four-vectors is written, the master {\tt
       KoralW$_2$} 
       suspends and waits for further input.
\item
       Simultaneously, the second, slave, {\tt KoralW$_3$} is executed,
       also in the 
       four-fermion mode, in the {\tt work} directory. 
       It reads a single set of four-vectors from the FIFO special file
       {\tt 4vect.data.special}, calculates the {\tt wtset} weights with 
       the CC03 and CCall matrix elements, writes
       {\tt wtset} down to another FIFO special file -- 
       {\tt wtext.data.special} and suspends itself.
\item
       Finally, the master {\tt KoralW$_2$} resumes execution and reads  
       from the FIFO file {\tt wtext.data.special} the weight
       {\tt wtset} calculated by the slave process. These new {\tt
       wtset}s {\em overwrite} the original values and the construction of the
       event is completed by the master process. 
\item
       After generation, the output file from the master program 
       is compared with the benchmark 
       one generated at the beginning. Apart from the value of a few 
       switches, responsible for read/write operations, there should be
       no numerical differences between the outputs.
\end{enumerate}
\item
       {\tt make KWspecialEXT} -- similar to the {\tt make KWspecial}
       demo with two exceptions: the master {\tt KoralW$_2$} runs in
       the CC03 mode and instead of the whole {\tt wtset} vector,
       only the single correction weight {\tt wtEXT} is passed back from
       the slave to the master process. This correction weight is
       then included in all {\tt wtset}s of the master process. 
       In particular, it means that all the so-called CC03 
       weights of the master process get {\em upgraded} to the
       four-fermion level.
       This is reflected in the comparison of the outputs, where the 
       ``upgraded CC03'' cross section is compared with the original
       four-fermion of the benchmark run.

\item
       {\tt make KWspecialDISK} -- similar to the {\tt make KWspecialEXT}
       demo with the only difference that the master {\tt KoralW$_2$}
       reads four-vectors from the disk instead of generating them.
       The four-vectors were generated by the introductory ``benchmark'' run
       of {\tt KoralW$_1$}. This demo shows how the CMC solution based on
       the FIFO mechanism can act as a single MC-like program, 
       reweighting events
       stored on the disk.
 
\end{itemize}

\subsection{Demo of the CMC {\tt KoralW$\&$YFSWW} with the FIFO mechanism}
\begin{itemize}
\item
       {\tt make KandY} -- this is the most important demo, showing how to
       generate four-fermion events with the ${\cal O}(\alpha)$ NL
       corrections  
       as a result of concurrent runs of {\tt KoralW} and {\tt YFSWW3}
       with the help of the FIFO (``named pipes'') mechanism. 
       If you work in the X environment, consider {\tt make XKandY} instead.
       The demo works as follows:
\begin{enumerate}
\item
          Two FIFO special files, {\tt 4vect.data.special} and {\tt
          wtext.data.special}, are created and linked to the appropriate
          input/output file names in the local working directories of
          {\tt KoralW} and {\tt YFSWW3}. The {\tt YFSWW3} distribution
          directory is located next to the {\tt KoralW} one and
          the working directory of {\tt YFSWW3} is the {\tt demo.koralw}
          one. 
          The default input data file common for both programs 
          is the one of {\tt KoralW}; two
          additional input data files, with the (optional) information
          specific to each program,
          are located in the corresponding working directories.
\item
          The master {\tt KoralW} run is started in the regular mode 
          that generates 
          four-fermion events and writes down the four-vectors of each event to 
          the FIFO special file {\tt 4vect.data.special}.
\item
          Simultaneously the slave run of {\tt YFSWW3} is started in 
          the mode that
          reads the four-vectors from the FIFO special file {\tt
          4vect.data.special} and then writes the ${\cal O}(\alpha)$ NL
          correction weight back into the other FIFO special file {\tt
          wtext.data.special}.
\item
          Finally, {\tt KoralW} reads the ${\cal O}(\alpha)$ NL 
          correction weight from the special file
          {\tt wtext.data.special}, combines it with all {\tt wtset} weights
          in an additive way and completes the generation of the events with
          both the four-fermion and ${\cal O}(\alpha)$ corrections included.
          Optionally, the rejection to the constant-weight events could be
          performed at this point.
\end{enumerate}
\end{itemize}

\subsection{X-based versions of some demos with FIFO}
\begin{itemize}
\item
       {\tt make XKWspecial} -- the version of {\tt make KWspecial} for
       the X Window System environment. 
\item
       {\tt make XKandY} -- the version of {\tt make KandY} for the X
       Window System environment. 
\end{itemize}

\subsection{Old tests from {\tt demo.14x} directory}
    After installing the package, it is also recommended to execute the 
    commands {\tt make KWdemoCC03}, {\tt make KWdemoGRCall}, 
    {\tt make KWdemo2HADR} and {\tt make KWdemo2SEMI}  
    from the {\tt demo.14x}
    directory, and to compare the outputs against the appropriate Linux 
    benchmark files. 

\section{Summary and Conclusions}

We described the new version {\tt 1.51} of the {\tt KoralW} Monte Carlo
event generator for all $e^+e^-\to f_1\bar f_2 f_3\bar f_4$ processes.
The basic physics content of the {\tt KoralW} program has not changed in
the version {\tt 1.51}, except for the reduction/extrapolation procedures.
However, a number of technical improvements have been added. 
They are motivated by the needs of the data analysis at LEP2. 
As the precision of the measurements related to the $W$-pair 
production has increased, 
it has become clear that the complete electroweak perturbative 
${\cal O}(\alpha)$ 
calculation for the $e^-e^+\to W^-W^+$ process  has to be included. 
We have found that the most economic way of including 
the missing EW ${\cal O}(\alpha)$ correction to the {\tt KoralW} Monte Carlo
event generator is to include them in the process
of generating each MC event 
with the help of an additional correcting weight provided by the {\tt YFSWW3}
program, i.e. to do it in an {\em event-per-event} manner.

This requires, of course, sending every MC event generated by {\tt KoralW}
to {\tt YFSWW3} in order to get a correction weight.
At the technical level this is organized in two ways:
\begin{enumerate}
\item
  The constant-weight events from {\tt KoralW} are stored on the disk/tape
  and reprocessed at a later time with the help of {\tt YFSWW3}
  version {\tt 1.16}
  in order to include the missing correction. 
  The appropriate facilities to read the events from disk and calculate weights are now
  implemented in {\tt KoralW} and {\tt YFSWW3}.
\item
  We also provide a new option in which {\tt KoralW} and {\tt YFSWW3}
  communicate in real time using the UNIX/Linux standard tool,
  the FIFO mechanism (named pipes), in order to correct
  each event generated by {\tt KoralW} with the weight provided by {\tt YFSWW3},
  accounting for the missing ${\cal O}(\alpha)$ corrections.
  The important advantage of this solution is that there is no need 
  to modify any part of the two source codes.
  From the user's point of view
  this solution, which we call the ``Concurrent Monte Carlo (CMC)
  {\tt KoralW$\&$YFSWW3}'', behaves as a regular Monte Carlo 
  event generator of constant-weight events with all of its normal features,
  a single input/output for example.
\end{enumerate}
It has to be stressed that the distributions and cross sections resulting
from our new CMC {\tt KoralW$\&$YFSWW3} include both
 the complete ${\cal O}(\alpha)$ corrections to the $WW$ production process
in the LPA and the corrections due to the so-called background diagrams.
The CMC {\tt KoralW$\&$YFSWW3}
fulfills almost completely the requirements of the LEP2 data analysis 
for the $WW$ process.
It is at the moment the only MC program with constant-weight events for
this kind of process.

In addition, we have made in {\tt KoralW}
certain modifications that are 
useful for the studies of the ``contamination''
of the two-fermion processes by the four-fermion processes.
As it was proposed in \cite{2f-LEP2YR:2000}, upon
adding the virtual-pair corrections from the $\cal KK$MC program,
one can obtain the QED \Order{\alpha^2} prediction (including 
the \Order{\alpha} EW corrections)
for the $e^-e^+\to f\bar{f}$ process,
for the realistic experimental event selection, 
based entirely on the Monte Carlo simulations,
without any use of the semi-analytical calculations.
To this end, in the present {\tt KoralW} version {\tt 1.51}
we introduce a new extrapolation/reduction procedure, better 
suited  to the relevant processes, which are dominated by the large $t$-channel
contributions. 
They are based on the LL concept of the independent emission
of photons and fermion pairs.

\section*{Acknowledgments}
\noindent
We would like to thank the members of the LEP2 WW/4f Working Group
(particularly R. Chierici, M. Gr\"unewald, A. Valassi, M. Verzocchi)
for numerous stimulating discussions. 
We acknowledge the kind support of the CERN TH and EP divisions
and of DESY-Zeuthen.

\newpage
\section*{Appendix:
New and Modified Program Parameters}
%
\noindent
\begin{table}[hp]
\begin{tabular}{|l|l|}
\hline
Variable & Position and meaning
\\
\hline \hline
\tt ifphot & {\tt xpar(1074)} (=1)
\\
& =2 {\tt PHOTOS} is ON, radiation from quarks and leptons (for tests)
\\
\tt KeyMix & {\tt xpar(1041)} (=1) -- default value changed to 1
\\
\tt KeyBra & {\tt xpar(1021)} (=1)
\\
& =1 arbitrary values from input, no consistency checks
\\
& =3 values pre-set in the source code
\\
\tt KeyCul & {\tt xpar(1014)} (=2)
\\
& =2 screened Coulomb correction, new default
\\
\tt KeyISR & {\tt xpar(1011)} (=1)
\\
& =2 ISR is ON, extrapolation procedure as in {\tt YFSWW3}
\\
& =3 ISR is ON, extrapolation procedure for $t$-channel-dominated
processes 
\\
\hline
\end{tabular}
\begin{caption}
{\sf The list of input parameters of the {\tt KoralW} generator 
modified in 
version {\tt 1.51}. Only the modified
settings are shown. The default values are in brackets.} 
\end{caption}
\end{table}
%
\newpage
\noindent
\begin{table}[hbp]
\begin{tabular}{|l|l|}
\hline
Variable & Position and meaning
\\
\hline \hline
\tt i\_pres & {\tt xpar(1079)} (=0)
\\
& =1 monitor of presampler probabilities is ON
\\
& =0 monitor of presampler probabilities is OFF
\\
\tt i\_yfs & ={\tt xpar(1080)} (=0)  previously dipswitch {\tt i\_yfs} 
in {\tt karludw.f}
\\
& =0 photonic internal tests OFF
\\
& =1 photonic internal tests ON
\\
\tt i\_file & {\tt xpar(1081)} (=1)  previously dipswitch {\tt i\_file} in
{\tt karludw.f}
\\
& =1 pretabulated spectra for photonic presampling
\\
& =0 spectra for photonic presampling from analytic function
\\
\tt i\_disk & ={\tt xpar(1082)} (=0)  previously dipswitch {\tt msdump} in
{\tt karludw.f}
\\
& =0 normal MC generation of events
\\
& =1 technical; 4f phase space debug mode (reads 4-vectors from disk)
\\
& =2 reading event in PDG-like format:
\\ &
four-momenta of final fermions, multiplicity and four-momenta of photons
\\
& =3 reading event in one of two possible formats: {\tt character*1} marker 
\\ &
different from ``E'' followed by PDG-like event record of {\tt i\_disk}=2, 
\\ &
or end-of-run marker ``E''.
\\
& =4 reading event in one of possible three formats:
\\ &
{\tt character*1} marker different from ``R'' and ``E'' followed by PDG-like 
\\ &
event record and MC weight, or marker ``R'' (rejected event) 
\\ &
followed by PDG-like event record and weight,
or end-of-run marker ``E''
\\
\hline
\end{tabular}
\begin{caption}
{\sf The list of new input parameters of the {\tt KoralW} generator 
version {\tt 1.51}. The default values are in brackets.} 
\end{caption}
\label{tab:5}
\end{table}

\noindent
\begin{table}[hp]
\begin{tabular}{|l|l|}
\hline
Variable & Position and meaning
\\
\hline \hline
\tt i\_prnt & ={\tt xpar(1083)} (=1)  previously dipswitch {\tt kardmp} in
{\tt KW.f}
\\
& =0 printout on weight over maximal weight OFF
\\
& =1 printout on weight over maximal weight ON
\\
\tt i\_sw4f & ={\tt xpar(1084)} (=1)
\\
& =--1 all approximations set in data set {\tt /isw4f/} in {\tt amp4f.f}
\\
& =0 CC03
\\
& =1 complete four-fermion 
\\
& =2 ISNS$_{\gamma+Z}$ $\tau^+\tau^-$ pair emission in
            $e^+e^-\to \mu^+\mu^-$ process.
\\
& =3 FSNS$_{\gamma+Z}$ $\tau^+\tau^-$ pair emission in
            $e^+e^-\to \mu^+\mu^-$ process.
\\
& =4 ISNS$_\gamma$ +FSNS$_\gamma$ $\tau^+\tau^-$ or $e^+e^-$
            pair emission in $e^+e^-\to \mu^+\mu^-$ process.
\\
& =5 ISNS$_\gamma$ $\tau^+\tau^-$ or $e^+e^-$ pair emission
            in $e^+e^-\to \mu^+\mu^-$ process.
\\
& =6 FSNS$_{\gamma+Z}$ $\mu^+\mu^-$ pair emission in
            $e^+e^-\to \tau^+\tau^-$ process.
\\
\tt i\_prwt & ={\tt xpar(1085)} (=4)  
            previously dipswitch {\tt i\_principal\_weight} in {\tt KW.f}
\\
& =2 first-order ISR in principal weight for rejection
\\
& =3 second-order ISR in principal weight for rejection
\\
& =4 third-order ISR in principal weight for rejection
\\
\tt i\_writ\_wt & ={\tt xpar(1086)} (=0) 
\\
& =0 writing weights on disk OFF
\\
& =1 {\tt wtext} written on disk
\\
& =2 {\tt wtset(1-9)} written on disk
\\
\tt i\_writ\_4v & ={\tt xpar(1087)} (=0)
\\
& =0 writing four-vectors on disk OFF 
\\
& =2 four-vectors written on disk, format as in {\tt
i\_disk=2}; 
\\ &
(readable by {\tt i\_disk=2}) 
\\
& =3 four-vectors written on disk, format as in {\tt
i\_disk=3};
\\ &
     (readable by {\tt i\_disk=3})
\\
& =4 four-vectors written on disk, format as in {\tt
i\_disk=4}; 
\\ &
     (readable by {\tt i\_disk=4})
\\
\tt i\_read\_wt &={\tt xpar(1088)} (=0) 
\\
& =0 reading weights from disk OFF
\\
& =1 reads {\tt wtext}, combines additively
\\
& =2 reads {\tt wtext}, combines multiplicatively
\\
& =3 reads {\tt wtset(1-9)}, overwrites original values
\\
& =4 reads {\tt wtset(1-9)}, puts them into {\tt wtset(91-99)}
\\
& =--3 tests: reads {\tt wtset(1-9)} and compares with original 
\\
& =--2 tests: reads {\tt wtext} and compares with original 4f weight 
\\
& =--1 tests: reads {\tt wtext} and compares with original CC03 weight 
\\
\hline
\end{tabular}
\begin{caption}
{\sf (cont.): The list of new input parameters of the {\tt KoralW} generator 
version {\tt 1.51}. The default values are in brackets. }
\end{caption}
\end{table}


\newpage

\newpage  

\newpage  

\newpage  

\newpage

\newpage  
\bibliographystyle{prsty}

\end{document}